\documentclass[11pt,a4]{article}
\usepackage{amsmath,amsthm,amssymb}
\usepackage{mathrsfs}
\usepackage[all]{xy}
\usepackage{graphicx}
\usepackage{float}
\usepackage{lscape}

\usepackage[top=25truemm,bottom=25truemm,left=25truemm,right=25truemm]{geometry}

\numberwithin{equation}{section}

\theoremstyle{definition}
\newtheorem{df}{Definition}[section]
\newtheorem{prop}[df]{Proposition}
\newtheorem{ex}[df]{Example}
\newtheorem{thm}[df]{Theorem}
\newtheorem{rem}[df]{Remark}
\newtheorem{cor}[df]{Corollary}
\newtheorem{lem}[df]{Lemma}
\newtheorem{conj}[df]{Conjecture}
\newtheorem{fact}[df]{Fact}
\newtheorem{nota}[df]{Notation}

\newcommand{\Proof}{\proof}

\newtheorem*{thm*}{Theorem}
\newtheorem*{conj*}{Conjecture}

\newcommand{\bra}[1]{\left\langle #1 \right| }
\newcommand{\ket}[1]{\left| #1 \right\rangle }
\newcommand{\braket}[2]{\left\langle #1 | #2  \right\rangle }
\newcommand{\tbra}[1]{\langle #1 | }
\newcommand{\tket}[1]{| #1 \rangle }
\newcommand{\tbraket}[2]{\langle #1 | #2 \rangle}
\newcommand{\NPb}{{\rlap{\raise.4ex\hbox{$\scriptscriptstyle\bullet$}}
\lower.4ex\hbox{$\scriptscriptstyle\bullet$}}}
\newcommand{\NPc}{{\rlap{\raise.4ex\hbox{$\scriptscriptstyle\circ$}}
\lower.4ex\hbox{$\scriptscriptstyle\circ$}}}
\newcommand{\seteq}{\mathbin{:=}}
\newcommand{\rseteq}{\mathbin{=:}}
\newcommand{\twopii}{2 \pi \sqrt{-1} }

\newcommand{\chargQ}{Q}

\newcommand{\vl}{\vec{\lambda}}
\newcommand{\vm}{\vec{\mu}}
\newcommand{\vn}{\vec{\nu}}
\newcommand{\lo}{\lambda^{(1)}}
\newcommand{\lt}{\lambda^{(2)}}
\newcommand{\lN}{\lambda^{(N)}}
\newcommand{\mo}{\mu^{(1)}}
\newcommand{\mt}{\mu^{(2)}}
\newcommand{\mN}{\mu^{(N)}}
\newcommand{\vu}{\vec{u}}
\newcommand{\vv}{\vec{v}}
\newcommand{\vw}{\vec{w}}
\newcommand{\Xo}{X^{(1)}}

\newcommand{\tX}{\tilde{X}}
\newcommand{\tXo}{\tilde{X}^{(1)}}
\newcommand{\tXt}{\tilde{X}^{(2)}}
\newcommand{\tL}{\tilde{\Lambda}}
\newcommand{\tLo}{\tilde{\Lambda}^1}
\newcommand{\tLt}{\tilde{\Lambda}^2}
\newcommand{\tP}{\tilde{P}}
\newcommand{\tK}{\tilde{K}}
\newcommand{\tN}{\tilde{N}}
\newcommand{\tM}{\widetilde{M}}
\newcommand{\tPhi}{\tilde{\Phi}}
\newcommand{\tT}{\tilde{T}}
\newcommand{\overstar}[1]{\mathop{\overset{*}{ #1 }}}
\newcommand{\overwstar}[1]{\mathop{\overset{**}{ #1 }}}

\newcommand{\bi}{b^{(i)}}
\newcommand{\bj}{b^{(j)}}

\newcommand{\ao}{a^{(1)}}
\newcommand{\at}{a^{(2)}}
\newcommand{\ai}{a^{(i)}}
\newcommand{\aN}{a^{(N)}}

\newcommand{\hi}{h^{(i)}}
\newcommand{\hj}{h^{(j)}}

\newcommand{\sa}{\mathsf{a}}

\newcommand{\sai}{\sa^{(i)}}
\newcommand{\saj}{\sa^{(j)}}

\newcommand{\Qi}{Q^{(i)}}

\newcommand{\QN}{Q^{(N)}}
\newcommand{\lc}{\mathsf{lt}}

\newcommand{\vr}{\vec{r}}
\newcommand{\vs}{\vec{s}}
\newcommand{\lrs}{\lambda_{\vr,\vs}}

\newcommand{\be}{\begin{equation}}
\newcommand{\ee}{\end{equation}}
\newcommand{\ba}{\begin{eqnarray}}
\newcommand{\ea}{\end{eqnarray}}
\date{}

\begin{document}

\thispagestyle{empty}

$\,$

\vspace{25mm}

\begin{center}

{\LARGE {\bf Singular Vector of Ding-Iohara-Miki Algebra and Hall-Littlewood Limit of
5D AGT Conjecture} \\ 
}

\end{center}

\vspace{15mm}

\begin{center}
{\Large {\bf Yusuke Ohkubo }}
\end{center}

\vspace{90mm}

\begin{center}
PhD thesis submitted to Nagoya University\\
Graduate School of Mathematics \\
March 2017
\end{center}

\newpage

\thispagestyle{empty}

\vspace*{\stretch{1}}

\begin{abstract}
In this thesis, 
we obtain the formula for the Kac determinant of the algebra 
arising from the level $N$ representation of the Ding-Iohara-Miki algebra. 
This formula can be proved by decomposing the level $N$ representation 
into the deformed $W$-algebra part and the $U(1)$ boson part, and 
using the screening currents of the deformed $W$-algebra. 
It is also discovered that 
singular vectors obtained by its screening currents 
correspond to the generalized Macdonald functions. 
Moreover, 
we investigate the $q \rightarrow 0$ limit of 
five-dimensional AGT correspondence. 
In this limit, 
the simplest 5D AGT conjecture is proved, 
that is, 
the inner product of the Whittaker vector of the deformed Virasoro algebra 
coincides with the partition function of the 5D pure gauge theory. 
Furthermore, 
the R-Matrix of the Ding-Iohara-Miki algebra is explicitly calculated, 
and its general expression in terms of the generalized Macdonald functions 
is conjectured.

\end{abstract}

\vspace{\stretch{2}}
\pagebreak

\setcounter{page}{1}

\tableofcontents

\newpage

\section{Introduction}

{\bf {\Large 1.1.}} 
The Jack symmetric polynomials \cite{
Jack:1970, Stanley:1989} 
are a system of orthogonal polynomials 
expressing the excited states 
of an integrable one-dimensional quantum many-body system 
with the trigonometric type potential 
called the Calogero-Sutherland model 
\cite{Sutherland:1971:version1, Sutherland:1971:version2}.%
\footnote{ \label{footnote:Hbeta} 
To be precise, 
the Jack polynomials are eigenfunctions of 
a Hamiltonian $H_{\beta}$ which is obtained by a 
certain transformation of the Calogero-Sutherland Hamiltonian. 
Here $\beta$ is a parameter appearing in the Calogero-Sutherland model. 
The excited states can be constructed from the Jack polynomials. 
} 
These are one-parameter deformations of the Schur symmetric polynomials. 
In general, being integrable 
means that the model has sufficiently many conserved quantities, 
and that system can be analytically solved. 
Like the Calogero-Sutherland model, 
many of the integrable systems are not physical models 
of particles existing in the real world. 
However, the mathematical structure of the integrable models, 
e.g., excellent solvability, 
can be used to advantage in many fields of mathematics.

Let us consider symmetric functions 
which are defined as a projective limit of 
symmetric polynomials with finite variables \cite[Chap.\ 1]{Macdonald}. 
In the case of the Jack polynomials, 
the infinite-variable limit exists and is called 
the Jack symmetric functions. 
The Jack functions are parametrized by partitions or Young diagrams, 
and has the complex parameter $\beta$ 
(see also Footnote \ref{footnote:Hbeta}). 
Actually we can consider the parameter $\beta$ 
as an indeterminate, 
and then the Jack functions 
are defined over the field $\mathbb{Q}(\beta)$. 
The surprising result due to Mimachi and Yamada is 
that the Jack functions 
associated to rectangular Young diagrams 
have a one-to-one correspondence with singular vectors of the Virasoro algebra 
\cite{MimachiYamada:1995}. 
The Virasoro algebra is constructed by the infinitesimal conformal transformations in two dimensions, 
and is the Lie algebra generated by $L_n$（$n\in \mathbb{Z}$） 
and the central element $c$ satisfying the relations 
\begin{equation}
[L_n,L_m]=(n-m)L_{n+m}+ c \: \frac{n(n^2 - 1)}{12}\delta_{n+m,0} ,\quad n,m \in \mathbb{Z}, 
\end{equation}
\begin{equation} 
[L_n,c]=0,\quad n \in \mathbb{Z}. 
\end{equation} 
This is an essential algebra 
to two-dimensional conformal field theories 
required for string theory and statistical mechanics. 
To obtain the irreducible representations of the Virasoro algebra 
is important not only in representation theory but also 
in the conformal field theories. 
The irreducibility of highest weight representations can be determined 
by special vectors called singular vectors 
in the highest weight representation. 
Although the singular vectors have an integral representation, 
the expression formula of the Jack functions by the Dunkl operator \cite{LapointeVinet:1996}
is more useful. 
Further, various properties of Jack functions are known.
Thus, 
the expression of the singular vectors by the Jack functions 
is very convenient and beneficial.

As a $q$-difference deformation of the Jack polynomials, 
there is a system of orthogonal polynomials 
with rich theory called the Macdonald polynomials \cite{Macdonald}. 
For later use 
let us introduce the notation for 
Macdonald symmetric function,
which is the infinite-variable version 
of the Macdonald polynomial. 
We denote by $P_{\lambda}(p_n;q,t)$ 
the Macdonald symmetric function 
associated to the partition $\lambda$. 
Here $q$ and $t$ are free parameters, 
and they can be considered as complex numbers 
or indeterminates. 
In this paper, 
we regard the power sum symmetric functions $p_n$ 
as variables of the Macdonald functions 
(for more detail, see Appendix \ref{sec: Macdonald and HL}). 
The Macdonald polynomials are also 
simultaneous eigen-functions of 
commuting $q$-difference operators, 
now called Macdonald difference operators.
Let us also mention that they are 
related to the Ruijsenaars model \cite{Ruijsenaars} 
which is a relativistic extension of the Calogero-Sutherland model. 
The $q$-deformation like the Macdonald functions makes 
theory clearer and often mathematically easier to handle. 
For example, 
the Jack functions can be characterized as the Hamiltonian $H_{\beta}$ 
(see Footnote \ref{footnote:Hbeta}), 
but they have ​​degenerate eigenvalues, 
and difficulties arise when we prove their orthogonality 
and ​​coincidence with the singular vectors. 
In the theory of the Macdonald functions, 
this degeneracy problem can be eliminated and the discussion is clearer. 
Also, 
the Hamiltonian $H_{\beta}$ 
has an infinite number of commuting operators. 
However, 
it is difficult to write down these operators explicitly \cite{Sekiguchi}, 
and in the Macdonald theory we have an explicit formula for the commuting 
family of difference operators having $P_{\lambda}(p_n;q,t)$ as simultaneous 
eigenfunctions. 
For the above reason, 
it can be said that Macdonald's theory is more beautiful. 

In the $q \rightarrow 1$ ($t=q^{\beta}$) limit with $\beta$ fixed, 
the Macdonald functions are reduced to the Jack functions. 
On the other hand, 
in $q \rightarrow 0$ limit with $t$ fixed, 
they are reduced to the symmetric functions called the Hall-Littlewood functions. 
The Hall-Littlewood functions have a close connection to 
the character of the general linear group over finite fields, 
and they are also a generalization of the Schur functions \cite[Chap.\ III]{Macdonald}. 
It is one of the advantages that it is possible to unify and generalize 
the two generalizations of the Schur functions. 
Some applications in knot invariants 
\cite{AS:2011:Knot, Cherednik:2013:Jones, GN:2013:Refined}
and stochastic processes \cite{BC:2014:Macdonald} 
are also known. 
The Macdonald functions are one of the important symmetric functions 
for modern mathematics.

\begin{figure}[H]
\begin{center}
\includegraphics[width=6cm]{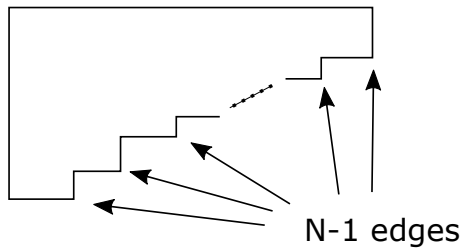}
\caption{Young diagram with $N-1$ edges}
\label{fig:YoungDiag_N-1edges}
\end{center}
\end{figure}

Awata, Kubo, Odake and Shiraishi
introduced in \cite{SKAO:1995:quantum}
a $q$-deformation of the Virasoro algebra,
which is named the deformed Virasoro algebra. 
This deformed algebra is designed so that singular vectors 
of Verma modules correspond to Macdonald symmetric functions 
$P_{\lambda}(p_n;q,t)$. 
The deformed Virasoro algebra is an associative algebra 
defined over the base field $\mathbb{Q}(q,t)$,
where $q$ and $t$ are the same parameters as in $P_{\lambda}(p_n;q,t)$. 
The generators are denoted by $T_n$ ($n \in \mathbb{Z}$),
and the defining relation is 
\begin{equation}
[T_n, T_m] = -\sum_{l=1}^{\infty}f_l(T_{n-l}T_{m+l}-T_{m-l}T_{n+l})
             -\frac{(1-q)(1-t^{-1})}{1-p}(p^n-p^{-n}) \delta_{n+m,0},
\end{equation}
where $p\seteq q/t$ and $f_l$ are the structure constants defined by 
\begin{equation}
f(z) = \sum_{l=0}^{\infty}f_l \:  z^l
 \seteq \exp \left( \sum_{n=1}^{\infty}\frac{1}{n}\frac{(1-q^n)(1-t^{-n})}{1+p^n}z^n \right). 
\end{equation}
It is shown that the singular vectors of the deformed Virasoro algebra 
coincide with the Macdonald functions 
associated with rectangular Young diagrams.   
It is also possible to obtain the Jack and Macodnald functions 
associated with general partitions  
from the singular vectors of the $W_N$-algebra 
and the deformed $W_N$-algebra 
(which is the (deformed) Virasoro algebra when $N = 2$) 
\cite{MimachiYamada:RIMSkokyuroku, AKOS:1995Excited, Awata:1995Quantum}. 
To be exact, 
singular vectors of the (deformed) $W_N$-algebra 
can be realized by $N-1$ families of bosons 
under the free field representation. 
By a certain projection to one of these bosons, 
we can obtain 
the Jack (or Macdonald) functions associated with Young diagrams 
with $N-1$ edges  
(see Figure \ref{fig:YoungDiag_N-1edges}).  

\noindent 
{\bf {\Large 1.2.}} 
The representation theory of the Virasoro algebra 
plays an essential role in the two-dimensional conformal field theories. 
In 2009, while studying the low energy effective theory of M5-branes, 
Alday, Gaiotto and Tachikawa discovered the correspondence between 
the correlation functions of two-dimensional conformal field theories and 
the partition functions of four-dimensional supersymmetric gauge theories (AGT conjecture) 
\cite{alday2010liouville}. 
Gauge theory has a long history and is an attractive theory 
studied by a lot of mathematicians and physicists. 
Although it is difficult to calculate the partition functions 
of gauge theories in general, 
Nekrasov gave an explicit formula (Nekrasov formula) 
for the instanton partition function of four-dimensional $\mathcal{N}=2$ 
supersymmetric gauge theory in 2002 \cite{nekrasov2003seiberg}. 
The Nekrasov formula $Z_{\mathrm{Nek}}(\Lambda)$ is written by 
the summation of the terms $Z_{\vec{Y}}$ parametrized 
by tuples of Young diagrams:
\begin{equation}\label{eq:Nek formula}
Z_{\mathrm{Nek}}(\Lambda)
= \sum_{n=0}^{\infty} \Lambda^n \sum_{\left| \vec{Y} \right|=n} Z_{\vec{Y}}.
\end{equation}
These terms are given in a factorized form, 
and as $n$ increases, 
the amount of calculation becomes enormous. 
However, it can be calculated by a simple combinatoric method. 
The discovery of \cite{alday2010liouville} is the following relation 
between two-dimensional and four-dimensional field theories. 
The Nekrasov formula for the four-dimensional $SU(2)$ gauge theory 
with four matters in (anti-)fundamental representation 
(actually, it is the Nekrasov formula of the $U(2)$ gauge theory 
divided by the $U(1)$ factor $Z_{\mathrm{Nek}}/Z^{U(1)}$ )
coincides with 
the four-point conformal block of the two-dimensional conformal field theory.

Basics of the conformal field theories were established 
by Belavin, Polyakov and Zamolodchikov (BPZ) in 1984 \cite{BPZ:1984}. 
They described the critical phenomenon of the two-dimensional Ising model 
which is a model of the ferromagnet, and so on. 
The primary fields $V(z)$ are operators on the representation space 
of the Virasoro algebra such that 
\begin{equation}
[L_n, V(z)]=z^n\left( z\frac{\partial}{\partial z} + h(n+1) \right) V(z),
\quad z, h\in \mathbb{C}. 
\end{equation}
The primary fields are the main research object 
in the conformal field theories. 
Here $h$ is called the conformal dimension of the primary field. 
Furthermore, 
in the conformal field theories, 
it is a fundamental problem to calculate the correlation functions 
of the primary fields
. 
Generally, in the quantum field theories, 
the calculations of correlation functions are difficult, 
and usually it is often solved by approximation. 
BPZ succeeded in determining the exact forms of correlation functions 
in the conformal field theories. 
In particular, 
they derived differential equations with regular singularities 
for the correlation functions.

However, the research by BPZ was performed 
mainly for primary fields with the special conformal dimension, 
i.e. the minimal models, 
and they did not investigate the correlation functions 
in general forms. 
Even if we derive the differential equations of the correlation functions, 
it is difficult to find their solutions. 
From the standpoint of conformal field theories, 
the AGT conjecture that states the agreement 
between the Nekrasov formulas and the conformal blocks 
(originally in the Liouville theory, that is 
the theory having the primary field 
with generic conformal dimensions 
\footnote{
Also the AGT conjecture using the Minimal models is studied in 
\cite{BF:2014:AGT}. 
To be exact, contribution of the Heisenberg algebra is added 
to the Minimal models. 
}) 
is studied under the expectation 
that general formulas for the correlation functions can be obtained.

Various extensions were made immediately 
after the AGT conjecture was discovered. 
First of all,  
the original AGT conjecture 
deals with the four-dimensional gauge theory 
in the case that the number of (anti-)fundamental matters $N_f$ 
is $4$. 
Immediately after this original conjecture \cite{alday2010liouville}, 
the cases with $N_f=0,1,2,3$ were studied in \cite{Gaiotto:2009}. 
These cases can be obtained from the case of $N_f=4$ 
by applying the same degenerate limits to the Nekrasov formula and the conformal block. 
Especially when $N_f=0$, 
the conformal block degenerates to the inner product 
of the vector $\ket{G_{\mathrm{vir}}}$ called the Whittaker vector  
of the Virasoro algebra.%
\footnote{ 
Also in the $N_f=1,2$ case, 
the degenerate conformal blocks 
can be realized by the inner product of certain vectors 
that are the general form of the vector $\ket{G_{\mathrm{vir}}}$. 
} 
Moreover, 
it is also expected that the four-dimensional gauge theories 
with the higher gauge group $SU(N)$ correspond with the $W_N$-algebra 
\cite{Wyllard:2009}.

The Jack functions and the Macdonald functions 
also play an important role in the AGT conjecture. 
For example, 
the expansion coefficients of the Whittaker vector $\ket{G_{\mathrm{vir}}}$ 
by the Jack functions 
are clarified \cite{Yanagida:2010Whittaker}. 
In addition, 
it is known that a good basis called AFLT basis 
\cite{alba2011combinatorial,FL:2011Integrable, belavin2011agt} 
can be regarded 
as a sort of generalization of the Jack functions. 
The AFLT basis is a basis in the representation space 
of the algebra $\mbox{(Virasoro algebra)}\otimes \mbox{(Heisenberg algebra)}$, 
which is first introduced by Alba, Fateev, Litvinov and Tarnopolskiy, 
and the conformal block can be combinatorially expanded by this basis. 
The AFLT basis is an orthogonal basis 
which parametrized by pairs of Young diagrams $\vec{Y}$. 
In the $W_N$ algebra case, it is parametrized by 
$N$-tuples of Young diagrams and exists 
in the representation space of the algebra 
$\mbox{($W_N$ algebra)}\otimes \mbox{(Heisenberg algebra)}$. 
By inserting the identity 
$1=\sum_{\vec{Y}} 
\frac{\tket{\vec{Y}}\tbra{\vec{Y}}}{\braket{\vec{Y}}{\vec{Y}}}$ 
with respect to the AFLT basis $\{\tket{\vec{Y}}\}$, 
the calculation of correlation functions 
$\langle V(z_1) \cdots V(z_n)\rangle$  
is attributed to that of the matrix element 
$\tbra{\vec{Y}} V(z) \tket{\vec{W}}$, 
where $V(z)$ is a sort of the primary field defined by some 
relations with generators of the Virasoro algebra and the Heisenberg algebra. 
Then the three-point functions $\tbra{\vec{Y}} V (z)\tket{\vec{W}}$ 
are factorized and coincide with the significant factors 
called the Nekrasov factors,  
which compose the Nekrasov formula. 
Namely, 
if we expand the correlation functions by using the AFLT basis, 
then the form of its expansion is quite the same 
as that of the Nekrasov formula (\ref{eq:Nek formula}). 
Further, the conformal block of the algebra 
$\mbox{(Virasoro algebra)} \otimes \mbox{(Heisenberg algebra)}$ 
coincides with the partition function $Z_{\mathrm{Nek}}(\Lambda)$ of $U(2)$ 
gauge theory. 
Actually, 
such a good basis does not exist in the representation space 
of the Virasoro algebra. 
Since the $U(1)$ factor contributes and complicates the AGT conjecture, 
we need the adjustment by the Heisenberg algebra. 
Since the AFLT basis correspond to the torus fixed points 
in the instanton moduli space, 
it is also called the fixed point basis.

In \cite{morozov2013finalizing},  
the original AGT conjecture is "proved" 
with the help of the AFLT basis (the generalized Jack functions) 
and the free field representation.%
\footnote{The AGT conjecture are proved in the case of $N_f=0,1,2$ in 
\cite{Hadasz:2010, Yanagida:2010Norms} by using Zamolodchikov reccursion relation. Some proofs from geometric representation theory are also 
given in \cite{MaulikOkounkov:2012, Schiffmann:2013,BFN:2014:Instanton}.} 
However, this "proof" is based on another conjecture. 
To explain it in more detail, 
recall that the free field representation of the conformal blocks 
can be written by the Dotsenko-Fateev integral
$\langle   F  \rangle$, 
where $\langle \quad \rangle$ means some integrals 
of the integrand $F$. 
Then $F$ can be expanded 
by a sum of the products of the generalized Jack functions $J_{\vec{Y}}$ 
and their dual functions $J^*_{\vec{Y}}$, 
which are parametrized by tuples of Young diagrams. 
This expansion formula is called the Cauchy formula. 
At that time,  
it was conjectured that the integral value of each term
$\langle J_{\vec{Y}} J^*_{\vec{Y}} \rangle$ 
directly corresponds to $Z_{\vec{Y}}$ in the Nekrasov formula. 
This is the scenario of the "proof." 
Although this proof is straightforward without using recurrence formulas etc, 
since the integral value of the generalized Jack functions 
is still a conjecture, 
it is necessary to prove it in order to complete this proof. 
For that, 
we need to investigate more properties of the generalized Jack functions.

$q$-deformed version of the AGT conjecture is also provided.%
\footnote{ 
Elliptic deformations of the AGT conjecture 
are also proposed in \cite{Nieri:2015:elliptic, IKY:2015Elliptic}. 
}
That is, the deformed Virasoro/$W$-algebra is related 
to five-dimensional gauge theories (5D AGT conjecture) 
\cite{AwataYamada1, AwataYamada2}. 
In the simplest case, 
it is shown that the inner product of the Whittaker vector 
of the deformed Virasoro algebra coincides with 
the instanton partition function (K-theoretical partition function) 
of the five-dimensional $\mathcal{N}=1$ pure $U(2)$ gauge theory. 
Also the same approach as \cite{morozov2013finalizing} is taken  
in the $q$-deformed case. 
In other words, 
it is conjectured that the $q$-deformed Dotsenko-Fateev integral 
corresponds to the partition function with $N_f=4$ matters, 
and this conjecture is checked 
by using the generalized Macdonald functions \cite{Zenkevich:2014}. 
The $q$-deformed version of the AFLT basis \cite{awata2011notes} 
(that is, the generalized Macdonald functions) 
exists in the representation space of the level $N$ representation 
of the Ding-Iohara-Miki algebra (DIM algebra).

The DIM algebra (explained in Appendix \ref{sec:Def of DIM}) 
has the face of a $q$-deformation of the $W_{1+\infty}$ algebra 
as introduced by Miki in \cite{Miki:2007}, 
and the deformed Virasoro/$W$-algebra appear  
in its representation \cite{FHSSY}. 
Since the DIM algebra has a lot of background, 
there are a lot of other names 
such as quantum toroidal $\mathfrak{gl}_1$ algebra 
\cite{FJMM:2015:Quantum, FJMM:2016:Finite}, 
quantum $W_{1+\infty}$ algebra \cite{Bourgine:2016Coherent}, 
elliptic Hall algebra \cite{BS:2012:I} and so on. 
The DIM algebra has a Hopf algebra structure 
which does not exist in the deformed Virasoro/$W$-algebra, 
and the DIM algebra is associated with the Macdonald functions 
having rich theory.%
\footnote{ 
In the study of the algebraic structure of the 
operator $\eta(z)$ 
that is free field representation of the Macdonald's difference operator, 
it is discovered that 
$\eta(z)$ form a part of representation of the DIM algebra \cite{FHHSY}. 
See also Fact \ref{fact:lv. 1 rep of DIM}. 
}
Unlike the case of the generalized Jack functions, 
the generalized Macdonald functions 
can be constructed by the coproduct of the DIM algebra 
\cite{awata2011notes}. 
It is a surprising phenomenon 
that the structure of the coproduct of the DIM algebra has information 
on the partition functions of the five-dimensional gauge theories. 
Furthermore, 
in the $q$-deformed case, 
Awata-Kanno's and Iqbal-Kozkaz-Vafa's refined topological vertices 
\cite{AK:2008Refined, IKV:2007} 
are also reproduced by the matrix elements of some intertwining operator 
of the DIM algebra, 
and the coincidence between the correlation function of the DIM algebra 
and the 5D Nekrasov formula is proved \cite{awata2012quantum}. 

The AGT conjecture with respect to the $q$-deformed AFLT basis 
\cite{awata2011notes} 
(recalled in Section \ref{sec:Reargument of DI alg and AGT}) 
is almost parallel to the undeformed case, 
and it suffices to consider the algebra 
$\langle X^{(i)}_n \rangle$, denoted by $\mathcal{A}(N)$, 
which is generated by certain operators 
$X^{(i)}_n$ ($i=1,\ldots , N$, $n\in \mathbb{Z}$) 
obtained by the level $N$ representation of the DIM algebra. 
The level $N$ representation is 
that on a Fock module $\mathcal{F}_{\vu}$ 
with the highest weight $\vu=(u_1, \ldots, u_N)$. 
The vertex operator 
$\Phi(z):\mathcal{F}_{\vu} \rightarrow \mathcal{F}_{\vv}$ on this Fock module 
is defined by the relation (Definition \ref{df:DIM vertex op})
\begin{equation}
(X^{(i)}_n- e_{N}(\vv) z X^{(i)}_{n-1})\Phi(z) 
= \Phi(z) (X^{(i)}_n-(t/q)^i e_{N}(\vv) z X^{(i)}_{n-1}), 
\end{equation}
where $e_{N}(\vv):=v_1 v_2 \cdots v_N$. 
$\Phi(z)$ can be regarded as 
an analog of the Virasoro primary field. 
The generalized Macdonald functions 
are defined to be the eigenfunctions of the generator $\Xo_0$ 
constructed by the copoduct of the DIM algebra. 
Then, 
it is conjectured that 
the matrix elements of $\Phi(z)$ with respect 
to the generalized Macdonald functions 
reproduce the five-dimensional Nekrasov factors. 
Under this conjecture, 
the four-point conformal block of $\Phi(z)$ corresponds to the 
5D $U(2)$ Nekrasov formula with $N_f=2N$ matters.

\vspace{1.5em}

\noindent 
{\bf {\Large 1.3.}} 
The first main theorem in this thesis is the formula for 
the Kac determinant of the algebra 
$\mathcal{A}(N)$ 
(Theorem \ref{thm:KacDet}): 

\begin{thm*}
\begin{align}\label{eq:Kac det in Intro}
\det\left( \braket{X_{\vl}}{X_{\vm}} \right)_{|\vl|=|\vm|=n}=
&\prod_{\vl \vdash n} \prod_{k=1}^N b_{\lambda^{(k)}}(q) b'_{\lambda^{(k)}}(t^{-1})
\\
& \times \prod_{\substack{1\leq r,s\\ rs\leq n}}
\left( 
(u_1 u_2 \cdots u_N)^{2}
 \prod_{1\leq i < j \leq N} (u_i-q^st^{-r}u_j)(u_i-q^{-r}t^s u_j) \right)^{P^{(N)}(n-rs)}, \nonumber
\end{align}
where 
$b_{\lambda}(q) \seteq \prod_{i\geq 1} \prod_{k=1}^{m_i} (1-q^k)$, 
$b'_{\lambda}(q) \seteq \prod_{i\geq 1} \prod_{k=1}^{m_i} (-1+q^k)$, and   
$P^{(N)}(n)$ 
denotes the number of the $N$-tuples of Young diagrams of the size $n$. 
For the definition of $m_i=m_i(\lambda)$, 
see Notations in the latter part of this section. 
\end{thm*}

This determinant can be proved 
by using the fact that the generators $X^{(i)}_n$ 
can be decomposed into the deformed $W$-algebra part and the $U(1)$ part 
by a linear transformation of the bosons, 
and using the screening currents of the deformed $W$-algebra. 
By this formula, 
we can solve the conjecture \cite[Conjecture 3.4]{awata2011notes} that 
the following PBW type vectors of the algebra $\mathcal{A}(N)$ 
(Definition \ref{df:ordinary PBW vct}) 
are a basis: 
\begin{align} 
 &\ket{X_{\vec{\lambda}}} \seteq 
  X^{(1)}_{-\lambda^{(1)}_1} X^{(1)}_{-\lambda^{(1)}_2} \cdots  X^{(2)}_{-\lambda^{(2)}_1} X^{(2)}_{-\lambda^{(2)}_2}\cdots X^{(N)}_{-\lambda^{(N)}_1} X^{(N)}_{-\lambda^{(N)}_2} \cdots \ket{\vec{u}}. 
\end{align}

We also discover that singular vectors of the algebra 
$\mathcal{A}(N)$ 
correspond to some generalized Macdonald functions 
as the second main theorem. 
By this result, 
we can get singular vectors from generalized Macdonald functions.%
\footnote{
Whether the singular vectors considered in this thesis, e.g., 
$\ket{\chi_{\vr,\vs}}$  
can express all singular vectors of the algebra $\mathcal{A}(N)$ 
is incompletely understood.
However, 
the Kac determinant can be proved by 
the only vanishing points given by the singular vectors 
$\tket{\chi^{(i)}_{r,s}}$ (see (\ref{eq:sing vct chi(i)rs})) 
corresponding to the simple roots, 
because the determinant has $\mathfrak{sl}_N$ weyl group invariance. 
}
The singular vectors are intrinsically 
the same as those of the deformed $W$-algebra. 
However, 
as the projection of the bosons is necessary 
for the correspondence with the ordinary Macdonald functions, 
the result of this thesis that does not need projections 
can be thought to be a 
generalization of \cite{Awata:1995Quantum}. 
As a corollary of this fact, 
we can find a new relation of the ordinary Macdonald functions 
and the generalized Macdonald functions 
by the projection of the bosons. 
Furthermore, since screening operators are written by integrals, 
we can also get an integral representation of generalized Macdonald  functions.

\begin{figure}[H]
\begin{center}
\includegraphics[width=14cm]{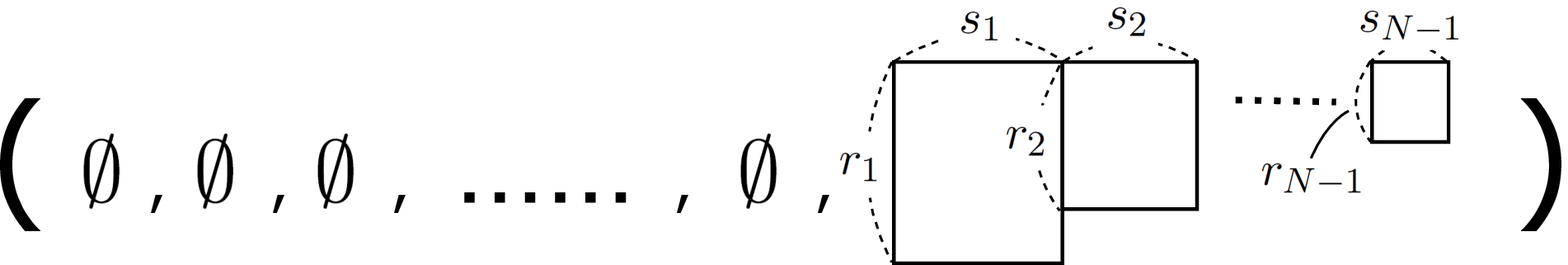}
\caption{Young diagram corresponding to singular vector. (A)}
\label{fig:YoungDiag_OnlyRightSide}
\end{center}
\end{figure}

Concretely, 
the vector $\ket{\chi_{\vec{r}, \vec{s}}}$ defined to be 
\begin{equation}
\ket{\chi_{\vec{r}, \vec{s}}} \seteq  
\oint \prod_{k=1}^{N-1} \prod_{i=1}^{r_k} dz^{(k)}_i 
S^{(1)}(z^{(1)}_1) \cdots S^{(1)}(z^{(1)}_{r_1}) \cdots 
S^{(N-1)}(z^{(N-1)}_1) \cdots S^{(N-1)}(z^{(N-1)}_{r_{N-1}}) 
\tket{\vv}
\end{equation}
is a singular vector. 
Here $S^{(i)}(z)$ denotes the screening operator, 
the $N$-tuple of parameters $\vv=(v_1, \ldots , v_N)$ is 
a function of $\tilde{\alpha}^{(k)}$, 
and for non-negative integers $s_k$, $r_k \geq r_{k+1} \geq 0$, 
\begin{equation}
\tilde{\alpha}^{(k)}=\tilde{\alpha}^{(k)}_{\vec{r}, \vec{s}}=\sqrt{\beta} (1-r_k +r_{k+1})-\frac{1}{\sqrt{\beta}}(1+s_k) ,\quad r_N=0 
\end{equation}
(for more details, see Section \ref{sec:sing vct and Gn Mac}). 
The singular vector $\ket{\chi_{\vec{r}, \vec{s}}}$ 
coincides with the generalized Macdonald function with the $N$-tuple of 
Young diagrams in Figure \ref{fig:YoungDiag_OnlyRightSide} 
(Theorem \ref{thm:Sing vct and Gn Mac 2}. (A). (main theorem)). 
In fact, Figure \ref{fig:YoungDiag_N-1edges} means 
the same Young diagram being on the rightmost side in Figure 
\ref{fig:YoungDiag_OnlyRightSide}. 
Hence the projection of this generalized Macdonald function 
corresponds to the ordinary Macdonald functions 
associated with the rightmost Young diagram with $N-1$ edges in Figure 
\ref{fig:YoungDiag_OnlyRightSide} 
(Corollary \ref{cor:projection of Gn Mac}).

\begin{figure}[H]
\begin{center}
\includegraphics[width=14cm]{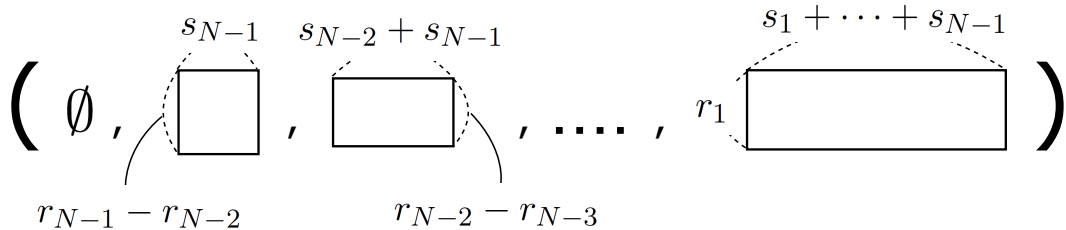}
\caption{Young diagram corresponding to singular vector. (B)}
\label{fig:YoungDiag_SomeRectangle}
\end{center}
\end{figure}

When the condition $r_k \geq r_{k+1}$ for 
the number of screening currents and parameter $\tilde{\alpha}^{(k)}$ 
in $\ket{\chi_{\vec{r}, \vec{s}}}$ 
is removed, 
the above figure is not a Young diagram. 
However it turns out that 
the vector $\ket{\chi_{\vec{r}, \vec{s}}}$ coincides with 
the generalized Macdonald function 
obtained by cutting off the protruding part 
and moving boxes to the Young diagrams on the left side.
For example, if $0\leq r_k < r_{k+1}$ for all $k$, 
the corresponding $N$-tuple of Young diagrams of the generalized Macdonald function is Figure \ref{fig:YoungDiag_SomeRectangle} 
(Theorem \ref{thm:Sing vct and Gn Mac 2}. (B). (main theorem)). 

\vspace{1.5em}

\noindent 
{\bf {\Large 1.4.}} 
Furthermore, 
we investigate behavior in the limit to the Hall-Littlewood functions, 
$q \rightarrow 0$, 
of the deformed Virasoro algebra and 
the algebra $\mathcal{A}(N)$. 
Also 5D AGT conjecture is studied in this limit. 
The reason of considering such a limit is that 
the situation becomes simple and some problems are solved. 
In particular, 
the simplest 5D AGT conjecture can be proved, 
and PBW type vectors can be expressed in terms of Hall-Littlewood functions. 
By virtue of the theory of Hall-Littlewood functions, 
we can obtain and prove an explicit formula (Theorem \ref{thm:main theorem}) for the four-point correlation function of a certain operator 
$\tPhi^{\vv}_{\vu}(z):\mathcal{F}_{\vu} \rightarrow \mathcal{F}_{\vv}$, 
which is the limit $q \rightarrow 0$ of the vertex operator $\Phi(z)$ 
associated with $\mathcal{A}(2)$.  
Here, $\mathcal{F}_{\vu}$ is 
the Fock module with the highest weight $\vu=(u_1, u_2)$.

\begin{thm*}
\begin{equation}
\bra{\vw} \tPhi^{\vw}_{\vv}(z_2) \tPhi^{\vv}_{\vu}(z_1) \ket{\vu}  
= \sum_{\lambda} \left( \frac{u_1u_2z_1}{w_1w_2z_2} \right)^{|\lambda|} 
\frac{ \prod_{k=1}^{\ell(\lambda)} \left(1- t^{k-1} \frac{w_1w_2}{v_1v_2} \right) }{t^{2n(\lambda)} b_{\lambda}(t^{-1}) }. 
\end{equation}
Here for a partition $\lambda$, 
$n(\lambda) \seteq \sum_{i\geq 1} (i-1) \lambda_i$, 
and $b_{\lambda}$ is the same one in (\ref{eq:Kac det in Intro}). 
\end{thm*}

The function $\bra{\vw} \tPhi(z_2) \tPhi(z_1) \ket{\vu} $ can be 
calculated by the generalized Hall-Littlewood functions in the same way as \cite{awata2011notes}. 
However, 
we can obtain this formula by inserting the identity 
with respect to the PBW type vectors.

We call this Hall-Littlewood limit $q \rightarrow 0$  `crystallization' 
after the use of the quantum groups \cite{Kashiwara:1989Crystalization}, 
where the parameter $q$ represents the temperature 
in the RSOS model \cite{LP:1996:Multipoint} 
which has symmetry of the deformed Virasoro algebra, 
and the limit $q\rightarrow 0$ can be considered 
as the zero temperature limit. 
Although our studies are mathematically different 
from the notion of the original crystal base of quantum groups, 
the physical meaning and the motivation to simplify phenomena are the same. 
To investigate their mathematical relationship is 
an interesting open problem. 
On the other hand, 
little is known about the physical meaning of the Hall-Littlewood limit 
in the gauge thoery at present.

\vspace{1.5em}

\noindent 
{\bf {\Large 1.5.}} 
In this thesis, 
the R-matrix of the DIM algebra is also investigated. 
The result with respect to the R-matrix 
is based on the collaborative researches 
\cite{AKMMMOZ:2016:Toric, AKMMMOZ:2016:Anomaly}, 
and only works of the author is described. 
In general, 
a R-matrix is defined as a solution of the Yang-Baxter equation, 
and is closely related to the solvable lattice models, 
the knot invariants and so on. 
Further, 
it is well-known that R-matrices can be constructed by 
Hopf algebras such as the quantum groups. 
In general, 
a Hopf algebra $H$ with the coproduct $\Delta$ 
is called quasi-cocommutative if 
there exists an invertible element $\mathcal{R}$ in the algebra $H \otimes H$ 
such that 
\begin{equation}
\Delta^{\mathrm{op}}(x) =\mathcal{R}\Delta(x) \mathcal{R}^{-1} \quad ( \forall x \in H). 
\end{equation}
This $\mathcal{R}$ is called the universal R-matrix. 
If $\mathcal{R}$ also satisfies the relations 
\begin{equation}
(\Delta \otimes \mathrm{id}) \mathcal{R}=\mathcal{R}_{13}\mathcal{R}_{23}, \quad 
(\mathrm{id}\otimes \Delta  ) \mathcal{R}=\mathcal{R}_{13}\mathcal{R}_{12}, 
\end{equation}
(see definition of $\mathcal{R}_{ij}$ in Section \ref{sec:R-Matrix}) 
then $H$ is called quasi-triangular and 
$\mathcal{R}$ satisfies the Yang-Baxter equation 
$\mathcal{R}_{12}\mathcal{R}_{13}\mathcal{R}_{23}
=\mathcal{R}_{23}\mathcal{R}_{13}\mathcal{R}_{12}$. 
The DIM algebra is known to be quasi-triangular \cite{FJMM:2015:Quantum}. 
In this thesis, 
the representation matrix of the universal R-matrix $\mathcal{R}$ 
is explicitly calculated. 
In the tensor product of the level 1 representation of the DIM algebra
(we denote it by $\rho_{u_1u_2}$), 
it is block-diagonalized at each level 
of the free boson Fock space. 
Also, it can be seen that 
the action of $\mathcal{R}$ on the generalized Macdonald functions 
corresponds to the exchange of spectral parameters, 
partitions, and variables in the generalized Macdonald functions. 
Moreover, 
by using the renormalized generalized Macdonald functions 
(the integral form $\ket{K}$), 
it can be conjectured that 
\begin{align}
\rho_{u_1u_2}(\mathcal{R}) \tket{K_{\vl}}  
= \tket{K^{\mathrm{op}}_{\vl}}, 
\end{align}
where $\ket{K^{\mathrm{op}}}$ 
is the vector obtained 
by exchanging partitions, variables and spectral parameters 
in $\ket{K_{\vl}}$ 
(see definitions in Section \ref{sec:Explicit cal of R}). 
As a consequence, 
we have conjecture (Conjecture \ref{conj:R-Matrix by int form}) 
of the explicit formula for the representation matrix $R_{\vl, \vm}$ 
of the universal R-matrix in the basis of $\ket{K_{\vl}}$: 
\begin{conj*}
\begin{equation}
R_{\vl, \vm} \overset{?}{=} 
\frac{1}{\braket{K_{\vm} }{K_{\vm}}}
\braket{K_{\vm}}{K^{\mathrm{op}}_{\vl}}.  
\end{equation}
\end{conj*}
In \cite{AKMMMOZ:2016:Toric, AKMMMOZ:2016:Anomaly}, 
the RTT relation of the DIM algebra is also studied using this R-matirx.

\vspace{1.5em}

\noindent 
{\bf {\Large 1.6.}} 
This thesis is organized as follows. 
In Section \ref{sec:5D AGT}, 
two examples of the 5D AGT conjecture are reviewed. 
One is the correspondence 
between the Whittaker vector of the deformed Virasoro algebra 
and the partition function of the 5D pure gauge theory. 
The other is the conjecture on the AFLT basis 
using the level $N$ representation of the DIM algebra. 
In Section \ref{sec:Kac det and sing vct}, 
we give a factorized formula for the Kac determinant 
of the algebra $\mathcal{A}(N)$. 
Its proof depends on some results of the deformed $W$-algebra. 
The relationship between the singular vectors 
and the generalized Macdonald functions is also revealed. 
In Section \ref{sec:Crystallization}, 
we investigated the $q\rightarrow 0$ limit 
of the deformed Virasoro algebra, 
the algebra $\mathcal{A}(N)$ and the 5D AGT conjecture. 
In particular, 
the simplest 5D AGT conjecture is proved in this limit. 
In Section \ref{sec:R-Matrix}, 
the explicit form of the representation 
of the universal R-matrix of the DIM algebra is calculated. 
Its general form is also conjectured 
in terms of the generalized Macdonald functions. 
In Section \ref{sec:Properties of Gn Mac}, 
properties of the generalized Macdonald functions are studied. 
First, 
to state the existence theorem of the generalized Macdonald functions, 
we need partial orderings 
among $N$-tuples of partitions. 
In this thesis, 
by using the partial orderings $\overstar{>}$ 
(see Definition \ref{def:ordering1}) 
and $\overstar{\succ}$ (see Definition \ref{df:ordering elaborated version}), 
the existence theorem is proved. 
However, 
in \cite{awata2011notes}, 
another ordering $>^{\mathrm{L}}$ is used 
and the proof of existence theorem \cite[Proposition 3.8]{awata2011notes} 
is omitted.  
We justify the theorem \cite[Proposition 3.8]{awata2011notes} 
by comparing $\overstar{\succ}$ and $>^{\mathrm{L}}$ in Subsection \ref{sec:partial orderings}. 
In Subsection \ref{sec:realization of rank N rep}, 
we also investigate the action of the generators $\Xo_{\pm 1}$ 
and higher rank Hamiltonians 
on the generalized Macdonald functions. 
Their actions are based on a conversion rule called spectral duality 
that exchanges the level $N$ representation and the rank $N$ representation 
of the DIM algebra. 
Furthermore, 
in Subsection \ref{sec:Limit to beta}, 
the $q\rightarrow 1$ limit is also studied. 
Since the generalized Jack functions have degenerate eigenvalues, 
their Cauchy formula used in the senario of proof of the AGT conjecture 
\cite{morozov2013finalizing} is non-trivial. 
By taking the limit from the Macdonald functions, 
we can justify the orthogonality of the generalized Jack functions 
and show the Cauchy formula. 
In Appendix \ref{sec: Macdonald and HL}, 
the definition and basic facts of the ordinary Macdonald functions 
and the Hall-Littlwood functions are briefly reviewed following  \cite{Macdonald}.  
In Appendix \ref{sec:Def of DIM}, 
the definition of the DIM algebra and the level $N$ representation 
are explained following mainly 
\cite{FHHSY, FFJMM:2011:semiinfinite, FT:2011:Equivariant}. 
Moreover we also describe the definition of another representation 
of the DIM algebra called level $(0,1)$ representation 
or the rank $N$ representation. 
In Appendix \ref{sec:proofs and chekc of crystal}, 
we present some proofs and checks of conjectures in Section \ref{sec:Crystallization}. 
At last in Appendix \ref{sec:Ex of R-matrix}, 
we give explicit examples of R-matrix at level 2.


\newpage

\section*{Notations}

\begin{itemize}
 \item $\mathbb{N},\: \mathbb{Z},\: \mathbb{Q},\: \mathbb{R},\: \mathbb{C}$ 
denote the set of positive integers, integers, rational numbers, real numbers, complex numbers, respectively. 

 \item $\mathbb{Z}_{\geq 0}$ denotes the set of non-negative integers. 

 \item $\mathbb{Z}_{\neq 0}$ denotes the set of integers except $0$. 

 \item $\delta_{i, j}$ denotes the Kronecker delta.  
 \item $\mathbb{K}[x_1,\ldots ,x_n]$ denotes the ring of polynomials in 
$x_1,\ldots ,x_n$ over a field $\mathbb{K}$. 

 \item $\#\{\qquad \}$ denotes the cardinality of set. 
 \item Functions $f(a_1,a_2, \ldots)$ depending on multiple variables $a_n$ ($n=1, 2, \ldots$) are occasionally written as $f(a)$ or $f(a_n)$ for abbreviation. 
 \item For a partition $\lambda$, 
 $p_{\lambda}$ and $m_{\lambda}$ denote 
 the power sum symmetric function and  
 the monomial symmetric function, respectively. 
 \item For $n \in \mathbb{N}$, $e_{n}$ denotes 
 the elementary symmetric function. 
\end{itemize}


\vspace{5mm}

Let us explain the notation of partitions and Young diagrams. 

A partition $\lambda=(\lambda_1,\ldots,\lambda_n)$ 
is a non-increasing sequence of integers 
$\lambda_1 \geq \ldots \geq \lambda_n \geq 0$. We write 
$\left| \lambda \right| \seteq \sum_{i} \lambda_i$.  
The length of $\lambda$, denoted by $\ell(\lambda)$, 
is the number of elements $\lambda_i$ with $\lambda_i \neq 0$. 
Partitions are identified if all elements except $0$ are the same.   
For example, $(3,2)=(3,2,0)$. 
$m_i=m_i(\lambda)$ denotes the number of elements that are equal to $i$ in $\lambda$, 
and we occasionally write partitions as 
$\lambda=(1^{m_1},2^{m_2},3^{m_3}, \ldots)$.
For example, $\lambda=(6,6,6,2,2,1)=(6^3,2^2,1)$.

The partitions are identified with the Young diagrams, 
which are the figures written by 
putting  $\lambda_i$ boxes on the $i$-th row and aligning the left side.
For example, 
if $\lambda=(6,4,3,3,1)$, its Young diagram is 
\setlength{\unitlength}{1mm}
\newsavebox{\yang}
\savebox{\yang}{
\begin{picture}(30,25)(0,0)
\put(0,25){\line(1,0){30}}
\put(0,20){\line(1,0){30}}
\put(0,15){\line(1,0){20}}
\put(0,10){\line(1,0){15}}
\put(0,5){\line(1,0){15}}
\put(0,0){\line(1,0){5}}
\put(30,25){\line(0,-1){5}}
\put(25,25){\line(0,-1){5}}
\put(20,25){\line(0,-1){10}}
\put(15,25){\line(0,-1){20}}
\put(10,25){\line(0,-1){20}}
\put(5,25){\line(0,-1){25}}
\put(0,25){\line(0,-1){25}}
\end{picture}}
\begin{center}
\usebox{\yang}.
\end{center}
The conjugate of a partition $\lambda$, 
denoted by $\lambda'$, 
is the partition whose Young diagram is 
the transpose of the diagram $\lambda$. 
For example, 
The conjugate of $\lambda=(6,4,3,3,1)$ is $\lambda'=(5,4,4,2,1,1)$. 
For a partition $\lambda$ and a coordinate $(i, j) \in \mathbb{N}^2$, 
define 
\begin{equation}
A_{\lambda}(i,j):=\lambda_i - j,\qquad \qquad L_{\lambda}(i,j):=\lambda'_j -i. 
\end{equation}
$A_{\lambda}(i,j)$ is called arm length and $L_{\lambda}(i,j)$ is called leg length. 
In the diagram, 
they mean the numbers of boxes in right side from or below 
the box being in the $i$-th row and $j$-th column. 
For example, 
if $\lambda=(8,8,5,3,3,3,1)$, then 
$A_{\lambda}(2,3)=5$, $L_{\lambda}(2,3)=4$. 
\newsavebox{\armleg}
\savebox{\armleg}{
\begin{picture}(40,35)(0,0)
\put(0,35){\line(1,0){40}}
\put(0,30){\line(1,0){40}}
\put(0,25){\line(1,0){40}}
\put(0,20){\line(1,0){25}}
\put(0,15){\line(1,0){15}}
\put(0,10){\line(1,0){15}}
\put(0,5){\line(1,0){15}}
\put(0,0){\line(1,0){5}}
\put(40,35){\line(0,-1){10}}
\put(35,35){\line(0,-1){10}}
\put(30,35){\line(0,-1){10}}
\put(25,35){\line(0,-1){15}}
\put(20,35){\line(0,-1){15}}
\put(15,35){\line(0,-1){30}}
\put(10,35){\line(0,-1){30}}
\put(5,35){\line(0,-1){35}}
\put(0,35){\line(0,-1){35}}
\put(11.3,26){$s$}
\put(17.5,27.5){\circle*{1.5}}
\put(22.5,27.5){\circle*{1.5}}
\put(27.5,27.5){\circle*{1.5}}
\put(32.5,27.5){\circle*{1.5}}
\put(37.5,27.5){\circle*{1.5}}
\put(12.5,22.5){\circle{1.5}}
\put(12.5,17.5){\circle{1.5}}
\put(12.5,12.5){\circle{1.5}}
\put(12.5,7.5){\circle{1.5}}
\put(22,5){$s=(2,3)$}
\end{picture}}
\begin{center}
\usebox{\armleg}
\end{center}
Note that they can take negative values
as $A_{\lambda}(3,7)=-2$, $L_{\lambda}(3,7)=-1$. 
For a partition $\lambda$, 
we define $n(\lambda):=\sum_{i \geq 1} (i-1) \lambda_i$. 
This means the sum of the numbers obtained by attaching a zero to box in 
the top row of the Young diagram of $\lambda$, 
a $1$ to each box in the second row, and so on.

For $N$-tuple of partitions $\vl=(\lo, \ldots, \lN)$, 
define $|\vl|:=|\lo|+\cdots+|\lN|$. 
If $|\vl|=m$, 
we occasionally use the symbol "$\vdash$" as $\vl \vdash m$.



\vspace{10mm}

\section*{Acknowledgments}
The author would like to express his deepest gratitude to his supervisor 
Hidetoshi Awata for a great deal of advice. 
Without his guidance and persistent help, 
this thesis would not have been possible. 
The author shows his greatest appreciation to Hiroaki Kanno 
for his insightful comments and suggestions, 
and H. Fujino, T. Matsumoto, A. Mironov, Al. Morozov, And. Morozov 
and Y. Zenkevich 
for the collaborative researches. 
Some of the results in this thesis are based on the collaborations with them. 
The author also would like to thank 
M. Hamanaka, K. Iwaki, T. Shiromizu, S. Yanagida and friends 
for valuable discussions and supports. 
The author is supported in part by Grant-in-Aid for JSPS Fellow 26-10187. 

\newpage 
\section{5D AGT conjecture}
\label{sec:5D AGT}

\subsection{Review of the simplest 5D AGT correspondence}\label{sec:Review of Simplest 5D AGT}

We start with recapitulating the result of the Whittaker vector 
of the deformed Virasoro algebra 
and the simplest five-dimensional AGT correspondence.

\begin{df}
Let $q$ and $t$ be independent parameters and $p \seteq q/t$. 
The deformed Virasoro algebra is the associative algebra over $\mathbb{Q}(q,t)$ 
generated by $T_n$ ($n \in \mathbb{Z}$) with the commutation relation
\begin{equation}\label{comm.rel.of qVir}
[T_n, T_m] = -\sum_{l=1}^{\infty}f_l(T_{n-l}T_{m+l}-T_{m-l}T_{n+l})
             -\frac{(1-q)(1-t^{-1})}{1-p}(p^n-p^{-n}) \delta_{n+m,0},
\end{equation}
where the structure constant $f_l \in \mathbb{Q}(q,t)$ is defined by 
\begin{equation}
f(z) = \sum_{l=0}^{\infty}f_l \:  z^l
 \seteq \exp \left( \sum_{n=1}^{\infty}\frac{1}{n}\frac{(1-q^n)(1-t^{-n})}{1+p^n}z^n \right). 
\end{equation}
The relation (\ref{comm.rel.of qVir}) can be written 
in terms of the generating function $T(z)\seteq \sum_{n\in \mathbb{Z}}T_nz^{-n}$ as 
\begin{equation}
f\left(\frac{w}{z}\right) T(z)T(w)-T(w)T(z)f\left(\frac{z}{w}\right) 
= -\frac{(1-q)(1-t^{-1})}{1-p}\left[ \delta\left( \frac{pw}{z} \right) - \delta\left( \frac{p^{-1}w}{z} \right) \right], 
\end{equation}
where $\delta(x)=\sum_{n\in \mathbb{Z}}x^n$.
\end{df}

The deformed Virasoro algebra is introduced in \cite{SKAO:1995:quantum}. 
Let $\ket{h}$ be the highest weight vector 
such that $T_0 \ket{h}=h \ket{h}$, $T_n\ket{h}=0$ ($n >0$), 
and 
$M_h$ be the Verma module generated by $\ket{h}$. 
Similarly, 
$\bra{h}$ is the vector satisfying the condition that $\bra{h}T_0=h \bra{h}$, $\bra{h} T_{n}=0$ ($n< 0$). 
$M^{*}_h$ is the dual module generated by $\bra{h}$. 
The PBW type vectors $\tket{T_{-\lambda}} \seteq T_{-\lambda_1} T_{-\lambda_2} \cdots \ket{h}$ for partitions $\lambda$ 
form a basis over $M_h$. 
Also, 
$\bra{T_{\lambda}} \seteq \bra{h} \cdots T_{\lambda_2} T_{\lambda_1}$ form a basis over $M^{*}_h$. 
Here $\lambda=(\lambda_1, \lambda_2, \ldots )$ 
is a partition or a Young diagram. 
The bilinear form $M^*_h \otimes M_h \rightarrow \mathbb{C}$ is 
uniquely defined by $\braket{h}{h}=1$. 
This bilinear form is called the Shapovalov form. 
The Whittaker vector $\ket{G}$ is defined as follows.

\begin{df}[\cite{AwataYamada1}]
For a generic parameter $\Lambda$, define the Whittaker vector
\footnote{The vector $\ket{G}$ is also called the Gaiotto state or the irregular vector. }
 $\ket{G}$ by the relations
\begin{equation}
T_1\ket{G}=\Lambda^2\ket{G}, \qquad T_n \ket{G}=0 \quad (n>1).
\end{equation}
Similarly, the dual Whittaker vector $\bra{G}\in M_h^*$ is 
defined by the condition that
\begin{equation}
\bra{G}T_{-1}=\Lambda^2\bra{G}, \qquad \bra{G} T_{n}=0 \quad (n<-1).
\end{equation}
\end{df}

This vector is in the form 
$\ket{G}=\sum_{\lambda}\Lambda^{2 |\lambda|} B^{\lambda, (1^n)} T_{-\lambda} \ket{h}$ 
and its norm is calculated as 
$\braket{G}{G} = \sum_{n=0}^{\infty} \Lambda^{4n} B^{(1^n),(1^n)}$, 
where $B^{\lambda, \mu}$ denotes the inverse matrix element of 
the Shapovalov matrix $B_{\lambda, \mu} := \braket{T_{\lambda}}{T_{-\mu}}$.

It is useful to consider the free field representation of the deformed Virasoro algebra. 
By the Heisenberg algebra generated by $a_n$ ($n\in \mathbb{Z}$) and $\chargQ$ with the relations 
\begin{equation}\label{eq:comm rel of qt-boson}
[a_n, a_m]=n\frac{1-q^{|n|}}{1-t^{|n|}} \delta_{n+m,0}, 
\qquad [a_n,\chargQ]=\delta_{n,0},
\end{equation}
the generating function $T(z)$ can be represented as 
\begin{align}\label{eq:free field rep.}
T(z) &= 
 \Lambda^+(z) + \Lambda^-(z), \\
\Lambda^{\pm}(z)
&:= \exp \left\{ \mp \sum_{n=1}^{\infty} \frac{1-t^{n}}{n(t^n+q^n)} (q/t)^{\mp\frac{n}{2}} a_{-n}z^n \right\} 
    \exp \left\{ \mp \sum_{n=1}^{\infty} \frac{1-t^n}{n} (q/ t)^{\pm \frac{n}{2}} a_{n}z^{-n} \right\} K^{\pm}. 
\end{align}
Here $K^{\pm} \seteq e^{\pm a_0}$. 
Let $\ket{0}$ be the highest weight vector 
in the Fock module of the Heisenberg algebra 
such that $a_n\ket{0}=0$ ($n\geq 0$), 
and $\ket{k} \seteq k^{\chargQ}\ket{0}$. 
Then $K\ket{k}=k\ket{k}$. 
Furthermore, $\ket{k}$ can be regarded as the highest weight vector $\ket{h}$ 
of the deformed Virasoro algebra with highest weight $h=k+k^{-1}$. 
In \cite{AwataYamada1}, 
Awata and Yamada conjectured an explicit formula 
for $\ket{G}$ in terms of Macdonald functions 
under the free field representation, 
and Yanagida proved it in \cite{Yanagida:2014:Whittaker}. 
The simplest five-dimensional AGT conjecture is 
that the inner product $\braket{G}{G}$ 
coincides with the five-dimensional (K-theoretic) Nekrasov formula 
for pure $SU(2)$ gauge theory \cite{AK:2008Refined, NakajimaYoshioka1, NakajimaYoshioka2} : 
\begin{align}\label{eq:Nek formula for pure}
Z^{\mathrm{inst}}_{\mathrm{pure}} &\seteq \sum_{\lambda, \mu} \frac{(\Lambda^4 t/q)^{|\lambda|+|\mu|}}{N_{\lambda \lambda}(1)N_{\lambda \mu}(Q)N_{\mu \mu}(1)N_{\mu \lambda}(Q^{-1})},  \\
N_{\lambda \mu}(Q) &\seteq \prod_{(i,j)\in \lambda} \left( 1- Q q^{A_{\lambda}(i,j)}t^{L_{\mu}(i,j)+1} \right)  \prod_{(i,j)\in \mu} \left( 1- Q q^{-A_{\mu}(i,j)-1} t^{-L_{\lambda}(i,j)} \right) \label{eq:def of Nek factor}, 
\end{align}
where $A_{\lambda}(i,j)$ and $L_{\lambda}(i,j)$ are 
the arm length and the leg length defined in Introduction, 
and $\lambda'$ is the conjugate of $\lambda$.

\begin{fact}\label{fact:simpleAGT}
 For $k=Q^{\frac{1}{2}}$,
\begin{equation}
\braket{G}{G}=Z^{\mathrm{inst}}_{\mathrm{pure}}.
\end{equation}
\end{fact}

This fact is conjectured in \cite{AwataYamada1} and proved 
in \cite{Yanagida:2010:FiveD, Yanagida:2014:Norm} when the parameter $q$ is generic. 

\subsection{Reargument of Ding-Iohara-Miki algebra and AGT correspondence}
\label{sec:Reargument of DI alg and AGT}

We now turn to the DIM algebra \cite{Ding-Iohara, Miki:2007}. 
Let us recall the AFLT basis in the 5D AGT correspondence 
of the $SU(N)$ gauge theory along \cite{awata2011notes}. 
In this section, 
we use $N$ kinds of bosons $a_n^{(i)}$ 
($n \in \mathbb{Z}_{\neq 0}$, $i=1,2, \ldots , N$) 
and $U_i$ with the relations 
\begin{equation}
[a^{(i)}_n, a^{(j)}_m ]= n \frac{1-q^{|n|}}{1-t^{|n|}} \, \delta_{i,j} \, \delta_{n+m, 0}, 
\end{equation}
\begin{equation}
[a^{(i)}_n, U_j]=0, 
\quad [U_i, U_j]=0 , \qquad (\forall i,j,n).
\end{equation}
Here $U_i$ is the substitution of zero mode $a^{(i)}_0$, 
which is realized in two different ways in Sections \ref{sec:Kac det and sing vct} and \ref{sec:Crystallization}, respectively. 
Let us define the vertex operators $\eta^{(i)}$ and $\varphi^{(i)}$.

\begin{df}Set
\begin{align}
\eta^{(i)} (z)
&:= \exp \left( \sum_{n=1}^{\infty} \frac{1-t^{-n}}{n}\, z^{n} a^{(i)}_{-n} \right) 
   \exp \left( -\sum_{n=1}^{\infty}\frac{(1-t^n)}{n} \, z^{-n} a^{(i)}_n \right), \\
\varphi^{(i)} (z)
&:= \exp \left(  \sum_{n=1}^{\infty} \frac{1-t^{-n}}{n} (1-p^{-n})  z^{n} a^{(i)}_{-n} \right). 
\end{align}
\end{df}

\begin{df}
Define generators $X^{(i)}(z) = \sum_{n} X^{(i)}_n z^{-n}$ by 
\begin{align}
 X^{(i)}(z) 
 &\seteq \sum_{1\leq j_1 <\cdots <j_i \leq N} \NPb \Lambda_{j_1}(z) \cdots \Lambda_{j_i}(p^{i-1}z) \NPb ,
\end{align}
where $\NPb \quad \NPb$ denotes the usual normal ordered product, and
\begin{equation}
\Lambda^{i}(z) 
:= \varphi^{(1)}(z) \varphi^{(2)}(z p^{-\frac{1}{2}}) \cdots \varphi^{(i-1)}(z p^{-\frac{i-2}{2}}) \eta^{(i)}(z p^{-\frac{i-1}{2}}) U_i.
\end{equation}
\end{df}

The generator $X^{(1)}(z)$ arises 
from the level $N$ representation of Ding-Iohara-Miki algebra \cite{FHHSY, FHSSY}, 
and is obtained by acting the coproduct of the DIM algebra 
to the vertex operator $\eta(z)$ $N$ times 
(see Appendix \ref{sec:Def of DIM}). 
The other generators $X^{(i)}_n$ appear in the commutation relations 
of generators $X^{(i-k)}_n$ ($k=1, \ldots i-1$).  
When we just consider the AGT conjecture, 
it suffices to deal with the subalgebra $\langle X^{(i)}_n \rangle$ 
in some completion of the endomorphism of the algebra 
of $N$-tensored Fock modules for our Heisenberg algebra.  

\begin{nota}
We denote the algebra $\langle X^{(i)}_n \rangle$ by $\mathcal{A}(N)$. 
\end{nota}

\begin{prop}
If $N=2$, 
the commutation relations of the generators are 
\begin{align}
&f^{(1)} \left( \frac{w}{z} \right) X^{(1)}(z) X^{(1)}(w) - X^{(1)}(w) X^{(1)}(z) f^{(1)} \left( \frac{z}{w} \right) \label{eq:rel. of generator X^1} \\ 
& \qquad \qquad \qquad = \frac{(1-q)(1-t^{-1})}{1-p} \left\{ \delta \left( \frac{w}{pz} \right) X^{(2)}(z ) -\delta \left( \frac{pw}{z} \right) X^{(2)}(w ) \right\} , \nonumber \\
&f^{(2)} \left( \frac{w}{z} \right) X^{(2)}(z) X^{(2)}(w) - X^{(2)}(w) X^{(2)}(z) f^{(2)} \left( \frac{z}{w} \right) =0,  \\
&f^{(1)} \left( \frac{pw}{z} \right) X^{(1)}(z) X^{(2)}(w) - X^{(2)}(w) X^{(1)}(z) f^{(1)} \left( \frac{z}{w} \right) =0,
\end{align}
where $\delta(x)=\sum_{n \in \mathbb{Z}} x^n$ is 
the multiplicative delta function 
and  the structure constant $ f^{(i)}(z) =\sum_{l =0}^{\infty} f^{(i)}_l z^{l}$ is defined by 
\begin{equation}
 f^{(1)}(z) \seteq 
 \exp \left\{ \sum_{n >0} \frac{(1-q^n)(1-t^{-n})}{n}z^n \right\}, 
\end{equation}
\begin{equation}
 f^{(2)}(z) \seteq 
 \exp \left\{ \sum_{n >0} \frac{(1-q^n)(1-t^{-n})(1+p^{n})}{n}z^n \right\}. 
\end{equation}
These relations are equivalent to 
\begin{align}
& [X^{(1)}_n, X^{(1)}_m]= -\sum_{l =1}^{\infty} f^{(1)}_l (X^{(1)}_{n-l} X^{(1)}_{m+l} - X^{(1)}_{m-l} X^{(1)}_{n+l}) +\frac{(1-q)(1-t^{-1})}{1-p}(p^m-p^n) X^{(2)}_{n+m},  \\
& [X^{(2)}_n, X^{(2)}_m]= -\sum_{l =1}^{\infty} f^{(2)}_l (X^{(2)}_{n-l} X^{(2)}_{m+l} - X^{(2)}_{m-l} X^{(2)}_{n+l}), \\
& [X^{(1)}_n, X^{(2)}_m]= -\sum_{l =1}^{\infty} f^{(1)}_l (p^{l} X^{(1)}_{n-l} X^{(2)}_{m+l} -  X^{(2)}_{m-l} X^{(1)}_{n+l}) .
\end{align}
\end{prop}

The proof is similar to the calculation of the deformed Virasoro algebra 
or the deformed $W$-algebra. 
In the formula (\ref{eq:rel. of generator X^1}), we use 
\begin{equation}
f^{(1)}(x) - f^{(1)}(1/px) = \frac{(1-q)(1-t^{-1})}{1-p} \left( \delta(x)- \delta(px) \right). 
\end{equation}

For an $N$-tuple of parameters $\vu= (u_1, \ldots, u_N)$, 
define $\ket{\vu}$ and $\bra{\vu}$ to be the highest weight vectors such that 
$a_n^{(i)} \ket{\vu}=\bra{\vu}a^{(i)}_{-n} =0$ ($n \geq 1$, $\forall i$),  
$U_i\ket{\vec{u}}=u_i \ket{\vec{u}}$ and 
$\bra{\vec{u}} U_i=u_i \bra{\vec{u}}$. 
$\mathcal{F}_{\vec{u}}$ is the highest weight module generated by 
$\ket{\vec{u}}$, 
and $\mathcal{F}_{\vec{u}}^*$ is the dual module generated by $\bra{\vec{u}}$. 
The bilinear form (Shapovalov form) 
$\mathcal{F}_{\vec{u}}^* \otimes \mathcal{F}_{\vec{u}} \rightarrow \mathbb{C}$
is uniquely determined by the condition $\braket{\vu}{\vu}$.

\begin{df}\label{df:ordinary PBW vct}
For an $N$-tuple of partitions 
$\vl=(\lo, \lt, \ldots, \lambda^{(N)})$, 
set
\begin{align}
 &\ket{X_{\vec{\lambda}}} \seteq 
  X^{(1)}_{-\lambda^{(1)}_1} X^{(1)}_{-\lambda^{(1)}_2} \cdots  X^{(2)}_{-\lambda^{(2)}_1} X^{(2)}_{-\lambda^{(2)}_2}\cdots X^{(N)}_{-\lambda^{(N)}_1} X^{(N)}_{-\lambda^{(N)}_2} \cdots \ket{\vec{u}}, \\
 & \bra{X_{\vl}} \seteq 
\bra{\vu} \cdots X^{(N)}_{\lambda^{(N)}_2} X^{(N)}_{\lambda^{(N)}_1} \cdots X^{(2)}_{\lambda^{(2)}_2} X^{(2)}_{\lambda^{(2)}_1} \cdots 
X^{(1)}_{\lambda^{(1)}_2} X^{(1)}_{\lambda^{(1)}_1} . 
\end{align}
\end{df}

The PBW theorem cannot be used 
because the algebra $\mathcal{A}(N)$ is not a Lie algebra, 
but in \cite{awata2011notes} 
it was conjectured that the PBW type vectors $\ket{X_{\vec{\lambda}}}$ and 
$\bra{X_{\vec{\lambda}}}$ are a basis over $\mathcal{F}_{\vu}$ 
and $\mathcal{F}_{\vec{u}}^*$, respectively. 
This conjecture can be solved by the Kac determinant of the algebra $\mathcal{A}(N)$, 
which is proved in Section \ref{sec:Kac det and sing vct}. 
In this section, 
we consider another type of the PBW basis, 
since it has good expression in $q\rightarrow 0$ limit 
in terms of the Hall-Littlewood functions  
(see Section \ref{sec:crystallization of N=2}).

\begin{df}\label{df:another PBW vct}
For $\vl=(\lo, \lt, \ldots, \lambda^{(N)})$, set
\begin{align}
 &\tket{X'_{\vec{\lambda}}} \seteq X^{(N)}_{-\lambda^{(N)}_1} X^{(N)}_{-\lambda^{(N)}_2} \cdots X^{(2)}_{-\lambda^{(2)}_1}X^{(2)}_{-\lambda^{(2)}_2}\cdots X^{(1)}_{-\lambda^{(1)}_1}X^{(1)}_{-\lambda^{(1)}_2}\cdots \ket{\vu}, \\
 & \tbra{X'_{\vec{\lambda}}} \seteq  \bra{\vec{u}} \cdots X^{(1)}_{\lambda^{(1)}_2} X^{(1)}_{\lambda^{(1)}_1} \cdots X^{(2)}_{\lambda^{(2)}_2} X^{(2)}_{\lambda^{(2)}_1} \cdots X^{(N)}_{\lambda^{(N)}_2} X^{(N)}_{\lambda^{(N)}_1}. 
\end{align}
\end{df}

Let us review the AFLT basis in $\mathcal{F}_{\vu}$, 
which is also called generalized Macdonald functions. 
In order to state its existence theorem, 
let us prepare the following ordering. 

\begin{df}\label{def:ordering1}
For $N$-tuple of partitions $\vl$ and $\vm$, 
\begin{align}
\vl \overstar{>} \vm \quad \overset{\mathrm{def}}{\Longleftrightarrow} \quad  
& |\vl| = |\vm|, \quad 
\sum_{i=k}^N |\lambda^{(i)}| \geq \sum_{i=k}^N |\mu^{(i)}| \quad (\forall k ) \quad \mathrm{and}   \\
&  (|\lo|,|\lt|,\ldots ,|\lN|) \neq (|\mo|,|\mt|,\ldots ,|\mN|).  \nonumber
\end{align}
Here $|\vl| \seteq |\lo|+\cdots + |\lambda^{(N)}|$. 
Note that the second condition can be replaced 
with $\sum_{i=1}^{k-1} |\lambda^{(i)}| \leq \sum_{i=1}^{k-1} |\mu^{(i)}|\quad  (\forall k )$. 
\end{df}

We can state the existence theorem of generalized Macdonald functions
in the basis of products of Macdonald functions 
$\prod_{i=1}^N P_{\lambda^{(i)}}(a^{(i)}_{-n};q,t) \ket{\vu}$, 
where $P_{\lambda}(a_{-n}^{(i)};q,t)$ are Macdonald symmetric functions 
defined in Appendix \ref{sec: Macdonald and HL} 
with substituting the bosons $a^{(i)}_{-n}$ 
for the power sum symmetric functions $p_n$.

\begin{prop}
For each $N$-tuple of partitions $\vl$, 
there exists a unique vector $\ket{P_{\vl}} \in \mathcal{F}_{\vu}$ 
such that 
\begin{align}
 &\ket{P_{\vec{\lambda}}} 
  = \prod_{i=1}^N P_{\lambda^{(i)}}(a^{(i)}_{-n};q,t) \ket{\vu} 
  + \sum_{\vm \mathop{\overset{*}{<}} \vl} c_{\vl, \vm} \prod_{i=1}^N P_{\mu^{(i)}}(a^{(i)}_{-n};q,t) \ket{\vu}, \\
 &X^{(1)}_0 \ket{P_{\vl}} = \epsilon_{\vl} \ket{P_{\vl}}, 
\end{align}
where $c_{\vl, \vm}=c_{\vl, \vm}(u_1, \ldots ,u_N;q,t)$ is a constant, 
$\epsilon_{\vl}=\epsilon_{\vl}(u_1,\ldots,u_N;q,t)$ is the eigenvalue of $X^{(1)}_0$. 
Similarly, there exists a unique vector 
$\bra{P_{\vl}} \in \mathcal{F}_{\vu}^*$ such that 
\begin{align}
 &\bra{P_{\vec{\lambda}}} 
  = \bra{\vu} \prod_{i=1}^N P_{\lambda^{(i)}}(a^{(i)}_{n};q,t) 
  + \sum_{\vec{\mu} \overstar{>} \vec{\lambda}} c_{\vl, \vm}^* \bra{\vu} \prod_{i=1}^N P_{\mu^{(i)}}(a^{(i)}_{n};q,t), \\
 & \bra{P_{\vec{\lambda}}} \Xo_0 = \epsilon_{\vec{\lambda}}^* \bra{P_{\vl}}.
\end{align}
Then the eigenvalues are 
\begin{equation}\label{eq:eigenvalue of gn Mac}
\epsilon_{\vl}= 
\epsilon_{\vl}^*= 
\sum_{k=1}^N u_k e_{\lambda^{(k)}}, \quad 
e_{\lambda}:= 1+(t-1) \sum_{i \leq 1} (q^{\lambda_i}-1)t^{-i}. 
\end{equation}
\end{prop}

Although the ordering of Definition \ref{def:ordering1} 
is different from the one in \cite{awata2011notes}, 
the eigenfunctions $\ket{P_{\vl}}$ are quite the same. 
The proof is similar to the one in Section \ref{sec:partial orderings}, 
which follows from triangulation of $\Xo_0$. 
By this proposition, 
it can be seen that $\ket{P_{\vl}}$ is a basis over $\mathcal{F}_{\vu}$, 
and the eigenvalues of $X^{(1)}_0$ are non-degenerate.  
In Section \ref{sec:partial orderings}, 
a more elaborated ordering is introduced 
and a relationship between these orderings is explained. 
In Section \ref{sec:Limit to beta}, 
it is shown that 
these vectors $\ket{P_{\vl}}$ correspond to the generalized Jack functions 
defined in \cite{morozov2013finalizing} in the $q \rightarrow 1$ limit. 
To use generalized Macdonald functions in the AGT correspondence, 
we need to consider its integral form. 
In this paper, 
we adopt the following renormalization, 
which is slightly different from that of \cite{awata2011notes}.

\begin{df}\label{df:integral form of Gn Macdonald}
Define the vectors $\ket{K_{\vec{\lambda}}}$ and $\bra{K_{\vec{\lambda}}}$, 
called the integral forms, 
by the condition that 
\begin{align}
&\ket{K_{\vec{\lambda}}}= \sum_{\vm} \alpha_{\vl \vm} \ket{X'_{\vm}} \propto \ket{P_{\vl}}, \quad 
\alpha_{\vl, (\emptyset,\ldots,\emptyset, (1^{|\vl|}))} =1, \\
&\bra{K_{\vec{\lambda}}}= \sum_{\vm} \beta_{\vl \vm} \bra{X'_{\vm}} \propto \bra{P_{\vl}}, \quad 
\beta_{\vl, (\emptyset,\ldots, \emptyset, (1^{|\vl|}))} =1.
\end{align}
\end{df}

\begin{conj}
The coefficients $\alpha_{\vl \vm}$ and $\beta_{\vl \vm}$ are 
polynomials in $q^{\pm 1}$, $t^{\pm 1}$ and $u_i$ with integer coefficients.  
\end{conj}

\begin{ex}
	If $N=2$, 
	the transition matrix $\alpha_{\vl, \vm}$ is as follows:  
	\begin{equation*}
	\begin{array}{c||c c} 
	\vl \setminus \vm & (\emptyset, (1)) & ((1), \emptyset) \\ \hline \hline
	(\emptyset, (1))  & 1 &-\frac{q u_2}{t} \\
	((1), \emptyset)  & 1 &-\frac{q u_1}{t}
	\end{array},
	\end{equation*}
	\begin{equation*}
	\begin{array}{c||c c c } 
	\vl \setminus \vm & (\emptyset, (2)) & (\emptyset, (1^2)) &((1), (1)) \\ \hline \hline
	(\emptyset, (2)) & \frac{(q-1) u_2
		\left(t u_1 q^2-u_1 q^2+t u_2 q^2-u_2 q^2-u_2 q+tu_1\right)}{t^2} & 1  & -\frac{q (q+1) u_2}{t} \\
	(\emptyset, (1^2)) & \frac{q (t-1) u_2
		\left(-u_1 t^2+q u_2 t+q u_1-u_1+q u_2-u_2\right)}{t^3} & 1 & -\frac{q (t+1) u_2}{t^2}  \\
	((1), (1)) & \frac{(q-1) q (t-1)
		\left(u_1^2+u_2 u_1+u_2^2\right)}{t^2} & 1 & -\frac{q (u_1+u_2)}{t} \\
	((2), \emptyset) & \frac{(q-1) u_1
		\left(t u_1 q^2-u_1 q^2+t u_2 q^2-u_2 q^2-u_1 q+t
		u_2\right)}{t^2} & 1 & -\frac{q (q+1) u_1}{t} \\
	((1^2), \emptyset) & \frac{q (t-1) u_1
		\left(-u_2 t^2+q u_1 t+q u_1-u_1+q u_2-u_2\right)}{t^3} & 1  & -\frac{q (t+1) u_1}{t^2}
	\end{array},
	\end{equation*}
	\begin{equation*}
	\begin{array}{c|| c c} 
	\vl \setminus \vm  &((2), \emptyset) &((1^2), \emptyset) \\ \hline \hline
	(\emptyset, (2)) & -\frac{(q-1) q^2 u_2^2 (-q u_1+q t u_1+t u_1-q u_2)}{t^3} & \frac{q^3 u_2^2}{t^2}  \\
	(\emptyset, (1^2)) &-\frac{q^2 (t-1) u_2^2 (q u_1-t u_1-u_1+q u_2)}{t^4} & \frac{q^2 u_2^2}{t^3}  \\
	((1), (1)) & -\frac{(q-1) q^2 (t-1) u_1 u_2 (u_1+u_2)}{t^3} & \frac{q^2 u_1 u_2}{t^2} \\
	((2), \emptyset) & -\frac{(q-1) q^2 u_1^2 (-q u_1-q u_2+q t u_2+t u_2)}{t^3} &\frac{q^3 u_1^2}{t^2} \\
	((1^2), \emptyset) &-\frac{q^2 (t-1) u_1^2 (q u_1+q u_2-t u_2-u_2)}{t^4} & \frac{q^2 u_1^2}{t^3}
	\end{array}.
	\end{equation*}
\end{ex}

By using these integral forms, 
the five dimensional AGT conjecture can be stated in the following form. 
(c.f. \cite[Conjecture 3.11 and Conjecture 3.13]{awata2011notes})

\begin{conj}
The norm of $\ket{K_{\vl}}$ reproduces the Nekrasov factor:
\begin{equation}
\tbraket{K_{\vl}}{K_{\vl}} 
\overset{?}{=} (-1)^N \, e_N(\vu)^{|\vl|} \, \prod_{i=1}^{N}t^{-N n(\lambda^{(i)})} q^{N n(\lambda^{(i)'})} u_i^{N |\lambda^{(i)}|}
 \prod_{i,j=1}^N N_{\lambda^{(i)}, \lambda^{(j)}} (qu_i/tu_j),
\end{equation}
where $e_N(\vu) = u_1 u_2 \cdots u_N$. 
\end{conj}

\begin{df}\label{df:DIM vertex op}
Call the linear operator 
$\Phi(z) = \Phi^{\vv}_{\vu}(z): \mathcal{F}_{\vec{u}} \rightarrow \mathcal{F}_{\vec{v}}$ 
the vertex operator if it satisfies  
\begin{equation}\label{eq:rel. of DIM vertex op. 1}
(1-e_N(\vv) w/z) X^{(i)}(z) \Phi(w)= (1- p^{-i} e_N(\vv) w/z) \Phi(w) X^{(i)}(z)
\end{equation}
and $\bra{\vec{v}} \Phi(w) \ket{\vec{u}} = 1$. 
Then the relations for the Fourier components are 
\begin{equation}\label{eq:rel. of DIM vertex op. 2}
(X^{(i)}_n- e_{N}(\vv) w X^{(i)}_{n-1})\Phi(w) 
= \Phi(w) (X^{(i)}_n-(t/q)^i e_{N}(\vv) w X^{(i)}_{n-1})
\end{equation}
for $i=1,2, \ldots , N$.
\end{df}

\begin{ex}
If $N=1$, it is known that $\Phi(z)$ exists and is given by 
\begin{equation}
\Phi(z)= \exp \left\{ -\sum_{n=1}^{\infty} \frac{1}{n} \frac{v^n-(t/q)^n u^n}{1-q^n}a_{-n}z^n \right\}
\exp \left\{ \sum_{n=1}^{\infty} \frac{1}{n} \frac{v^{-n}- u^{-n}}{1-q^{-n}}a_{n}z^{-n} \right\} 
\mathcal{Q}, 
\end{equation} 
where $\mathcal{Q}$ is the operator from 
$\mathcal{F}_{u}$ to $\mathcal{F}_{v}$ satisfying 
the relation $U  \mathcal{Q}= (v/u) \, \mathcal{Q} \, U$.   
\end{ex}

\begin{conj}
The matrix elements of $\Phi(w)$ with respect to generalized Macdonald functions are
\begin{align}
\bra{K_{\vl}} \Phi^{\vv}_{\vu}(w) \ket{K_{\vm}}  
& \overset{?}{=} (-1)^{|\vl| +(N-1)|\vm|} (t/q)^{N(|\vl|-|\vm|)} e_N(\vu) ^{|\vl|} e_N(\vv)^{|\vl|-|\vm|} w^{|\vl|-|\vm|} \\ 
& \quad \times \prod_{i=1}^N u_i^{N |\mu^{(i)}|} q^{N\, n(\mu^{(i)'})} t^{-N\, n(\mu^{(i)})}
\times \prod_{i,j=1}^N N_{\lambda^{(i)},\mu^{(j)}}(qv_i/tu_j). \nonumber 
\end{align}
\end{conj}

Under these conjectures, 
we can obtain a formula for multi-point correlation functions of $\Phi(z)$ 
by inserting the identity $1=\sum_{\vl}\frac{\ket{K_{\vl}} \bra{K_{\vl}}}{\braket{K_{\vl}}{K_{\vl}}} $. 
In particular, 
the formula for the four-point functions 
agrees with the 5D $U(N)$ Nekrsov formula with $N_f=2N$ matters. 
An M-theoretic derivation of this formula 
is also given by \cite{Tan2013}. 

\section{Kac determinant and singular vecter of the algebra $\mathcal{A}(N)$}
\label{sec:Kac det and sing vct}

\subsection{Kac determinant of the algebra $\mathcal{A}(N)$}
\label{sec:Kac det of DIM alg}

In this section, 
we give the formula for the Kac determinant of the algebra 
$\mathcal{A}(N)$ and 
prove it. 
Moreover, 
it is shown that singular vectors correspond to the generalized Macdonald functions. 
In order to prove the Kac determinant, 
we need screening currents of the algebra $\mathcal{A}(N)$. 
To construct them, 
it is necessary to realize the operator $U_i$ 
and the highest wight vector $\ket{\vu}$ 
in terms of the charge operator $Q^{(i)}$ and $a^{(i)}_0$ ($i=1,\ldots N$). 
Let $Q^{(i)}$ be the operator satisfying the relation 
\begin{equation}
[a^{(i)}_n,Q^{(j)}]=\delta_{n,0}\delta_{i,j}, 
\end{equation}
$\ket{0}$ be the highest weight vector in the Fock module 
of the Heisenberg algebra such that $a^{(i)}_n \ket{0}=0$ for $n \geq0$. 
For an $N$-tuple of complex parameters $\vu=(u_1,\ldots ,u_N)$ 
with $u_i = q^{\sqrt{\beta}w_i}p^{-\frac{N+1}{2}+i}$,  
we realize the highest wight vector $\ket{\vu}$ and $U_i$ as 
\begin{equation}
 U_i= q^{\sqrt{\beta} a^{(i)}_0} p^{-\frac{N+1}{2}+i}, 
\qquad 
\ket{\vu}= e^{\sum_{i=1}^N w_i Q^{(i)}} \ket{0}, 
\end{equation}
where $\beta$ is defined by $t=q^{\beta}$. 
Then they satisfy the required relation $U_i \ket{\vu}= u_i \ket{\vu}$. 
Similarly, 
let $\bra{0}$ be the dual highest weight vector, and 
$\bra{\vu}=\bra{0}e^{-\sum_{i=1}^N w_i Q^{(i)}}$. 
These highest wight vectors are normalized by $\braket{0}{0}=1$, 
and satisfy the condition of the Shapovalov form $\braket{\vu}{\vu}=1$.%
\footnote{
The parameters $q$ and $t$ are assumed to be generic in this section.}

We obtain the formula for the Kac determinant with respect to 
the PBW type vectors $\ket{X_{\vl}}$.%
\footnote{The formulas for the Kac determinant of 
the deformed Virasoro and the deformed $W_N$-algebra are proved in 
\cite{BouwknegtPilch:1998:Virasoro, BouwknegtPilch:1998:W}. }

\begin{thm}\label{thm:KacDet}
Let $\mathrm{det}_n \seteq 
\det \left( \braket{X_{\vl}}{X_{\vm}} \right)_{\vl, \vm \vdash n}$. 
Then  
\begin{align}\label{eq:KacDet for DIM}
\mathrm{det}_n=
&\prod_{\vl \vdash n} \prod_{k=1}^N b_{\lambda^{(k)}}(q) b'_{\lambda^{(k)}}(t^{-1})
\\
& \times \prod_{\substack{1\leq r,s\\ rs\leq n}}
\left( 
(u_1 u_2 \cdots u_N)^{2}
 \prod_{1\leq i < j \leq N} (u_i-q^st^{-r}u_j)(u_i-q^{-r}t^s u_j) \right)^{P^{(N)}(n-rs)}, 
\end{align}
where  
$b_{\lambda}(q) \seteq \prod_{i\geq 1} \prod_{k=1}^{m_i} (1-q^k)$, 
$b'_{\lambda}(q) \seteq \prod_{i\geq 1} \prod_{k=1}^{m_i} (-1+q^k)$,  
and $m_i$ is the number of entries in $\lambda$ equal to $i$.  
$P^{(N)}(n)$
denotes the number of $N$-tuples of Young diagrams of size $n$, i.e.,
$\# \big\{ \vl=(\lo, \ldots , \lN) \big| \vl \vdash n\big\}$. 
In particular, if $N=1$, 
\begin{equation}
\mathrm{det}_n=\prod_{\lambda \vdash n} 
b_{\lambda}(q) b'_{\lambda}(t^{-1})
\times u_1^{2\sum_{\lambda \vdash n}\ell(\lambda)}. 
\end{equation}
\end{thm}

\begin{cor}
If $u_i \neq 0$ and $u_i \neq q^rt^{-s} u_j$ for any numbers $i$, $j$ and 
integers $r$, $s$, 
then the  PBW type vectors $\ket{X_{\vl}}$ (resp. $\bra{X_{\vl}}$) are a basis 
over $\mathcal{F}_{\vu}$ (resp. $\mathcal{F}^*_{\vu}$). 
\end{cor}

It can be seen that 
the representation of the algebra $\mathcal{A}(N)$ 
on the Fock Module $\mathcal{F}_{\vu}$ is irreducible 
if and only if the parameters $\vu$ satisfy the condition 
that $u_i \neq 0$ and $u_i \neq q^rt^{-s} u_j$. 
The proof of Theorem \ref{thm:KacDet} is given in the next section. 

\subsection{Proof of Theorem \ref{thm:KacDet}}
\label{sec:Proof of Kac det thm}

It is known that 
the level $N$ representation of the DIM algebra introduced in the last section
can be regarded as the tensor product of the deformed $W_N$-algebra and 
the Heisenberg algebra associated with the $U(1)$ factor \cite{FHSSY}. 
This fact is obtained by a linear transformation of bosons. 
The point of proof of Theorem \ref{thm:KacDet} is 
to construct singular vectors by using screening currents 
of the deformed $W_N$-algebra 
under the decomposition of the generators $X^{(i)}(z)$ into 
the deformed $W_N$-algebra part and the $U(1)$ part. 
In general, a vector $\ket{\chi}$ in the Fock module $\mathcal{F}_{\vu}$ 
is called the singular vector of the algebra $\mathcal{A}(N)$
if it satisfies 
\begin{equation}
X^{(i)}_n \ket{\chi}=0
\end{equation}
for all $i$ and $n>0$. 
The singular vectors obtained by the screening currents are 
intrinsically the same one of the deformed $W$-algebra. 
From this singular vector, 
we can get the vanishing line of the Kac determinant 
in the similar way of the deformed $W_N$-algebra.

At first, in the $N\geq 2$ case, 
we introduce the following bosons. \\
{\bf U(1) part boson}
\begin{align}
&b'_{-n} \seteq \frac{(1-t^{-n})(1-p^n)}{n(1-p^{Nn})}p^{(N-1)n}\sum_{k=1}^N p^{(\frac{-k+1}{2})n} a^{(k)}_{-n}, \\
&b'_{n} \seteq -\frac{(1-t^{n})(1-p^n)}{n(1-p^{Nn})}p^{(N-1)n}\sum_{k=1}^N p^{\left( \frac{-k+1}{2} \right) n} a^{(k)}_{n}  
\quad (n>0), 
\end{align}
\begin{equation}
b'_{0}\seteq \ao_0 +\cdots +\aN_0, \qquad Q' \seteq \frac{Q^{(1)}+\cdots + Q^{(N)} }{N}.
\end{equation}
\\
{\bf Orthogonal component of $a^{(i)}_n$ for $b'$}
\begin{equation}
b^{(i)}_{-n} \seteq \frac{1-t^{-n}}{n} a^{(i)}_{-n}-p^{(\frac{-i+1}{2})n} b'_{-n}, \quad
b^{(i)}_{n} \seteq \frac{1-t^n}{n}a^{(i)}_{n}+p^{(\frac{-i+1}{2})n} b'_{n} 
\quad (n>0).
\end{equation}
\\
{\bf Fundamental boson of the deformed $W_N$-algebra part}
\begin{equation}
h^{(i)}_{-n}\seteq (1-p^{-n}) \left( \sum_{k=1}^{i-1} p^{\frac{-k+1}{2}n} b^{(k)}_{-n}\right)+  p^{\frac{-i+1}{2}n} b^{(i)}_{-n}, \qquad
h^{(i)}_{n}\seteq -p^{\frac{i-1}{2}n} b^{(i)}_n \quad (n>0),
\end{equation}
\begin{equation}
\hi_0\seteq \ai_0 -\frac{b'_0}{N}, \qquad Q^{(i)}_h\seteq Q^{(i)} - Q', \qquad 
Q^{(i)}_{\Lambda}\seteq \sum_{k=1}^i Q^{(k)}_h.
\end{equation}
Then they satisfy the following relations
\begin{equation}
[b'_n,b'_{m}]=-\frac{(1-q^{|n|})(1-t^{-|n|})(1-p^{|n|})}{n(1-p^{N|n|})}
\delta_{n+m,0}, \qquad
[\bi_n , b'_m]=[\hi_n,b'_m]=[\Qi_h,b'_m]=0,
\end{equation}
\begin{equation}
[\bi_n,\bj_{m}]=\frac{(1-q^{|n|})(1-t^{-|n|})}{n}\delta_{i,j}\delta_{n+m,0} 
-p^{(N-\frac{i+j}{2})|n|}\frac{(1-q^{|n|})(1-t^{-|n|})(1-p^{|n|})}{n(1-p^{N|n|})}\delta_{n+m,0},
\end{equation}
\begin{equation}
[\hi_n,\hj_{m}]= -\frac{(1-q^n)(1-t^{-n})(1-p^{(\delta_{i,j}N-1)n})}{n(1-p^{Nn})}p^{Nn \,\theta(i>j)} \delta_{n+m,0},
\end{equation}
\begin{equation}
[\hi_0,Q^{(j)}_h]=\delta_{i,j}-\frac{1}{N}, \qquad 
\sum_{i=1}^N p^{-in} \hi_n =0, \qquad 
\QN_{\Lambda}=\sum_{i=1}^N  Q^{(i)}_h=0 .
\end{equation}
where $\theta(P)$ is $1$ or $0$ if the proposition $P$ is true or false,  respectively.%
\footnote{Note that $\hi_n$ correspond to the fundamental bosons $h^{N-i+1}_n$ in \cite{Awata:1995Quantum}} 
Using these bosons, 
we can decompose the generator $X^{(i)}(z)$ into 
the U(1) part and the deformed $W$-algebra part. 
That is to say, 
\begin{equation}
\Lambda_i(z)=\Lambda'_i(z) \Lambda''(z)
\end{equation}
\begin{eqnarray}
\Lambda'_i(z) \seteq \NPb \exp\left( \sum_{n \in \mathbb{Z}_{\neq 0}} \hi_n z^{-n} \right) \NPb q^{\sqrt{\beta} \hi_0} p^{-\frac{N+1}{2}+i}, \quad 
\Lambda''(z) \seteq \NPb  \exp\left( \sum_{n \in \mathbb{Z}_{\neq 0}} b'_{-n}z^n \right) \NPb
q^{\sqrt{\beta} \frac{b'_0}{N}}, 
\end{eqnarray}
and
\begin{equation}
X^{(i)}(z) 
= W^{(i)}(z) Y^{(i)}(z)
\end{equation}
\begin{equation}
W^{(i)}(z) 
\seteq \sum_{1\leq j_1 <\cdots <j_i \leq N} 
\NPb \Lambda'_{j_1}(z) \cdots \Lambda'_{j_i}(p^{i-1}z) \NPb, \quad 
Y^{(i)}(z) \seteq \NPb \exp\left( \sum_{n\in \mathbb{Z}_{\neq 0}} 
\frac{1-p^{i n}}{1-p^n} b'_{-n}z^n \right) \NPb
q^{\sqrt{\beta}\frac{i b'_0}{N}}. 
\end{equation}
$W^{(i)}(z)$ is the generator of the deformed $W_N$-algebra. 
Let us introduce the new parameters $u'_i$ and $u''$ defined by 
\begin{equation}
\prod_{i=1}^N u'_i=1, \quad 
u'' u'_i = u_i \quad (\forall i).
\end{equation}
Then the inner product of PBW type vectors can be written as 
\begin{equation}
\braket{X_{\vl}}{X_{\vm}}=(u'')^{\sum_{k=1}^N k (\ell(\lambda^{(k)})+\ell(\mu^{(k)}))}
\times \left( \mbox{polynomial in $u_1', \ldots , u'_N$  } \right). 
\end{equation}
Hence, its determinant is also in the form 
\begin{equation}
\mathrm{det}_n=(u'')^{2\sum_{\vl \vdash n} \sum_{i=1}^N i \ell(\lambda^{(i)})}
\times F(u'_1,\ldots, u'_N), 
\end{equation}
where 
$F(u'_1,\ldots, u'_N)$ is 
a polynomial in $u'_i$ ($i=1,\ldots,N$) which is independent of $u''$. 
Now in \cite{Awata:1995Quantum}, 
the screening currents of the deformed $W_N$-algebra are introduced: 
\begin{equation}
S^{(i)}(z) \seteq 
\NPb \exp \left( \sum_{n\neq 0} \frac{\alpha^{(i)}_n }{1-q^n} \right) \NPb 
e^{\sqrt{\beta} Q^{(i)}_{\alpha}} z^{\sqrt{\beta} \alpha^{(i)}_0}, 
\end{equation}
where $\alpha^{(i)}_n$ is the root boson 
defined by $\alpha^{(i)}_n \seteq h^{(N-i+1)}_n-h^{(N-i)}_n$ and 
$\Qi_{\alpha} \seteq Q_h^{(N-i+1)}-Q_h^{(N-i)}$. 
The bosons $\alpha^{(i)}_n$ and $\Qi_{\alpha}$ commute with $b'_n$, 
and it is known that 
the screening charge $\oint dz S^{(i)}(z)$ commutes with 
the generators $W^{(j)}(z)$. 
Therefore, 
$\oint dz S^{(i)}(z)$ commutes with any generator $X^{(j)}_n$, 
and it can be considered as the screening charges of the algebra 
$\mathcal{A}(N)$. 
Define parameters $\hi$, $\alpha'$ 
by $h^{(i)}+ \frac{\alpha'}{N}=w_i$ and $\sum_{i=1}^N\hi=0$, 
and set $\alpha^{(i)}\seteq h^{(N-i+1)} -h^{(N-i)}$. 
Then 
$\ket{\vu}= e^{\sum_{i=1}^{N-1} \alpha^{(i)} Q^{(i)}_{\Lambda} 
+ \alpha' Q'} \ket{0}$. 
For any number $i=1,\dots , N$, 
the vector arising from the screening current $S^{(i)}(z)$, 
\begin{equation}\label{eq:sing vct chi(i)rs}
\tket{\chi^{(i)}_{r,s}} = 
\oint dz \prod_{k=1}^r S^{(i)}(z_k) \ket{\vv}, \quad 
v_{N-i}=q^{s} t^{r} v_{N-i+1} \quad  
(r, s \in \mathbb{Z}_{>0})
\end{equation}
is a singular vector. 
$\tket{\chi^{(i)}_{r,s}}$ is in the Fock module $F_{\vu}$ 
with the parameter $\vu$ satisfying $u_k=v_k$ for $k \neq N-i, N-i+1$ and 
$u_{N-i+1}=t^r v_{N-i+1}$, $u_{N-i}=t^{-r} v_{N-i}$. 
The $P^{(N)}(n-rs)$ vectors obtained by this singular vector 
\begin{eqnarray}
X_{-\vl} \tket{\chi^{(i)}_{r,s}}, \qquad |\vl|=n-rs
\end{eqnarray}
contribute the vanishing point 
$(u'_i - q^st^{-r} u'_{i+1})^{P^{(N)}(n-rs)}$ in the polynomial $F$. 
Similarly to the case of the deformed $W_N$-algebra 
(see \cite{BouwknegtPilch:1998:W}), 
by the $\mathfrak{sl}_N$ Weyl group invariance of 
the eigenvalues of $W^{(i)}_0$,  
the polynomial $F$ has the factor 
$(u'_i - q^{s}t^{-r} u'_{j})^{P^{(N)}(n-rs)}$ 
($\forall i,j$). 
Considering the degree of polynomials $F(u'_1,\ldots , u'_N)$, 
we can see that when $N \geq 2$, 
the Kac determinant is 
\begin{align}
\mathrm{det}_n
& = g_{N,n}(q,t) 
    \times (u'')^{2\sum_{\vl \vdash n} \sum_{i=1}^N i\,  \ell(\lambda^{(i)})}
    \times \prod_{1 \leq i<j\leq N}\prod_{\substack{1\leq r,s \\ rs \leq n}}
     \left( (u'_{i}-q^{s} t^{-r}u'_{j})(u'_{i}-q^{-r} t^{s}u'_{j}) \right)^{P^{(N)}(n-rs)} \nonumber \\ 
&= g_{N,n}(q,t) 
    \times\prod_{\substack{1\leq r,s \\ rs \leq n}} 
    \left( (u_1u_2\cdots u_N)^2 
      \prod_{1 \leq i<j\leq N}
      (u_{i}-q^{s} t^{-r}u_{j})(u_{i}-q^{-r} t^{s}u_{j}) \right)^{P^{(N)}(n-rs)}, \label{eq:Kacdet1}
\end{align}
where $g_{N,n}(q,t)$ is a rational function in parameters $q$ and $t$
and independent of the parameters $u_i$. 
If $N=1$, $\mathrm{det}_n$ is clearly in the form 
\begin{equation}
\mathrm{det}_n= g_{1,n}(q,t) \times 
(u_1 \cdots u_N)^{2\sum_{\lambda \vdash n} \ell(\lambda)}. 
\end{equation}

Next, 
the prefactor $g_{N,n}(q,t)$ can be computed in general $N$ case 
by introducing another boson 
\begin{equation}
\sa^{(i)}_{n} \seteq 
-\frac{1-t^{n}}{n}p^{\left( \frac{-i+1}{2}\right) n} a^{(i)}_{n}, 
\quad n \in \mathbb{Z}. 
\end{equation}
The commutation relation of the boson $\sai_n$ is
\begin{equation}
[\sai_n, \saj_{-n}] = -\frac{(1-t^{-n})(1-q^n)}{n}\delta_{i,j}, \quad n>0.
\end{equation}
Define the determinants  
$H^{(n,-)}\seteq \det (H^{(n,-)}_{\vl,\vm})$ and  
$H^{(n,+)}\seteq \det (H^{(n,+)}_{\vl,\vm})$ with 
$H^{(n,\pm)}_{\vl,\vm}$ given by the expansions 
\begin{equation}
\ket{X_{\vl}}=\sum_{\vm \vdash n} H^{(n,-)}_{\vl,\vm} \sa_{-\vm}\ket{\vu}, \quad 
\bra{X_{\vl}}=\sum_{\vm \vdash n} H^{(n,+)}_{\vl,\vm} \bra{\vu} \sa_{\vm} \quad (\vl \vdash n), 
\end{equation}
where $\sa_{-\vm}$ and $\sa_{\vm}$ are 
\begin{align}
&\sa_{-\vl}
:=\sa^{(1)}_{-\lo_1}\sa^{(1)}_{-\lo_2} \cdots 
\sa^{(2)}_{-\lt_1}\sa^{(2)}_{-\lt_2} \cdots 
\sa^{(N)}_{-\lN_1}\sa^{(N)}_{-\lN_2} \cdots,\\
&\sa_{\vl}
:=\cdots \sa^{(N)}_{\lN_2} \sa^{(N)}_{\lN_1}
\cdots \sa^{(2)}_{\lt_2} \sa^{(2)}_{\lt_1}
\cdots \sa^{(1)}_{\lo_2} \sa^{(1)}_{\lo_1}. 
\end{align}
By using these determinants, 
the Kac determinant can be written as
\begin{equation}
\mathrm{det}_n= H^{(n,+)}\, G_n(q,t)\, H^{(n,-)}. 
\end{equation}
Here $G_n(q,t)$ is the determinant of the diagonal matrix 
$(\bra{\vu}\sa_{\vl}\sa_{-\vm}\ket{\vu})_{\vl,\vm \vdash n}$. 
This factor is independent of the parameters $u_i$, 
and we have $G_n(q,t)=\prod_{\vl \vdash n} \prod_{k=1}^N b_{\lambda^{(k)}}(q) b'_{\lambda^{(k)}}(t^{-1})$. 
In (\ref{eq:Kacdet1}), 
the factor depending on $u_i$ in $\mathrm{det}_n$ 
was already clarified. 
Hence, we can determine the prefactor $g_{N,n}(q,t)$ 
by computing the leading term  in $H^{(n,+)}\times H^{(n,-)}$. 
That is, 
the prefactor $g_{N,n}(q,t)$ can be written as 
\begin{equation}
g_{N,n}(q,t) = G_n(q,t) \times 
\left( \mbox{coefficient of } \lc(H^{(n,+)}\times H^{(n,-)};u_1,\ldots,u_N) \right),  
\end{equation}
where we introduce the function $\lc(f;u)$ 
which gives the leading term of $f$ as the polynomial in $u$, 
and 
$\lc(f;u_1,\ldots,u_N)\seteq \lc( \cdots \lc(\lc(f;u_1);u_2)\cdots ;u_N)$. 
To calculate this leading term, 
define the operators $A^{(k)}(z)=\sum_{n\in \mathbb{Z}} A^{(k)}_n z^{-n}$,  
$B^{(k)}(z)=\sum_{n\in \mathbb{Z}}B^{(k)}_n z^{-n}$ by 
\begin{equation}
A^{(k)}(z)=\exp \left\{ \sum_{n>0} \left(\sa^{(k)}_{-n}+\sum_{i=1}^{k-1}p^{(k-i)n} \sai_{-n}\right)z^{n} \right\}, \quad 
B^{(k)}(z)=\exp \left( \sum_{n>0} \sum_{i=1}^k \sai_n z^{-n} \right). 
\end{equation}
$\lc(H^{(n,-)};u_1,\ldots,u_N)$ is arising from only the operator 
$\mathcal{L}^{(k)}(z) \seteq 
\NPb \Lambda_1(z)\Lambda_2(pz) \cdots \Lambda_k(p^{k-1z}) \NPb$ 
in $X^{(k)}(z)$. 
Then 
\begin{equation}
\mathcal{L}^{(k)}(z)
=U_1\cdots U_k  A^{(k)}(z)B^{(k)}(z). 
\end{equation}
Let the matrices $L^{(n,-)}_{\vl,\vm}$ 
and $C^{(n,-)}_{\vl,\vm}$ be given by 
\begin{equation}
\mathcal{L}_{-\vl}\ket{\vu}=\sum_{\vm \vdash n} L^{(n,-)}_{\vl,\vm}\, A_{-\vm}\ket{\vu}, \quad 
A_{-\vl}\ket{\vu}=\sum_{\vm \vdash n} C^{(n,-)}_{\vl,\vm}\, \sa_{-\vm}\ket{\vu} 
\quad (\vl \vdash n), 
\end{equation}
where $\mathcal{L}_{-\vl}$ and $A_{-\vl}$ are defined in the usual way: 
\begin{align}
& \mathcal{L}_{-\vl}
=\mathcal{L}^{(1)}_{-\lo_1}\mathcal{L}^{(1)}_{-\lo_2} \cdots 
\mathcal{L}^{(2)}_{-\lt_1}\mathcal{L}^{(2)}_{-\lt_2} \cdots 
\mathcal{L}^{(N)}_{-\lN_1}\mathcal{L}^{(N)}_{-\lN_2} \cdots,  \\
& A_{-\vl}
=A^{(1)}_{-\lo_1}A^{(1)}_{-\lo_2} \cdots 
A^{(2)}_{-\lt_1}A^{(2)}_{-\lt_2} \cdots 
A^{(N)}_{-\lN_1}A^{(N)}_{-\lN_2} \cdots.
\end{align}
Then $\lc(H^{(n,-)};u_1,\ldots,u_N)$ is expressed as  
\begin{equation}
\lc(H^{(n,-)};u_1,\ldots,u_N)
=\det(L^{(n,-)}_{\vl,\vm})
 \det (C^{(n,-)}_{\vl,\vm}). 
\end{equation}
Since the matrix $L^{(n,-)}_{\vl,\vm}$ is lower triangular 
with respect to the partial ordering $\overset{**}{>}^{\mathrm{R}}$ 
\footnote{ 
Here the partial orderings $\overset{**}{>}^{\mathrm{R}}$ 
and $\overset{**}{>}^{\mathrm{L}}$ 
are defined as follows: 
\begin{align}
\vl \overset{**}{\geq}^{\mathrm{R}} \vm \quad \overset{\mathrm{def}}{\Leftrightarrow} \quad
& |\vl|=|\vm|, \quad 
\sum_{i=1}^k |\lambda^{(i)}|  \geq  \sum_{i=1}^k |\mu^{(i)}| \quad (\forall k) \\
& \mathrm{or} \quad "(|\lo|,\ldots, |\lN|)=(|\mo|,\ldots, |\mN|)  
\quad \mathrm{and } \quad 
\lambda^{(i)} \geq \mu^{(i)} \quad (\forall i)" ,
\end{align}
\begin{align}
\vl \overset{**}{\geq}^{\mathrm{L}} \vm \quad \overset{\mathrm{def}}{\Leftrightarrow} \quad
& |\vl|=|\vm|, \quad 
\sum_{i=K}^N |\lambda^{(i)}|  \geq  \sum_{i=k}^N |\mu^{(i)}| \quad (\forall k) \\
& \mathrm{or} \quad "(|\lo|,\ldots, |\lN|)=(|\mo|,\ldots, |\mN|)  
\quad \mathrm{and } \quad 
\lambda^{(i)} \geq \mu^{(i)} \quad (\forall i)" .
\end{align}
Then we have $L^{(n,-)}_{\vl,\vm}=0$ 
unless $\vl \overset{**}{<}^{\mathrm{R}} \vm$.
}
and its diagonal elements are 
\begin{equation}
L_{\vl, \vl}=u_1^{\sum_{i=1}^{N} \ell(\lambda^{(i)})} u_2^{\sum_{i=2}^{N}\ell(\lambda^{(i)})} \cdots u_N^{\ell(\lambda^{(N)}) }, 
\end{equation}
we have 
\begin{align}
\det(L^{(n,-)}_{\vl,\vm})
&=u_1^{\sum_{\vl}\sum_{i=1}^{N} \ell(\lambda^{(i)})} u_2^{\sum_{\vl}\sum_{i=2}^{N}\ell(\lambda^{(i)})} \cdots u_N^{\sum_{\vl}\ell(\lambda^{(N)}) } \\
&=\prod_{k=1}^N u_k^{k \sum_{\vl}\ell(\lambda^{(N)})} \\
&=\prod_{k=1}^N \prod_{\substack{1\leq r,s\\ rs \leq n}} u_k^{k P^{(N)}(n-rs)}. 
\end{align}
The transition matrix $C^{(-)}_{\vl,\vm}$ is upper triangular 
with respect to the partial ordering $\overset{**}{>}^{\mathrm{L}}$, 
and all diagonal elements are $1$. 
Thus $\det (C^{(-)}_{\vl,\vm})_{\vl,\vm \vdash n}=1$. 
Similarly by considering the base transformation to 
$\bra{\vu} B_{\vl}$, 
it can be seen that 
\begin{equation}
\lc (H^{(n,+)}; u_1, \ldots , u_N)
=\prod_{k=1}^N \prod_{\substack{1\leq r,s\\ rs \leq n}} u_k^{k P^{(N)}(n-rs)}. 
\end{equation}
Therefore the prefactor $g_{N,n}(q,t)$ is  
\begin{equation}
g_{N,n}(q,t)=G_n(q,t)=\prod_{\vl \vdash n} \prod_{k=1}^N b_{\lambda^{(k)}}(q) b'_{\lambda^{(k)}}(t^{-1}). 
\end{equation}
Hence Theorem \ref{thm:KacDet} is proved. 

\subsection{Singular vectors and generalized Macdonald functions}
\label{sec:sing vct and Gn Mac}

In this subsection, 
the singular vectors of the algebra $\mathcal{A}(N)$ are discussed. 
Trivially, 
when $u_i=0$, the Kac determinant (\ref{eq:KacDet for DIM}) 
degenerates, 
and it can be easily seen that 
the vectors $a^{(i)}_{-\lambda}\ket{\vu}$ are singular vectors. 
Since the screening operator $S^{(i)}(z)$ is 
the same one of the deformed $W_N$-algebra, 
the situation of the singular vectors of $\mathcal{A}(N)$ 
except contribution arising when $u_i=0$ 
is the same as the deformed $W_N$-algebra.

We discover that singular vectors 
obtained by the screening currents $S^{(i)}(z)$ 
correspond to generalized Macdonald functions.%
\footnote{
The relation between singular vectors of the $SH^{c}$ algebra 
and the AFLT basis is investigated in \cite{FNMZ:2015:SHc}.}
First, we have the following simple theorem.

\begin{thm}\label{thm:Sing vct and Gn Mac 1}
For a number $i\in \{1, \ldots ,N-1 \}$, 
if $u_{N-i} = q^st^{-r} u_{N-i+1}$ and the other $u_j$ are generic, 
there exists a unique singular vector 
$\tket{\chi^{(i)}_{r,s}}$ in $\mathcal{F}_{\vu}$, 
and it corresponds to the generalized Macdonald function $\ket{P_{\vl}}$ 
with 
\begin{equation}
\vl= (\emptyset , \ldots , \emptyset , \overbrace{(s^r), \emptyset, \ldots , \emptyset}^{i}).
\end{equation}
That is, 
\begin{eqnarray}
\tket{\chi^{(i)}_{r,s}} \propto \ket{P_{(\emptyset , \ldots , \emptyset , (s^r), \emptyset, \ldots , \emptyset)}}. 
\end{eqnarray}
\end{thm}

\proof
Existence and uniqueness are understood by the formula 
for the Kac determinant (\ref{eq:KacDet for DIM}) in the usual way. 
Actually, 
the unique singular vector $\tket{\chi^{(i)}_{r,s}}$ is 
the one of (\ref{eq:sing vct chi(i)rs}). 
Since the screening charges commute with $X^{(1)}_0$, 
the singular vector is an eigenfunction of $\Xo_0$ of the eigenvalue 
$\sum_{i=1}^N v_i$. 
Using the relations 
$u_k=v_k$ for $k \neq N-i, N-i+1$ and 
$u_{N-i+1}=t^r v_{N-i+1}$, $u_{N-i}=t^{-r} v_{N-i}$, 
we have 
\begin{equation}
\sum_{i=1}^N v_i 
= \epsilon_{(\emptyset , \ldots , \emptyset , (s^r), \emptyset, \ldots , \emptyset)}(u_1, \ldots , u_N), 
\end{equation}
where $\epsilon_{\vl}=\epsilon_{\vl}(u_1, \ldots , u_N)$ is the eigenvalue
of the generalized Macdonald functions 
introduced in (\ref{eq:eigenvalue of gn Mac}). 
Thus, 
the singular vector $\tket{\chi^{(i)}_{r,s}}$ 
and the generalized Macdonald function 
$\ket{P_{(\emptyset , \ldots , \emptyset , (s^r), \emptyset, \ldots , \emptyset)}}$ 
are in the same eigenspace of $\Xo_{0}$. 
Moreover, 
by comparing the eigenvalues $\epsilon_{\vl}$, 
it can be shown that the dimension of the eigenspace of the eigenvalue 
$\epsilon_{(\emptyset , \ldots , \emptyset , (s^r), \emptyset, \ldots , \emptyset)}$ 
is $1$ even when $u_{N-i} = q^st^{-r} u_{N-i+1}$. 
Therefore, 
this theorem follows. 
\qed

Let us consider more complicated cases. 
For variables $\alpha^{(k)}$ ($k=1,\ldots, N-1$), 
define the function $\bar{h}^{(i)}$ by 
\begin{equation}
\bar{h}^{(i)}(\alpha^{(k)}) \seteq 
\frac{1}{N} \left( -\alpha^{(1)}-2\alpha^{(2)}- \cdots -(i-1) \alpha^{(i-1)}
+(N-i)\alpha^{(i)} +(N-i-1)\alpha^{(i+1)} +\cdots + \alpha^{(N-1)} \right). 
\end{equation}
Then it satisfies 
$\alpha^{(i)}=\bar{h}^{(i)}(\alpha^{(k)})-\bar{h}^{(i+1)}(\alpha^{(k)})$. 
We focus on the following singular vectors 
\begin{equation}
\ket{\chi_{\vec{r}, \vec{s}}} := 
\oint \prod_{k=1}^{N-1} \prod_{i=1}^{r_k} dz^{(k)}_i 
S^{(1)}(z^{(1)}_1) \cdots S^{(1)}(z^{(1)}_{r_1}) \cdots 
S^{(N-1)}(z^{(N-1)}_1) \cdots S^{(N-1)}(z^{(N-1)}_{r_{N-1}}) 
\tket{\vv}, 
\end{equation}
where 
the parameter $\vv=(v_1,\ldots,v_N)$ is 
$v_i=v'' v'_i$, 
$v'_i=q^{\sqrt{\beta}\bar{h}^{(N-i+1)}(\widetilde{\alpha}^{k}_{\vec{r}, \vec{s}}) } p^{-\frac{N+1}{2}+i}$, and 
for non-negative integers $s_k$ and $r_k$ ($k=1,\ldots, N-1$), 
\begin{equation}
\tilde{\alpha}^{(k)}_{\vec{r}, \vec{s}}\seteq 
\sqrt{\beta} (1-r_k +r_{k+1})-\frac{1}{\sqrt{\beta}}(1+s_k) ,\quad 
r_N:=0. 
\end{equation}
Then the singular vector $\ket{\chi_{\vec{r},\vec{s}}}$ 
is in the Fock module $\mathcal{F}_{\vu}$ 
of the highest weight $\vu=(u_1,\ldots,u_N)$ defined by $u_i=u'' u'_i$, 
$u'_i=q^{\sqrt{\beta}\bar{h}^{(N-i+1)}(\alpha^{k}_{\vec{r}, \vec{s}}) } p^{-\frac{N+1}{2}+i}$, $u''=v''$ and 
\begin{equation}
\alpha^{(k)}_{\vec{r},\vec{s}} \seteq 
\sqrt{\beta} (1+r_k -r_{k-1})-\frac{1}{\sqrt{\beta}}(1+s_k) ,\quad 
r_0:=0. 
\end{equation} 
Now we obtain the following main theorem with respect to 
the generalized Macdonald functions and the singular vectors of the DIM algebra. 
This theorem can be regarded as a generalization of the result in \cite{Awata:1995Quantum}.

\begin{thm}\label{thm:Sing vct and Gn Mac 2}
Let parameters $u_i$ satisfy 
$u_{i} = q^{s_{N-i}} t^{-r_{N-i}+r_{N-i-1}} u_{i+1}$ for all $i$. 

\textbf{(A).} 
If $r_k \geq r_{k+1} \geq 0$ for all $k$, then 
the singular vector $\ket{\chi_{\vec{r},\vec{s}}}$ 
coincides with the generalized Macdonald function 
$\ket{P_{(\emptyset,\ldots ,\emptyset, \lrs)}}$
with $\lrs= ((s_1+\cdots +s_{N-1})^{r_{N-1}}, (s_1+\cdots+s_{N-2})^{r_{N-2}-r_{N-1}},\ldots, s_1^{r_1-r_2} )$: 
\begin{equation}
\ket{\chi_{\vec{r},\vec{s}}} \propto \ket{P_{(\emptyset,\ldots ,\emptyset, \lrs)}}. 
\end{equation}
See Figure \ref{fig:YoungDiag_OnlyRightSide} in Introduction. 

\textbf{(B).} 
If $0 \leq r_k < r_{k+1}$ for all $k$, the singular vector 
$\ket{\chi_{\vec{r},\vec{s}}}$ coincides with 
the generalized Macdonald function associated with the tuple of Young diagrams 
$\varTheta_{\vr,\vs}= (\emptyset, (s_{N-1}^{r_{N-1}-r_{N-2}}), 
((s_{N-2}+s_{N-1})^{r_{N-2}-r_{N-3}}),\ldots, 
((s_1+\cdots+s_{N-1})^{r_1})  )$: 
\begin{equation}
\ket{\chi_{\vr,\vs}} \propto 
\ket{P_{\varTheta_{\vr,\vs}}}. 
\end{equation}
See Figure \ref{fig:YoungDiag_SomeRectangle} in Introduction. 
\end{thm}

\proof
The proof is quite similar to that of Theorem \ref{thm:Sing vct and Gn Mac 1}. 
The eigenvalue of this singular vector is 
\begin{equation}
\sum_{i=1}^N v_i = 
v''\sum_{i=1}^{N} q^{-\bar{h}^{(i)}(s_k)} t^{-r_i +\frac{1}{N}\sum_{k=1}^{N-1}r_k}. 
\end{equation}
On the other hand, 
the eigenvalue of the generalized Macdonald function 
in the case \textbf{(A)} 
is calcurated as follows. 
Firstly, 
\begin{equation}
e_{(\lambda_{\vec{r},\vec{s}})}=
t^{-r_1} +
\sum_{l=1}^{N-1}q^{s_1+\cdots + s_{N-l}} t^{-r_{N-l+1}} 
- \sum_{l=1}^{N-1}q^{s_1+\cdots + s_{N-l}} t^{-r_{N-l}}, 
\end{equation}
and 
\begin{equation}
u'_i=q^{-\bar{h}^{(N-i+1)}(s_k)} t^{-r_{N-i} +\frac{1}{N}\sum_{k=1}^{N-1}r_k}. 
\end{equation}
Hence, by using the equation 
$- \bar{h}^{(1)}(s_k)+s_1+\cdots + s_{i} = -\bar{h}^{(i+1)}(s_k)$, 
we can see that 
\begin{align}
\epsilon_{\emptyset,\ldots \emptyset, \lambda_{\vec{r}, \vec{s}}}
& = u''\sum_{i=1}^{N-1}u'_i + u'_N e_{\lambda_{\vec{r},\vec{s}}} \\
& = u''\sum_{i=1}^{N} q^{-\bar{h}^{(i)}(s_k)} t^{-r_i +\frac{1}{N}\sum_{k=1}^{N-1}r_k}. 
\end{align}
This is equal to the eigenvalue of the singular vector. 
Also, 
it can be seen that the dimension of the eigenspace of the eigenvalue 
$\epsilon_{(\emptyset,\ldots \emptyset, \lambda_{\vec{r}, \vec{s}})}$ 
is $1$ even when $u_{i} = q^{s_{N-i}} t^{-r_{N-i}+r_{N-i-1}} u_{i+1}$.

If the condition $r_k \geq r_{k+1}$ does not hold, 
Figure \ref{fig:YoungDiag_OnlyRightSide} is not a Young diagram. 
In this case, the singular vector $\ket{\chi_{\vr,\vs}}$ corresponds 
to the generalized Macdonald function with the $N$-tuple of Young diagram 
obtained by cutting off the protruding parts and 
moving the boxes to the Young diagram in the left side. 
That is, the case \textbf{(B)}. 
The proof in the case \textbf{(B)} is 
exactly the same as the case \textbf{(A)}, 
so it is omitted. 
\qed

It is known that projections of the singular vectors $\ket{\chi_{\vr,\vs}}$ 
in the case \textbf{(A)} 
onto the diagonal components of the boson $h^{(N)}_n$ 
correspond to ordinary Macdonald functions \cite[(35)]{Awata:1995Quantum}.  
Hence, 
ordinary Macdonald functions are obtained 
by the projection of generalized Macdonald functions. 
\begin{cor}\label{cor:projection of Gn Mac}
When $u_{i} = q^{s_{N-i}} t^{-r_{N-i}+r_{N-i-1}} u_{i+1}$ for all $i$, 
\begin{equation}
P_{\lrs}(p_n;q,t) \propto 
\bra{\vu} \exp \left\{ -\sum_{n>0}p_n \frac{h^{(N)}_n}{1-q^{n}} \right\} \tket{P_{(\emptyset,\ldots ,\emptyset, \lrs)}}. 
\end{equation}
Here, $p_n$ denotes the ordinary power sum symmetric functions. 
\end{cor}

\section{Crystalization of 5D AGT conjecture}
\label{sec:Crystallization}

\subsection{Crystallization of the deformed Virasoro algebra and AGT correspondence.}
\label{sec:crystal of qVir}

Next, we consider a crystallization of the results of 
Subsection \ref{sec:Review of Simplest 5D AGT}, 
namely the behavior in the $q \rightarrow 0$ limit 
of the deformed Virasoro algebra 
and the simplest 5D AGT correspondence.%
\footnote{The results of this section are based on the sub-thesis 
\cite{Ohkubo:2015:Crystallization}. } 
In this limit, 
the scaled generators 
\begin{equation}
\tilde{T}_n \seteq (q/t)^{\frac{|n|}{2}}T_n 
\end{equation}
satisfy the commutation relation
\begin{align}
[\tilde{T}_n,\tilde{T}_m] =&
-(1-t^{-1})\sum_{\ell =1}^{n-m}{\tilde T }_{n-\ell}{\tilde T }_{m+\ell}
\quad (n > m > 0 \quad  \mbox{or}\quad  0>n>m),  
\\ 
[{\tilde T }_n,{\tilde T }_0 ] =&
 -(1-t^{-1})\sum_{\ell =1}^{n}{\tilde T }_{n-\ell}{\tilde T }_{\ell}
-(t-t^{-1})\sum_{\ell =1}^{\infty} t^{-\ell}
{\tilde T }_{-\ell}{\tilde T }_{n+\ell} \quad  (n > 0),  
\\ 
[{\tilde T }_0,{\tilde T }_m] =&
 -(1-t^{-1})\sum_{\ell =1}^{-m}{\tilde T }_{-\ell}{\tilde T }_{m+\ell}
-(t-t^{-1})\sum_{\ell =1}^{\infty} t^{-\ell}
{\tilde T }_{m-\ell}{\tilde T }_{\ell} \quad (0 > m),  \allowdisplaybreaks[4]
\\ 
[{\tilde T }_n,{\tilde T }_m] =&
-(1-t^{-1}){\tilde T }_{m}{\tilde T }_{n} 
-(t-t^{-1})\sum_{\ell =1}^{\infty} t^{-\ell}
{\tilde T }_{m-\ell}{\tilde T }_{n+\ell} \nonumber
\\ 
 & +(1-t^{-1})\delta_{n+m,0}
\quad (n> 0> m).
\end{align}

In \cite{Awata:1996Virasoro}, 
the above algebra is introduced 
and its free field representation is given. 
Let the bosons $b_n$ ($n \in \mathbb{Z}$) satisfy the relations 
$[b_n, b_m]=n \frac{1}{1-t^{|n|}}\delta_{n+m,0}$, 
$[b_n, \chargQ] = \delta_{n,0}$. 
These bosons can be regarded as the $q\rightarrow 0$ 
limit of the bosons $a_n$ and $Q$ in (\ref{eq:comm rel of qt-boson}), 
i.e., $b_n= \lim_{q \rightarrow 0} a_n$, $\chargQ=\lim_{q \rightarrow 0} \chargQ$. 
Then $ {\tilde T }_n$ is represented as
\begin{equation}
{\tilde T }_n =\oint\frac{dz}{2 \pi \sqrt{-1} z}
\left( \theta[\,n\leq0\,] \tilde{\Lambda}^+(z) + 
           \theta[\,n\geq0\,] \tilde{\Lambda}^-(z)
\right) z^n,
\end{equation}
where
\begin{equation}
\tilde{\Lambda}^{\pm}(z)
 :=   \exp\left\{ \pm \sum_{n=1}^{\infty} \frac{1-t^{-n}}{n} b_{-n} z^{n} \right\}
       \exp\left\{ \mp \sum_{n=1}^{\infty} \frac{1-t^{n}}{n} b_{n} z^{-n} \right\} K^{\pm}
   = \lim_{q \rightarrow 0} \Lambda^{\pm}(p^{\pm 1/2} z) 
\end{equation}
and $\theta[P]$ is 1 or 0 if the proposition $P$ is true or false, respectively. 
By this free field representation, 
we can write the PBW type vectors in terms of Hall-Littlewood functions 
$Q_{\lambda}$ defined in Appendix \ref{sec: Macdonald and HL} : 
\begin{align}
\tT_{-\lambda} \ket{k} &= k^{\ell(\lambda)} Q_{\lambda}(b_{-n};t^{-1}) \ket{k}, \\
\bra{k} \tT_{\lambda} &= k^{-\ell(\lambda)} t^{|\lambda|}\bra{k} Q_{\lambda}(-b_n; t^{-1}).  
\end{align}
Here $Q_{\lambda}(b_{-n};t^{-1})$ is an abbreviation for $Q_{\lambda}(b_{-1}, b_{-2}, \ldots;t^{-1})$, 
and 
$\ket{k}$ and $\bra{k}$ are the same highest weight vectors 
in Section \ref{sec:Review of Simplest 5D AGT} 
such that $K^{\pm}\ket{k}=k^{\pm 1}\ket{k}$ and 
$\bra{k}K^{\pm}=k^{\pm 1}\bra{k}$.

These expressions are the consequences of Jing's operators 
(Fact \ref{fact:Jing's operator}). 
Because of (\ref{eq:inner prod of HL poly}), they are diagonalized as 
\begin{equation}
\tilde{B}_{\lambda, \mu} 
\seteq \tbraket{\tT_{\lambda}}{ \tT_{\mu}} 
= \frac{1}{b_{\lambda}(t^{-1})} \delta_{\lambda, \mu}, 
\end{equation}
where $b_{\lambda}(t)$ is defined in Appendix \ref{sec: Macdonald and HL}. 
Since $\tilde{B}_{\lambda, \mu} $ is non-degenerate, 
there is no singular vector in the limit $q \rightarrow 0$. 
The disappearance of singular vectors can be understood by the fact 
that the highest weight which has singular vectors diverges at $q=0$. 
The Whittaker vector of this algebra is similarly defined.

\begin{df}
Define the Whittaker vector $\tket{\tilde{G}}$ by the relation
\begin{equation}
\tilde{T}_1\tilde{\ket{G}}=\tilde{\Lambda}^2 \tilde{\ket{G}}, \qquad \tilde{T}_n \tilde{\ket{G}}=0 \quad (n>1).
\end{equation}
Similarly, the dual Whittaker vector $\tilde{\bra{G}}\in M_h^*$ is defined by 
\begin{equation}
\tilde{\bra{G}}\tilde{T}_{-1}=\tilde{\Lambda}^2\tilde{\bra{G}}, \qquad \tilde{\bra{G}} \tilde{T}_{n}=0 \quad (n<-1).
\end{equation}
\end{df}

Then the crystallized Whittaker vector is in the simple form
\begin{equation}
\tket{\tilde{G}} 
= \sum_{\lambda} \tL^{2 |\lambda|}\, \tilde{B}^{\lambda, \mu} \, \tket{\tT_{\lambda}}
= \sum_{n=0}^{\infty} \tL^{2n}\, \frac{1}{ b_{(1^n)}(t^{-1}) } \, \tket{\tT _{-(1^n)}}, 
\end{equation}
and its inner product is 
\begin{equation}
\tbraket{\tilde{G}}{\tilde{G}} 
= \sum_{n=0}^{\infty} \tL^{4n} \tilde{B}^{(1^n), (1^n)}
= \sum_{n=0}^{\infty} \tL^{4n} \, \frac{1}{b_{(1^n)} (t^{-1})}. 
\end{equation}

On the other hand, 
recalling the Nekrasov formula $Z^{\mathrm{inst}}_{\mathrm{pure}}$ 
given in (\ref{eq:Nek formula for pure}) of Subsection \ref{sec:Review of Simplest 5D AGT}, 
we can take the crystal limit with the following trick.

\begin{prop}
The renormalization $\tL^2\seteq \Lambda^2 (q/t)^{\frac{1}{2}}$ 
controls divergence in the $q \rightarrow 0$ limit ($\Lambda \rightarrow \infty$, $\tL$ : fixed):   
\begin{equation}
\lim_{\substack{\Lambda^2=\tL^2(t/q)^{\frac{1}{2}} \\ q \rightarrow 0}}
Z^{\mathrm{inst}}_{\mathrm{pure}}
=\tilde{Z}^{\mathrm{inst}}_{\mathrm{pure}},
\end{equation}
\begin{equation}\label{eq:tildeZ}
\tilde{Z}^{\mathrm{inst}}_{\mathrm{pure}} 
\seteq \sum_{n,m\geq 0} 
{
\tilde \Lambda^{4(n+m)}
\over
\prod_{s=1}^n (1-t^{-s})(1-Q^{-1}t^{n-m-s})
\prod_{s=1}^m (1-t^{-s})(1-Q     t^{m-n-s})
}.
\end{equation}
\end{prop}

\proof
Removing parts which have singularity in the Nekrasov factor, we have
\begin{align}
N_{\lambda \mu}(Q) &=q^{-\sum_{(i,j)\in \mu}j} N'_{\lambda \mu}(Q),  \\
N'_{\lambda \mu}(Q) &:= 
  \prod_{(i,j)\in \lambda} \left( 1- Q q^{A_{\lambda}(i,j)}t^{L_{\mu}(i,j)+1} \right)  
  \prod_{(i,j)\in \mu} \left( q^{A_{\mu}(i,j)+1}- Q  t^{-L_{\lambda}(i,j)} \right).
\end{align}
Hence, 
\begin{equation}
Z^{\mathrm{inst}}_{\mathrm{pure}} 
= \sum_{\lambda, \mu} \frac{(\tilde{\Lambda}^4 t^2)^{|\lambda|+|\mu|} q^{E_{\lambda \mu}}}
{N'_{\lambda \lambda}(1)N'_{\lambda \mu}(Q)N'_{\mu \mu}(1)N'_{\mu \lambda}(Q^{-1})},
\end{equation}
\begin{equation}
E_{\lambda \mu} := 2\left( \sum_{(i,j) \in \lambda}j+ \sum_{(i,j) \in \mu} j -|\lambda|-|\mu| \right).
\end{equation}
If $\lambda \neq (1^n)$ or $\mu \neq (1^m)$ for any  integer $n$, $m$, 
then $q^{E_{\lambda \mu}} \rightarrow 0$ at $q \rightarrow 0$. 
Therefore, 
the sum with respect to partitions $\lambda$, $\mu$ can be rewritten 
as the sum with respect to integers $n$, $m$, 
i.e., 
\begin{align}
\tilde{Z}^{\mathrm{inst}}_{\mathrm{pure}} 
&= \sum_{n, m} \frac{(\tilde{\Lambda}^4 t^2)^{m+n} }
{\tilde{N}'_{nn}(1)\tilde{N}'_{nm}(Q)\tilde{N}'_{mm}(1)\tilde{N}'_{mn}(Q^{-1})}, \\
\tilde{N}'_{nm}(Q) &=(-1)^m Q^m t^{-nm+\frac{1}{2}m(m+1)} \prod_{s=1}^n \left( 1-Q t^{m-s+1} \right). 
\end{align}
After some simple calculation, we get (\ref{eq:tildeZ}). 
\qed

Using these calculations, 
we can get the following theorem 
which is an analog of the simplest 5D AGT relation 
(Fact \ref{fact:simpleAGT}), 
and prove it more easily than the generic case. 
\begin{thm}
\begin{equation} 
\tbraket{\tilde{G}}{\tilde{G}}
= \tilde{Z}^{\mathrm{inst}}_{\mathrm{pure}}. 
\end{equation} 
Note that the left hand side is independent of $k$. 
\end{thm}

\Proof
$\tilde{Z}^{\mathrm{inst}}_{\mathrm{pure}}= \tilde{Z}^{\mathrm{inst}}_{\mathrm{pure}}(Q)$ can be rewritten as 
\begin{equation}
\tilde{Z}^{\mathrm{inst}}_{\mathrm{pure}}(Q)
=\sum_{n,m\geq 0} 
{
\tilde \Lambda^{4(n+m)} Q^n t^{(n+1)m}
\over
\prod_{s=1}^n (1-t^{-s})(Q-t^{n-m-s})
\prod_{s=1}^m (1-t^{ s})(Q-t^{n-m+s})
},
\end{equation}
which has simple poles at
$Q=t^M$
with 
$-m\leq M\leq n$, $M\neq n-m$ and $M\in \mathbb{Z}$.
Then 
\begin{eqnarray}
{\rm Res}_{Q=t^M} \tilde{Z}^{\mathrm{inst}}_{\mathrm{pure}}(Q)
&=&
\sum_{n,m\geq 0} 
\tilde \Lambda^{4(n+m)} Z_{(n,m)}^{(M)}
,
\cr
Z_{(n,m)}^{(M)}
&:=&
{
t^{nM+(n+1)m} (t^M-t^{n-m})
\over
\prod_{s=1}^n (1-t^{-s})
\prod_{s=-m, (s\neq M)}^n (t^M-t^s)
\prod_{s=1}^m (1-t^{ s})
}
\cr
&=&
{
t^{(n-M)m} (t^{m+M}-t^{n})
\over
\prod_{s=1}^n (1-t^{-s})
\prod_{s=1}^{m+M} (1-t^{-s})
\prod_{s=1}^m (1-t^{ s})
\prod_{s=1}^{n-M} (1-t^{ s})
}.
\end{eqnarray}
Note that
\begin{equation}
Z_{(n,m)}^{(M)}
+
Z_{(m+M,n-M)}^{(M)} 
=0.
\end{equation}
Thus
\begin{equation}
Z_{\left( {N+M\over 2}+ r, {N-M\over 2}-r \right)}^{(M)}
=
{
t^{\left({N-M\over 2}\right)^2 + {N+M\over 2}-r^2 } (t^{-r}-t^{r})
\over
\prod_{s=1}^{{N+M\over 2}+r} (1-t^{-s})
\prod_{s=1}^{{N+M\over 2}-r} (1-t^{-s})
\prod_{s=1}^{{N-M\over 2}-r} (1-t^{ s})
\prod_{s=1}^{{N-M\over 2}+r} (1-t^{ s})
}
\end{equation}
is an odd function in $r$.
Therefore
\begin{equation}
{\rm Res}_{Q=t^M} \tilde{Z}^{\mathrm{inst}}_{\mathrm{pure}}(Q)
=
\sum_{N\geq 0} 
\tilde \Lambda^{4N} \sum_{r={|M|-N\over 2},(r\neq 0)}^{{N-|M|\over 2}}
Z_{\left( {N+M\over 2}+ r, {N-M\over 2}-r \right)}^{(M)}
=0.
\end{equation}
Residues at all singularities in $Q$ of 
$\tilde{Z}^{\mathrm{inst}}_{\mathrm{pure}}(Q)$ vanish, 
but $|\tilde{Z}^{\mathrm{inst}}_{\mathrm{pure}}(\infty)|<\infty$.
Hence $\tilde{Z}^{\mathrm{inst}}_{\mathrm{pure}}(Q)$ is independent of $Q$.
Therefore,
\begin{equation}
\tilde{Z}^{\mathrm{inst}}_{\mathrm{pure}}(Q)
=
\tilde{Z}^{\mathrm{inst}}_{\mathrm{pure}}(0)
=
\sum_{m\geq 0} 
{
\tilde \Lambda^{4m}
\over
\prod_{s=1}^m (1-t^{-s})
}.
\end{equation}

\qed

In this paper, 
we discuss the crystallization only of the deformed Virasoro algebra. 
It is expected that the limit can be taken 
for the general deformed $W_N$-algebra. 
However in the case of $W_3$, 
an essential singularity seems to appear, 
and at present we do not know how to take an appropriate limit. 
To find an appropriate limit procedure 
and apply the AGT conjecture for the deformed $W_N$-algebra 
\cite{Taki:2014:AGTW}
we need further studies. 
In the crystallized case, 
the screening current diverges, 
which is one of the reasons why in this limit singular vectors disappear. 
Hence 
it may be difficult to apply the AGT correspondence studied by \cite{AwataYamada2}. 

\subsection{Crystallization of $N=1$ case of DIM algebra}
\label{sec:crystallization of N=1}

Next, 
we discuss a crystallization of the results of 
Subsection \ref{sec:Reargument of DI alg and AGT}. 
In this subsection and the next subsection, 
unlike Section \ref{sec:Kac det and sing vct}, 
the operators $U_i$ are assumed to be independent of the parameter $q$
in order to avoid difficulty in taking the $q\rightarrow 0$ limit. 
Let us realize the operators $U_i$ and the vector $\ket{\vu}$ as 
\begin{equation}
U_i \seteq e^{a_0^{(i)}}, \quad 
\ket{\vec{u}} \seteq \prod_{i=1}^{N} u_i^{\chargQ^{(i)}} \ket{0}. 
\end{equation}
Then they also satisfy relation $U_i\ket{\vec{u}}=u_i \ket{\vec{u}}$. 
Similarly $\bra{\vu} \seteq \bra{0} \prod_{i=1}^{N} u_i^{-\chargQ^{(i)}}$. 
Moreover, we consider the case 
that the parameters $u_i$ are independent of $q$. 
The case where the parameters $u_i$ depend on $q$ is 
briefly described in Section \ref{sec:Another type of limit}.

At first, let us demonstrate the $q\rightarrow 0$ limit in the $N=1$ case. 
In this subsection, we use the same bosons $b_n$ and $\chargQ$ 
as subsection \ref{sec:crystal of qVir}. 
Since singularity in $\Phi(z)$ can be removed by normalization $\Phi(p z)$, 
define the vertex operator $\tPhi(z)$ by 
\begin{equation}
\tPhi(z)\seteq \lim_{q \rightarrow 0} \Phi(pz) 
= \exp \left\{ \sum_{n=1}^{\infty} \frac{u^n}{n} b_{-n}z^n \right\}
\exp \left\{ \sum_{n=1}^{\infty} \frac{1}{n} \frac{u^{-n}- v^{-n}}{t^{-n}}b_{n}z^{-n} \right\} 
(v/u)^{\chargQ}.  
\end{equation}
If $N=1$, $\tket{P_{\vl}}$ are ordinary Macdonald functions, 
and their integral forms $\ket{K_{\lambda}}$ have, at $q=0$, the relation 
\begin{align}
 &\tket{\tK_{\lambda}} \seteq \lim_{q \rightarrow 0} \ket{K_{\lambda}} 
  = (-u/t)^{|\lambda|} t^{-n(\lambda)} Q_{\lambda}(b_{-n};t) \ket{u},  \\
 &\tbra{\tK_{\lambda}} \seteq \lim_{q \rightarrow 0} \bra{K_{\lambda}} 
  = (-u)^{|\lambda|} t^{-n(\lambda)} \bra{u}Q_{\lambda}(b_{n};t). 
\end{align}
Hence, the matrix elements $\tbra{\tK_{\lambda}}\tPhi(x) \tket{\tK_{\mu}}$ 
can be written in terms of integrals 
by virtue of Jing's operators $H(z)$ and $H^{\dagger}(z)$ 
defined in (\ref{eq:Jing's op}) and (\ref{eq:dual Jing's op}). 
Using the usual normal ordered product $\NPb \quad \NPb$ with respect to the bosons $b_n$,%
\footnote{
	Let $\mathcal{H}$ be the Heisenberg algebra generated by the bosons $b_n$ ($n\in \mathbb{Z}$), 
	$\chargQ$ and $1$.  
	$\mathcal{H}_c$ is the algebra obtained by making $\mathcal{H}$ commutative. 
	The normal ordered product $\NPb \quad \NPb$ is defined to be 
	the linear map from $\mathcal{H}_c$ to $\mathcal{H}$ 
	such that  
	for $\mathcal{P} \in \mathcal{H}_c$, 
	\begin{equation}
	\NPb \mathcal{P} b_n \NPb =
	  \left\{
	  \begin{array}{l}
	  \NPb \mathcal{P} \NPb \, b_n   , \quad n \geq 0,   \\
	  b_{n} \, \NPb \mathcal{P} \NPb , \quad n<0,
	  \end{array}
	  \right.
	\quad 
	\NPb \mathcal{P} \chargQ \NPb = \chargQ \NPb \mathcal{P} \NPb, 
	\end{equation}
	and $\NPb 1 \NPb =1$. 
	In the next subsection, 
	the same symbol $\NPb \quad \NPb$ denotes 
	the normal ordered product with respect to the bosons $b^{(i)}_n$ 
	which is defined similarly. 
	}
we have 
\begin{align}
& H^{\dagger}(w_{\ell(\lambda)}) \cdots H^{\dagger}(w_{1}) \tPhi(x) H(z_{1}) \cdots H(z_{\ell(\mu)}) \nonumber \\ 
& \qquad \qquad=\mathfrak{I}(w,x,z) \NPb H^{\dagger}(w_{\ell(\lambda)}) \cdots H^{\dagger}(w_{1}) \tPhi(x) H(z_{1}) \cdots H(z_{\ell(\mu)}) \NPb,\\
&\mathfrak{I}(w,x,z) \seteq \prod_{1 \leq j< i\leq \ell(\lambda)} \left( \frac{w_i-w_j}{w_i-tw_j} \right) 
    \prod_{1 \leq i< j\leq \ell(\mu)} \left( \frac{z_i-z_j}{z_i-tz_j} \right) 
    \prod_{\substack{1 \leq i \leq \ell(\lambda) \\ 1 \leq j \leq \ell(\mu)}} \left( \frac{w_i-tz_j}{w_i-z_j} \right)  \nonumber \\
 & \qquad \qquad \qquad \times \prod_{1 \leq i \leq \ell(\mu)} \left( \frac{x-(t/v)z_i}{x-(t/u)z_i} \right) 
    \prod_{1 \leq i \leq \ell(\lambda)} \left( \frac{w_i}{w_i-ux} \right). 
\end{align}
Thus
\begin{equation}
\tbra{\tK_{\lambda}}\tPhi(x) \tket{\tK_{\mu}} =
(-u)^{|\lambda|+|\mu|} t^{-n(\lambda)-n(\lambda)-|\mu|}
\oint\frac{dz}{2 \pi \sqrt{-1} z} \frac{dw}{2 \pi \sqrt{-1} w} \mathfrak{I}(w,x,z) z^{-\mu} w^{\lambda}, 
\end{equation}
where 
$\displaystyle \oint \frac{dz}{\twopii z} \frac{dw}{\twopii w} 
\seteq \oint \prod_{i=1}^{\ell(\mu)} \frac{dz_i}{2\pi \sqrt{-1}z_i} 
\prod_{i=1}^{\ell(\lambda)} \frac{dw_i}{2\pi \sqrt{-1}w_i}$, 
$z^{-\mu}\seteq z_1^{-\mu_1} \cdots z_{\ell(\mu)}^{-\mu_{\ell(\mu)}}$, 
$w^{\lambda}\seteq w_1^{\lambda_1} \cdots w_{\ell(\lambda)}^{\lambda_{\ell(\lambda)}}$, 
and  
the integration contour is $|w_{\ell(\lambda)}|> \cdots >|w_1|>|x|>|z_1|>\cdots |z_{\ell(\mu)}|$. 
This integral reproduces the $q \rightarrow 0$ limit of the Nekrasov factor.

\begin{df}\label{def:crystal Nek factor}
Set
\begin{align}
	\tilde{N}_{\lambda \mu}(Q) & \seteq \lim_{q \rightarrow 0} q^{n(\mu')} N_{\lambda \mu}((q/t)Q) \\
	&= (-Qt^{-1})^{|\check{\mu}|} \, t^{-\sum_{(i,j) \in \check{\mu}} L_{\lambda}(i,j)} 
	\prod_{(i,j)\in \mu \smallsetminus \check{\mu}} \left( 1- Q t^{-L_{\lambda}(i,j)-1} \right), 
	\nonumber
\end{align}
where $\check{\mu}$ is the set of boxes in $\mu$ whose arm length $A_{\mu}(i,j)$ is not zero. 
For example, 
if $\mu=(5,3,3,1)$, $\check{\mu}=(4,2,2)$. 
This Nekrasov factor has 
the property $\tilde{N}_{\lambda \emptyset}(Q)=1$ for any $\lambda$.
\end{df}

Therefore, the conjecture in the crystallized case of $N=1$ is 
\begin{equation}\label{eq:N=1 conjecture}
\oint\frac{dz}{\twopii z} \frac{dw}{\twopii w} \mathfrak{I}(w,x,z) z^{-\mu} w^{\lambda}
\overset{?}{=} \tN_{\lambda, \mu} (v/u) x^{|\lambda|-|\mu|} u^{|\lambda|} (-v)^{|\mu|} t^{|\mu|+n(\lambda)}.
\end{equation}
The case of some particular partitions 
can be checked by calculating the contour integral 
(Appendix \ref{sec:check of N=1 conjecture}). 

\subsection{Crystallization of $N=2$ case of DIM algebra}
\label{sec:crystallization of N=2}

Next, let us consider the $q\rightarrow 0$ limit in the case of $N=2$. 
In this case, the generator $a^{(1)}_n$ of the Heisenberg algebra 
is renormalized as 
\begin{equation}
a^{(1)}_n \mapsto p^{-n/2} a^{(1)}_n, 
\end{equation}
and the generator $a^{(2)}_n$ is used as it is. 
By this normalization, 
it is possible to take the limit. 
Also, the algebraic structure of $\Lambda^1$ and $\Lambda^2$ does not change. 
Then $\Lambda^1$ and $\Lambda^2$ have the form 
\begin{align}
 &\Lambda^{1} (z) \seteq 
 \exp \left( \sum_{n=1}^{\infty} \frac{1-t^{-n}}{n}\, p^{n/2}\, z^{n} a^{(1)}_{-n} \right) 
 \exp \left( -\sum_{n=1}^{\infty}\frac{(1-t^n)}{n} p^{-n/2} z^{-n} a^{(1)}_n \right) U_1, \\
 &\Lambda^{2}(z) \seteq 
 \exp \left( - \sum_{n=1}^{\infty} \frac{1-t^{-n}}{n} (1-p^{n}) p^{-n/2} z^{n} a^{(1)}_{-n} \right) \\
 &\qquad \qquad \times
 \exp \left( \sum_{n=1}^{\infty} \frac{1-t^{-n}}{n}\, p^{-n/2}\, z^{n} a^{(2)}_{-n} \right) 
 \exp \left( -\sum_{n=1}^{\infty}\frac{(1-t^n)}{n} p^{n/2} z^{-n} a^{(2)}_n \right) U_2. 
\end{align}
Moreover, 
let us use the bosons $b^{(i)}_n$ ($n\in \mathbb{Z},\, i=1,2$) and 
$\chargQ^{(i)}$ with the relation
\begin{equation}
[b^{(i)}_n,b^{(j)}_m]=n\frac{1}{1-t^{|n|}} \delta_{i,j} \, \delta_{n+m, 0}, \quad 
[b_n^{(i)}, \chargQ^{(j)}] =0, 
\end{equation}
and regard $b^{(i)}_n=\lim_{q \rightarrow 0} a^{(i)}_n$, $\chargQ^{(i)} = \lim_{q \rightarrow 0} \chargQ^{(i)}$. 
Let us define the generator at $q \rightarrow 0$.

\begin{df}\label{def:crystal first generator}
Set
\begin{equation}
\tilde{X}^{(1)}_n \seteq \lim_{q \rightarrow 0} p^{\frac{|n|}{2}} X^{(1)}_n. 
\end{equation}
\end{df}

\begin{prop}\label{prop:free field rep of X^1}
Definition \ref{def:crystal first generator} is well-defined, 
i.e., $p^{\frac{|n|}{2}} X^{(1)}_n$ has no singularity at $q=0$, 
and its free field representation is 
\begin{equation}\label{eq:free field rep of tXo}
\tilde{X}^{(1)}_n=\oint \frac{dz}{2 \pi \sqrt{-1} z} 
\left\{ \theta [n\geq 0] \tilde{\Lambda}^{1}(z) + \theta [n\leq 0] \tilde{\Lambda}^{2}(z) \right\} z^n, 
\end{equation}
where $\theta$ is defined in Section \ref{sec:crystal of qVir} and
\begin{align}
&\tilde{\Lambda}^{1}(z) \seteq 
\exp \left\{ \sum_{n>0} \frac{1-t^{-n}}{n}z^n b^{(1)}_{-n} \right\}
\exp \left\{-\sum_{n>0} \frac{1-t^{n}}{n}z^{-n} b^{(1)}_{n} \right\}U_1, \\
 &\tilde{\Lambda}^{2}(z)
\seteq 
\exp \left\{-\sum_{n>0} \frac{1-t^{-n}}{n}z^n b^{(1)}_{-n} \right\}
\exp \left\{\sum_{n>0} \frac{1-t^{-n}}{n}z^n b^{(2)}_{-n} \right\}
\exp \left\{-\sum_{n>0} \frac{1-t^{n}}{n}z^{-n} b^{(2)}_{n} \right\}U_2. 
\end{align}
\end{prop}

\Proof
Define $\Lambda^1_n$ and $\Lambda^2_n$ by 
\begin{equation}
\Lambda^1(z) \rseteq \sum_{n \in \mathbb{Z}} \Lambda^1_n p^{-n/2}z^{-n} , \quad
\Lambda^2(z) \rseteq \sum_{n \in \mathbb{Z}} \Lambda^2_n p^{n/2}z^{-n}, 
\end{equation}
we can see $\Lambda^i_n$ is well-behaved in the limit $q \rightarrow 0$ by the form of $\Lambda^i(z)$. 
If $n>0$, 
\begin{equation}
\tilde{X}^{(1)}_n = \lim_{q \rightarrow 0}( \Lambda^1_n +\Lambda^2_n p^n )
=\lim_{q \rightarrow 0} \Lambda^1_n
= \oint \frac{dz}{2 \pi \sqrt{-1} z} \tilde{\Lambda}^1(z) z^n,  
\end{equation}
if $n<0$, 
\begin{equation}
\tilde{X}^{(1)}_n = \lim_{q \rightarrow 0}( \Lambda^1_n p^{-n} +\Lambda^2_n )
=\lim_{q \rightarrow 0} \Lambda^2_n
=\oint \frac{dz}{2 \pi \sqrt{-1} z} \tilde{\Lambda}^2(z) z^n,  
\end{equation}
and if $n=0$, 
\begin{equation}
\tilde{X}^{(1)}_n = \lim_{q \rightarrow 0}( \Lambda^1_0  +\Lambda^2_0 )
=\oint \frac{dz}{2 \pi \sqrt{-1} z} (\tilde{\Lambda}^1(z)+\tilde{\Lambda}^2(z) ).  
\end{equation}
Thus $\tilde{X}^{(1)}_n$ is well-defined 
and (\ref{eq:free field rep of tXo}) is the natural free field representation. 
\qed

For the second generator, the following rescale is suitable.

\begin{df}
Set
\begin{equation}
\tilde{X}^{(2)}_n \seteq \lim_{q \rightarrow 0} p^{\frac{n}{2}} X^{(2)}_n. 
\end{equation}
\end{df}

\begin{prop}
The free field representation of $\tilde{X}^{(2)}_n$ is given by 
\begin{equation}
\tilde{X}^{(2)}_n  = \oint \frac{dz}{2 \pi \sqrt{-1} z} \tXt(z) z^n,
\end{equation}
where
\begin{align}
& \tXt(z) \seteq \NPb \tLo(z) \tLt(z) \NPb \\
 &= \exp \left\{ \sum_{n>0} \frac{1-t^{-n}}{n}z^n b^{(2)}_{-n} \right\}
\exp \left\{-\sum_{n>0} \frac{1-t^{n}}{n}z^{-n} b^{(1)}_{n} \right\}
\exp \left\{-\sum_{n>0} \frac{1-t^{n}}{n}z^{-n} b^{(2)}_{n} \right\}U_1 U_2. \nonumber
\end{align}
\end{prop}
This proposition is easily obtained 
by calculating $\lim_{q \rightarrow 0} X^{(2)}(p^{-1/2}z)$. 
We can calculate the commutation relation of these generators as follows.

\begin{prop}
The generators $\tilde{X}^{(1)}_n$ and $\tilde{X}^{(2)}_n$ satisfy the relations 
\begin{align}
[\tilde{X}^{(1)}_n, \tilde{X}^{(1)}_m] 
  &= -(1-t^{-1}) \sum_{l=1}^{n-m} \tilde{X}^{(1)}_{n-l} \tilde{X}^{(1)}_{m+l} 
  \qquad (n>m>0 \;\; \mathrm{or} \;\; 0>n>m), \\
[\tilde{X}^{(1)}_n, \tilde{X}^{(1)}_0] 
  &= -(1-t^{-1}) \sum_{l=1}^{n-1} \tilde{X}^{(1)}_{n-l} \tilde{X}^{(1)}_{l}  
     -(1-t^{-1}) \sum_{l=1}^{\infty} \tilde{X}^{(1)}_{-l} \tilde{X}^{(1)}_{n+l}  
     +(1-t^{-1}) \tilde{X}^{(2)}_{n}  
  \quad (n>0 ), \\
[\tilde{X}^{(1)}_n, \tilde{X}^{(1)}_m] 
  &= -(1-t^{-1})\sum_{l=0}^{\infty} \tilde{X}^{(1)}_{m-l} \tilde{X}^{(1)}_{n+l} + (1-t^{-1}) \tilde{X}^{(2)}_{n+m}  
   \quad (n>0>m ), \\
[\tilde{X}^{(1)}_0, \tilde{X}^{(1)}_m] 
  &= -(1-t^{-1}) \sum_{l=1}^{-m-1} \tilde{X}^{(1)}_{-l} \tilde{X}^{(1)}_{m+l}  
     -(1-t^{-1}) \sum_{l=1}^{\infty} \tilde{X}^{(1)}_{m-l} \tilde{X}^{(1)}_{l}  
     +(1-t^{-1}) \tilde{X}^{(2)}_{m}  
 \quad (0>m ),
\end{align}
\begin{align}
[\tilde{X}^{(1)}_n, \tilde{X}^{(2)}_m]  
  &= (1-t^{-1}) \sum_{l=1}^{\infty} \tilde{X}^{(2)}_{m-l} \tilde{X}^{(1)}_{n+l}
  \qquad (n>0,\; \forall m), \\
[\tilde{X}^{(1)}_0, \tilde{X}^{(2)}_m]  
  &= -(1-t^{-1}) \sum_{l=1}^{\infty} ( \tilde{X}^{(1)}_{-l} \tilde{X}^{(2)}_{m+l}- \tilde{X}^{(2)}_{m-l} \tilde{X}^{(1)}_{l}) 
  \quad ( \forall m),  \\
[\tilde{X}^{(1)}_n, \tilde{X}^{(2)}_m]  
  &= -(1-t^{-1}) \sum_{l=1}^{\infty} \tilde{X}^{(1)}_{n-l} \tilde{X}^{(2)}_{m+l}
  \qquad (n<0,\; \forall m),  
\end{align}
\begin{equation}
[\tilde{X}^{(2)}_n, \tilde{X}^{(2)}_m]
 =-(1-t^{-1}) \sum_{l=1}^{\infty} ( \tilde{X}^{(2)}_{n-l} \tilde{X}^{(2)}_{m+l}- \tilde{X}^{(2)}_{m-l} \tilde{X}^{(2)}_{n+l}) 
 \qquad (\forall n,m).
\end{equation}
\end{prop}

\Proof
These are obtained by the following relation of generating functions: 
\begin{align}
 \displaystyle g\left( \frac{w}{z} \right) \tLo(z) \tLo(w)- g \left( \frac{z}{w} \right) \tLo(w) \tLo(z)  &= 0,\\
 \displaystyle g\left( \frac{w}{z} \right) \tLt(z) \tLt(w)- g \left( \frac{z}{w} \right) \tLt(w) \tLt(z)  &= 0,\\
 \displaystyle \tLo(z) \tLt(w) + \left( g\left( \frac{z}{w} \right) -1-t^{-1} \right) \tLt(w) \tLo(z) 
&=   (1-t^{-1}) \delta \left( \frac{w}{z} \right) \NPb \tLo(z) \tLt(w) \NPb, \label{relation of tLo tLt}  
\\
 \displaystyle \tLo(z) \tXt(w) - g\left( \frac{z}{w} \right)\tXt(w) \tLo (z) &= 0, \\
 \displaystyle g\left( \frac{w}{z} \right) \tLt(z) \tXt(w) - \tXt(w) \tLt (z) &= 0, \\
 \displaystyle g\left( \frac{w}{z} \right) \tXt(z) \tXt(w) - g\left( \frac{z}{w} \right)\tXt(w) \tXt (z) &=0, 
\end{align}
where
\begin{equation}
g\left( x \right)=\exp \left\{ \sum_{n>0} \frac{1-t^{-n}}{n}x^n \right\} =1+ (1-t^{-1})\sum_{l=1}^{\infty} x^l, 
\end{equation}
and for (\ref{relation of tLo tLt}) we used the formula 
\begin{equation}
g(x)+g(x^{-1}) -1-t^{-1}= +(1-t^{-1}) \delta(x).
\end{equation}
\qed

The algebra generated by $\tXo_n$ and $\tXt_n$ is closely related to 
the Hall-Littlewood functions. 
In particular,  
the PBW type vectors can be written as the product of two Hall-Littlewood functions.

\begin{df}
For a pair of partitions $\vec{\lambda}=(\lo, \lt)$, set
\begin{align}
 &\tket{\tX_{\vec{\lambda}}} \seteq \tX^{(2)}_{-\lambda^{(2)}_1} \tX^{(2)}_{-\lambda^{(2)}_2}\cdots \tXo_{-\lambda^{(1)}_1} \tXo_{-\lambda^{(1)}_2} \cdots \ket{\vec{u}},\\
 &\tbra{\tX_{\vec{\lambda}}} \seteq \bra{\vec{u}} \cdots \tX^{(1)}_{\lambda^{(1)}_2} \tX^{(1)}_{\lambda^{(1)}_1} \cdots \tX^{(2)}_{\lambda^{(2)}_2} \tX^{(2)}_{\lambda^{(2)}_1}.  
\end{align}
\end{df}

We have the expression of these vectors 
in terms of the Hall-Littlewood functions.

\begin{prop}\label{prop:PBW vct Rep by Hall} 
\begin{align}
 &\tket{\tX_{\lambda, \mu}} = (u_1 u_2)^{\ell(\mu)} u_2^{\ell(\lambda)} Q_{\mu}(b^{(+)}_{-n};t^{-1}) Q_{\lambda}(b^{(-)}_{-n};t^{-1}) \ket{\vu}, \\
 &\tbra{\tX_{\lambda, \mu}} = u_1^{\ell(\lambda)} (u_1 u_2)^{\ell(\mu)} t^{|\lambda|+|\mu|} \bra{\vu} Q_{\lambda}(b^{(-)}_{n};t^{-1}) Q_{\mu}(b^{(+)}_{n};t^{-1}), 
\end{align}
where 
\begin{align}
 b^{(+)}_{n} & \seteq b^{(1)}_{n} + b^{(2)}_{n}, \quad  b^{(+)}_{-n} \seteq  b^{(2)}_{-n} \quad (n>0),  \\
 b^{(-)}_{n} & \seteq b^{(1)}_{n}, \quad  b^{(-)}_{-n} \seteq -b^{(1)}_{-n} + b^{(2)}_{-n} \quad (n>0). 
\end{align}
\end{prop}

The vectors $\tXo_{-\lambda^{(1)}_1} \tXo_{-\lambda^{(1)}_2} \cdots  \tX^{(2)}_{-\lambda^{(2)}_1} \tX^{(2)}_{-\lambda^{(2)}_2}\cdots \ket{\vec{u}}$ 
do not have such a good expression. 
This proposition is proved by the theory of Jing's operator. 
Then the vectors $\tket{\tX_{\vec{\lambda}}}$ are partially diagonalized as the following proposition. 
Furthermore, 
with the help of Hall-Littlewood functions, 
we can calculate the Shapovalov matrix $S_{\vl,\vm} \seteq \tbraket{\tX_{\vl}}{\tX_{\vm}}$ 
and its inverse $S^{\vl,\vm}$.

\begin{prop}\label{prop:Shapovalov in terms of HL poly}
We can express $S_{\vl, \vm}$ by the inner product $\langle -,- \rangle_{0,t}$ 
of Hall-Littlewood functions defined in Appendix \ref{sec: Macdonald and HL} : 
\begin{align}
S_{\vl,\vm} =& (u_1 u_2)^{\ell (\lt) +\ell(\mt)} u_1^{\ell(\lo)} u_2^{\ell(\mo)} \label{eq:Shapovalov} \\
& \times \frac{1}{b_{\lo}(t^{-1})}  \left\langle Q_{\lt}(p_n;t^{-1}), Q_{\mt}(-p_n;t^{-1}) \right\rangle_{0,t^{-1}} 
\delta_{\lo, \mo}, \nonumber \\
S^{\vl,\vm} =&  (u_1 u_2)^{-\ell(\mt) -\ell(\lt)} u_1^{-\ell(\mo)} u_2^{-\ell(\lo)} \label{eq:inverse Shapovalov} \\
 &\times  \frac{b_{\mo}(t^{-1})}{b_{\lt}(t^{-1}) b_{\mt}(t^{-1})} 
 \left\langle Q_{\lt}(-p_n;t^{-1}), Q_{\mt}(p_n;t^{-1}) \right\rangle_{0,t^{-1}} \delta_{\lo, \mo}. \nonumber
\end{align}
\end{prop}

\Proof
(\ref{eq:Shapovalov}) follows from Proposition \ref{prop:PBW vct Rep by Hall}. 
(\ref{eq:inverse Shapovalov}) can be obtained by the equation
\begin{equation}
\sum_{\mu} \frac{
	\left\langle Q_{\lambda}(p_n;t), Q_{\mu}(-p_n; t) \right\rangle_{0,t} 
	\left\langle Q_{\mu}(-p_n; t), Q_{\nu}(p_n;t) \right\rangle_{0,t}}
    {b_{\mu}(t)}
=b_{\lambda}(t) \delta_{\lambda, \nu}
\end{equation}
which is shown by inserting the complete system with respect to $Q_{\mu}(-p_n;t)$ 
into the equation 
$\left\langle Q_{\lambda}(p_n;t), Q_{\nu}(p_n;t) \right\rangle_{0,t} = b_{\lambda}(t) \delta_{\lambda, \nu}$. 
\qed

Existence of the inverse matrix $S^{\vl \vm}$ leads 
linear independence of $\tket{\tX_{\vl}}$. 
Since there are the same number of linear independent vectors as the dimension of each level of $\mathcal{F}_{\vu}$, 
we can see that $\tket{\tX_{\vl}}$ forms a basis over $\mathcal{F}_{\vu}$.

\begin{prop}
If $t$ is not a root of unity and $u_1, u_2 \neq 0$, 
$\tket{\tX_{\vl}}$ (resp. $\tbra{\tX_{\vl}}$) 
is a basis of $\mathcal{F}_{\vu}$ (resp. $\mathcal{F}_{\vu}^*$). 
\end{prop}

Next, let us introduce generalized Hall-Littlewood functions 
which are specialization of generalized Macdonald functions 
and give some crystallized versions of the AGT conjecture.

\begin{df}\label{def:Gn Hall-Littlewood}
Define the vectors $\tket{\tP_{\vl}}$ and $\tbra{\tP_{\vl}}$ as the $q \rightarrow 0$ limit of generalized Macdonald functions, i.e., 
\begin{equation}
\tket{\tP_{\vl}} \seteq \lim_{q \rightarrow 0} \tket{P_{\vl}}, \qquad 
\tbra{\tP_{\vl}} \seteq \lim_{q \rightarrow 0} \tbra{P_{\vl}}. 
\end{equation}
We call the vectors $\tket{\tP_{\vl}}$ 
generalized Hall-Littlewood functions. 
\end{df}

These are the eigenvectors of $\tilde{X}^{(1)}_0$: 
\begin{equation}
\tilde{X}^{(1)}_0 \tket{\tilde{P}_{\vec{\lambda}}} 
=\tilde{e}_{\vec{\lambda}} \tket{\tilde{P}_{\vec{\lambda}}}, \quad 
\tbra{\tilde{P}_{\vec{\lambda}}} \tilde{X}^{(1)}_0  
=\tilde{e}^*_{\vec{\lambda}} \tbra{\tilde{P}_{\vec{\lambda}}}
\end{equation}
Moreover the eigenvalues are
\begin{equation}
\tilde{e}_{\vl} =\tilde{e}^*_{\vl} 
=\sum_{k=1}^2 u_k \left( 1+(1-t)\sum_{i\geq 1}^{\ell(\lambda^{(k)})}t^{-i}  \right). 
\end{equation}
However there are too many degenerate eigenvalues 
to ensure the existence of generalized Hall-Littlewood functions. 
It is difficult to characterize $\tket{\tilde{P}_{\vl}}$ 
as the eigenfunction of only $\tXo_0$. 
For example, 
$\vec{\lambda}=((1),(2))$ and $\vec{\mu}=((2),(1))$ 
have the relation $\vec{\lambda} \overstar{>} \vec{\mu}$, but
$\tilde{e}_{\vec{\lambda}} = \tilde{e}_{\vec{\mu}}$.%
\footnote{
Definition \ref{def:Gn Hall-Littlewood} is given under the hypothesis 
that the vector $\tket{P_{\vl}}$ has no singulality in the limit $q \rightarrow 0$. 
If we can show the existence therem of both generalized Macdonald 
and generalized Hall-Littlewood functions 
by using the same partial ordering and the same basis, 
this hypothesis is guaranteed. 
}

\begin{ex} 
Let us define the transition matrices $(\tilde{c}_{\vl, \vm})$ 
and $(\tilde{c}^*_{\vl, \vm})$ 
by the expansions  
\begin{align}
&\tket{\tP_{\vl}}= \sum_{\vm} \tilde{c}_{\vl, \vm} 
\prod_{i=1}^2 P_{\mu^{(i)}}(b_{-n}^{(i)};t) \ket{\vu}, \\
& \tbra{\tP_{\vl}} = \sum_{\vm} \tilde{c}^*_{\vl, \vm} 
\bra{\vu}\prod_{i=1}^2 P_{\mu^{(i)}}(b_{n}^{(i)};t).
\end{align}
Then up to the degree 2 the matrix elements $\tilde{c}_{\vl, \vm}$ 
are given by 
\begin{equation*}
\begin{array}{c||c c} 
\vl \setminus \vm & (\emptyset, (1)) & ((1), \emptyset) \\ \hline \hline
(\emptyset, (1)) & 1 & \frac{u_2}{u_1-u_2} \\ 
((1), \emptyset) & 0 & 1 \\ 
\end{array} ,
\end{equation*}
\begin{equation*}
\begin{array}{c||c c c c c} 
 \vl \setminus \vm & (\emptyset, (2)) & (\emptyset, (1^2)) &((1), (1)) &((2), \emptyset) &((1^2), \emptyset) \\ \hline \hline
(\emptyset, (2)) & 1 & 0 & 0 & \frac{u_2}{u_1-u_2} & 0 \\ 
(\emptyset, (1^2)) & 0 & 1 & \frac{u_2}{tu_1-u_2} & -\frac{u_2}{tu_1-u_2} & \frac{tu_2^2}{(u_1-u_2)(t u_1-u_2)} \\ 
((1), (1)) & 0 & 0 & 1 & -1 & -\frac{t(1+t)u_2}{-u_1+tu_2} \\ 
((2), \emptyset) & 0 & 0 & 0 & 1 & 0 \\ 
((1^2), \emptyset) & 0 & 0 & 0 & 0 & 1\\ 
\end{array}.
\end{equation*}
Up to the degree 2 the matrix elements $\tilde{c}^*_{\vl, \vm}$ 
are given by 
\begin{equation*}
\begin{array}{c||c c} 
\vl \setminus \vm  & ((1), \emptyset) &(\emptyset, (1)) \\ \hline \hline
((1),\emptyset) & 1 & -\frac{u_2}{u_1-u_2} \\
(\emptyset, (1)) & 0 & 1 \\ 
\end{array} ,
\end{equation*}
\begin{equation*}
\begin{array}{c||c c c c c} 
\vl \setminus \vm   &((2), \emptyset) &((1^2), \emptyset)&((1), (1)) &(\emptyset, (2)) & (\emptyset, (1^2)) \\ \hline \hline
((2), \emptyset) &1 & 0 & 1-t & -\frac{u_2}{u_1-u_2} & 0 \\
((1^2), \emptyset) & 0 & 1 & \frac{t u_2}{t u_2-u_1} & 0 & \frac{t u_2^2}{(u_1-u_2)(u_1-t u_2)} \\
((1), (1)) & 0 & 0 & 1 & 0 & -\frac{(t+1) u_2}{t u_1-u_2} \\
(\emptyset, (2)) & 0 & 0 & 0 & 1 & 0 \\ 
(\emptyset, (1^2)) & 0 & 0 & 0 & 0 & 1\\ 
\end{array}.
\end{equation*}

\end{ex}

Similarly to the case of generic $q$, 
we define the integral forms of generalized Hall-Littlewood functions 
and give a conjecture of their norms.

\begin{df}
The integral forms $\tket{\tilde{K}_{\vec{\lambda}}}$ and 
$\tbra{\tK_{\vl}}$ are defined by
\begin{align}
&\tket{\tK_{\vl}}= \sum_{\vm} \tilde{\alpha}_{\vl \vm} \tket{\tX_{\vm}} 
\propto \tket{\tilde{P}_{\vl}}, \quad
\tilde{\alpha}_{\vl, (\emptyset ,(1^{|\vl|}))}=1, \\
&\tbra{\tK_{\vl}} 
= \sum_{\vm} \tilde{\beta}_{\vl \vm} \tbra{\tX_{\vm}} 
\propto \tbra{\tilde{P}_{\vl}}, \quad
\tilde{\beta}_{\vl, (\emptyset ,(1^{|\vl|}))}=1. 
\end{align}
Note that the coefficients $\tilde{\alpha}_{\vl, ((1^{|\vl|}), \emptyset )}$ and 
$\tilde{\beta}_{\vl, ( (1^{|\vl|}), \emptyset )}$ can be zero at $q=0$. 
\end{df}

\begin{conj}\label{conj:inner prod of Gn HL}
\begin{equation}
\tbraket{\tilde{K}_{\vec{\lambda}}}{\tilde{K}_{\vec{\lambda}}} 
\overset{?}{=} (u_1 u_2)^{|\vec{\lambda}|} u_1^{2|\lambda^{(1)}|} u_2^{2|\lambda^{(2)}|} t^{-2\left( n(\lambda^{(1)})+n(\lambda^{(2)}) \right)}
\prod_{i,j=1}^{2} \tilde{N}_{\lambda^{(i)}, \lambda^{(j)}}(u_i/u_j).
\end{equation}
\end{conj}

Next, 
let us define the vertex operator at crystal limit. 
 
\begin{df} 
The vertex operator 
$\tPhi(z)=\tPhi^{\vv}_{\vu}(z) 
: \mathcal{F}_{\vu} \rightarrow \mathcal{F}_{\vv}$ 
is the linear operator satisfying the relations 
\begin{align} 
&\tXo_n \tPhi(z)= \tPhi(z) \tXo_n -v_1\, v_2\, z\, \tPhi(z)\tXo_{n-1} \quad (n\leq 0), \\
&\tXo_n \tPhi(z)= \tPhi(z) \tXo_n \quad (n\geq 1), \\
&\tXt_n \tPhi(z)= \tPhi(z) \tXt_n -v_1\, v_2\, z\, \tPhi(z)\tXt_{n-1} \quad (\forall n), \label{eq:comm rel of tPhi and tXt}
\end{align}
\begin{equation}
\bra{\vv}\tPhi \ket{\vu}=1. 
\end{equation} 
\end{df}

The existence of such an operator is shown 
by the renormalization $\tPhi(z)=\Phi(p^{\frac{3}{2}}z)$. 
In the relation of $\Phi(z)$ and $X^{(i)}_n$, 
it is understood that this renormalization is appropriate 
by considering the shift of $z$ 
such that $\tPhi(z)$ and not all $\tX^{(i)}_n$ are commutative 
and the relation does not diverge. 
We give some simple properties of the vertex operator $\tPhi(z)$. 

\begin{prop}
\begin{equation}\label{eq:simple property1}
\tbra{\tX_{\vl}} \tPhi(z) \ket{\vu}=
\left\{
\begin{array}{ll}
(-v_1 v_2 u_1 u_2 z)^n  , &\quad \vl=(\emptyset, (1^n)) \quad \mbox{for some $n$}, \\
0 , &\quad \mbox{otherwise.}
\end{array}
\right.
\end{equation}
For any $n \geq 1$, 
\begin{equation}
\tbra{\vv} \tPhi(z) \tX^{(i)}_{-n} \ket{\vu}=
\left\{
\begin{array}{ll}
\left(  \frac{1}{v_1v_2z} \right)^n (v_1+v_2+u_1+u_2) , &\quad i=1, \\
\left(  \frac{1}{v_1v_2z} \right)^n  (u_1 u_2-v_1v_2), &\quad i=2.
\end{array}
\right.
\end{equation}
\end{prop}

These follow from the commutation relations of $\tPhi(z)$.
Especially note that the three-point function 
which has generators $\tX^{(i)}_n$ on the left side, 
i.e., $\tbra{\tX_{\vl}} \tPhi(z) \ket{\vu}$,  
remains only in the case of special Young diagrams $(\emptyset, (1^n))$
with only one vertical column.

\begin{conj}\label{conj:mat element of phi wrt Gn HL}
The matrix elements of $\tPhi(z)$ with respect to the integral form $\tket{\tK_{\vl}}$ are
\begin{align}
\tbra{\tK_{\vl}} \tPhi(z) \tket{\tK_{\vm}} 
&\overset{?}{=} (-1)^{|\vl|+|\vm|} (u_1 u_2 v_1 v_2 z)^{|\vl|-|\vm|} u_1^{2|\mo|} u_2^{2|\mt|} (u_1u_2)^{|\vm|} t^{-2(n(\mo)+n(\mt))}\\
&\quad \times \prod_{i,j=1}^2 \tN_{\lambda^{(i)},\mu^{(j)}} (v_i/u_j). \nonumber
\end{align}
\end{conj}

Under these conjectures \ref{conj:inner prod of Gn HL} and
\ref{conj:mat element of phi wrt Gn HL}, 
we obtain the formula for correlation functions of the vertex operator $\tPhi(z)$. 
For example, the function corresponding to the four-point conformal block is  
\begin{align}
\bra{\vw} \tPhi_{\vv}^{\vw}(z_2) \tPhi_{\vu}^{\vv}(z_1) \ket{\vu}  
&= \sum_{\vl} \frac{\tbra{\vw} \tPhi(z_2) \tket{\tK_{\vl}}   \tbra{\tK_{\vl}} \tPhi(z_1) \tket{\vu}}{\tbraket{\tK_{\vl}}{\tK_{\vl}}} \nonumber \\
&\overset{?}{=} \sum_{\vl} \left( \frac{u_1u_2z_1}{w_1w_2z_2} \right)^{|\vl|}
\prod_{i,j=1}^{2} \frac{\tN_{\emptyset , \lambda^{(j)}} (w_i/v_j)   \tN_{\lambda^{(i)}, \emptyset} (v_i/u_j) }{\tN_{\lambda^{(i)}, \lambda^{(j)}} (v_i/v_j)} \nonumber \\
&= \sum_{\vl} \left( \frac{u_1u_2z_1}{w_1w_2z_2} \right)^{|\vl|}
\prod_{i,j=1}^{2} \frac{\tN_{\emptyset , \lambda^{(j)}} (w_i/v_j) }{\tN_{\lambda^{(i)}, \lambda^{(j)}} (v_i/v_j)}. \label{eq:expansion by AFLT} 
\end{align}
(\ref{eq:expansion by AFLT}) is the AGT conjecture 
in the limit $q \rightarrow 0$ with help of the AFLT basis. 
However, 
in the crystallized case, 
we can prove another formula for this four-point correlation function 
by using the PBW type basis. 
At first, let us show the following two lemmas. 

\begin{lem}\label{lem:matrix el. of tPhi wrt PBW}
The matrix elements with respect to PBW type vector $\tket{\tX_{\emptyset,\lambda}}$ 
and $\bra{\vv}$ are
\begin{equation}
\bra{\vv} \tPhi(z) \tket{\tX_{\emptyset,\lambda}}
=(-1)^{\ell (\lambda)} \left( \frac{1}{v_1v_2z} \right)^{|\lambda|} t^{-n(\lambda)} 
\prod_{k=1}^{\ell(\lambda)} (t^{k-1}v_1v_2-u_1u_2). 
\end{equation}
\end{lem}

\Proof
For $i \geq 2$, by (\ref{eq:comm rel of tPhi and tXt}) 
and the relation $\tXt_{-n+1} \tXt_{-n} = t^{-1} \tXt_{-n} \tXt_{-n+1}$, 
\begin{align}
\bra{\vv} \tPhi(z) \left( \tXt_{-i} \right)^m 
&= \left( \frac{1}{v_1v_2z} \right) \bra{\vv} \tPhi(z) \tXt_{-i+1} \left( \tXt_{-i} \right)^{m-1}    \\
&= \left( \frac{1}{v_1v_2z} \right) t^{-m+1} \bra{\vv} \tPhi(z) \left( \tXt_{-i} \right)^{m-1} \tXt_{-i+1}  \nonumber \\
&= \left( \frac{1}{v_1v_2z} \right)^m t^{-\frac{1}{2}m(m-1)} \bra{\vv} \tPhi(z) \left( \tXt_{-i+1} \right)^{m}.   \nonumber 
\end{align}
Repeating this calculation, we get
\begin{equation}
\bra{\vv} \tPhi(z) \left( \tXt_{-i} \right)^m 
= \left( \frac{1}{v_1v_2z} \right)^{mk} t^{-\frac{1}{2}m(m-1)k} 
\bra{\vv} \tPhi(z) \left( \tXt_{-i+k} \right)^{m}, \label{eq:preparation for i>1}
\end{equation}
where  $0 \leq k\leq i-1$. 
When $i=1$, by similar calculation 
\begin{align}
\bra{\vv} \tPhi(z) \left( \tXt_{-1} \right)^m \tket{\vu} 
&=\left( -\frac{1}{v_1v_2z} \right) \left( v_1v_2 - t^{-m+1}u_1u_2 \right) \bra{\vv} \tPhi(z) \left( \tXt_{-1} \right)^{m-1} \tket{\vu}   \nonumber \\
&= \left( -\frac{1}{v_1v_2z} \right)^m \prod_{k=1}^{m}(v_1v_2-t^{-k+1}u_1u_2).  \label{eq:preparation for i=1}
\end{align}
By using above two formulas (\ref{eq:preparation for i>1}) and (\ref{eq:preparation for i=1}), 
if we write $\lambda=(i_1^{m_1}, i_2^{m_2}, \ldots , i_l^{m_l}) \; \; (i_1> i_2>\cdots >i_l)$, 
\begin{align}
\bra{\vv} \tPhi(z) \tket{\tX_{\emptyset, \lambda}}  
&= \left( \frac{1}{v_1v_2z} \right)^{m_1(i_1-i_2)} t^{-\frac{1}{2}m_1(m_1-1)(i_1-i_2)} \nonumber \\
& \quad \times \bra{\vv} \tPhi(z) \left( \tXt_{-i_2} \right)^{m_1+m_2} \left( \tXt_{-i_3} \right)^{m_3} \cdots \left( \tXt_{-i_l} \right)^{m_l} \ket{\vu} \nonumber \\
&= \left( \frac{1}{v_1v_2z} \right)^{m_1(i_1-i_3)+m_2(i_2-i_3)} t^{-\frac{1}{2}m_1(m_1-1)(i_1-i_2)-\frac{1}{2}(m_1+m_2)(m_1+m_2-1)(i_2-i_3)} \nonumber \\
& \quad \times \bra{\vv} \tPhi(z) \left( \tXt_{-i_3} \right)^{m_1+m_2+m_3} \left( \tXt_{-i_4} \right)^{m_4} \cdots \left( \tXt_{-i_l} \right)^{m_l} \ket{\vu} \nonumber \\
&= (-1)^{\ell(\lambda)} \left( \frac{1}{v_1v_2z} \right)^{|\lambda|} t^{-n(\lambda)} 
\prod_{k=1}^{\ell(\lambda)} (t^{k-1}v_1v_2-u_1u_2).
\end{align}
\qed

We have explicit form of formulas 
for some parts of the inverse Shapovalov matrix.

\begin{lem}\label{lem:exlicit form of inv. shapovalov} 
\begin{equation}
S^{(\emptyset, (1^{|\lambda|})), (\emptyset,\lambda)}
=S^{(\emptyset,\lambda), (\emptyset, (1^{|\lambda|}))} 
= \frac{(-1)^{|\lambda|} t^{-n(\lambda)} (u_1u_2)^{-|\lambda|-\ell (\lambda)} }{b_{\lambda}(t^{-1})}. 
\end{equation}
\end{lem}

\Proof
In this proof, we put $s=|\lambda|$. 
Hall-Littlewood function $Q_{(1^{s})}(p_n;t)$ 
is the elementary symmetric function $e_{s}(p_n)$ times $b_{(1^{s})}(t)$. 
Elementary symmetric functions have the generating function 
\begin{equation}\label{eq:gen fn of elementary sym fn}
\sum_{k=0}^{\infty} z^{k} e_k(p_n) = \exp \left\{ -\sum_{n >0} \frac{(-z)^n}{n} p_n \right\}. 
\end{equation}
Hence by the $r=0$ case of Fact \ref{fact:spetialization of HL poly}, 
\begin{align}
\left\langle e_{s}(-p_n), Q_{\lambda}(p_n;t) \right\rangle_{0,t} 
&= \left. (-1)^s \exp \left\{ \sum_{n>0} \frac{z^n}{1-t^n} \frac{\partial}{\partial p_n} \right\} Q_{\lambda}(p_n;t) \right|_{\mathrm{coefficient}\; \mathrm{of}\; z^s} \\
 &= \left. (-1)^s Q_{\lambda}(p_n;t) \right|_{p_n \mapsto \frac{1}{1-t^n}}  \nonumber \\
 &= (-1)^s t^{n(\lambda)}. \nonumber
\end{align}
Therefore, 
the lemma follows from Proposition \ref{prop:Shapovalov in terms of HL poly}. 
\qed

We give other proofs of this lemma 
in Appendix \ref{seq:Another proof of Lemma }, 
and the form of $S^{(\emptyset, (|\lambda|)), (\emptyset,\lambda)}$ 
can be found in  
Appendix \ref{sec:explicit form of <Q,Q>}. 
By the property (\ref{eq:simple property1}), Proposition \ref{prop:Shapovalov in terms of HL poly} and  
Lemmas \ref{lem:matrix el. of tPhi wrt PBW} and \ref{lem:exlicit form of inv. shapovalov}, 
we can show the following theorem.

\begin{thm}\label{thm:main theorem}
\begin{align}\label{eq:expansion formula by PBW}
\bra{\vw} \tPhi^{\vw}_{\vv}(z_2) \tPhi^{\vv}_{\vu}(z_1) \ket{\vu}
&= \sum_{\vl} \tbra{\vw} \tPhi(z_2) \tket{\tX_{\vl}} S^{\vl \vm}  \tbra{\tX_{\vm}} \tPhi(z_1) \tket{\vu} \nonumber \\
&= \sum_{\lambda} \tbra{\vw} \tPhi(z_2) \tket{\tX_{\emptyset, \lambda}} S^{(\emptyset,\lambda), (\emptyset,(1^{|\lambda|}))}  \tbra{\tX_{\emptyset, (1^n)}} \tPhi(z_1) \tket{\vu} \nonumber \\
&= \sum_{\lambda} \left( \frac{u_1u_2z_1}{w_1w_2z_2} \right)^{|\lambda|} 
   \frac{ \prod_{k=1}^{\ell(\lambda)} \left(1- t^{k-1} \frac{w_1w_2}{v_1v_2} \right) }{t^{2n(\lambda)} b_{\lambda}(t^{-1}) }. 
\end{align}
\end{thm}

In this way, 
the explicit formula for the correlation function can be obtained, 
where we don't use any conjecture. 
The formulas (\ref{eq:expansion by AFLT}) and (\ref{eq:expansion formula by PBW}) 
are compared in Appendix \ref{Comparison of two formula}. 
We expect that these works will be generalized to $N \geq3$ case. 

\subsection{Other types of limit}
\label{sec:Another type of limit}
Finally, we present other types of the crystal limit. 
In this paper, 
we investigated the crystal limit while the parameters $u_i$, $v_i$ and $w_i$ are fixed. 
However, 
it is also important to study the cases when these parameters depend on $q$. 
For example, 
let us consider the case that $u_i=p^{-M_i} u'_i$, $v_i= p^{-A_i}v_i'$, $w_i=p^{-M_{i+2}} w_i'$ 
($M_i, A_i \in \mathbb{R}$) 
and $u_i'$, $v_i'$, $w_i'$ are independent of $q$ or fixed in the limit $q \rightarrow 0$. 
Let $M_i+1 > A_i > M_{j+2}$ for all $i, j\in \{1,2 \}$ 
and $A_1 =A_2$. 
Then the Nekrasov formula for generic $q$ case ($N=2$) 
\begin{equation}
Z^{\mathrm{inst}}_{N_f=4} \seteq 
\sum_{\vl} \left( \frac{u_1 u_2 z_1}{w_1 w_2 z_2} \right)^{|\vl|}
\prod_{i,j=1}^2 
\frac{N_{\emptyset, \lambda^{(j)}}(qw_i/t v_j) N_{\lambda^{(i)}, \emptyset}(qv_i/t u_j)}
{N_{\lambda^{(i)}, \lambda^{(j)}}(qv_i/t v_j)}
\end{equation}
depends only on the partitions of the shape $ \vl= ((1^n),(1^m))$ 
in the limit $q \rightarrow 0$, 
where $\left( \frac{u_1 u_2 z_1}{w_1 w_2 z_2} \right) =\tilde{\Lambda}$ is fixed, 
and coincides with the partition function of the pure gauge theory (\ref{eq:tildeZ}):  
\begin{equation}
Z^{\mathrm{inst}}_{N_f=4} \underset{q \rightarrow 0}{\longrightarrow} \tilde{Z}^{\mathrm{inst}}_{\mathrm{pure}},  
\end{equation}
where $Q=v_1'/v_2'$. 
Hence, we expect that the vector 
\begin{equation}
\Phi(z) \ket{\vu}
\end{equation}
corresponds to the Whittaker vector in the section \ref{sec:crystal of qVir} in this limit, 
though we were not able to properly explain it. 
In this way, by considering the various other values of $M_i$ and $A_i$, 
we can find special behavior of $Z^{\mathrm{inst}}_{N_f=4}$ and the conformal block $\bra{\vec{w}} \Phi(z_2) \Phi(z_1)\ket{\vu}$ and may prove the relation. 
These are our future studies. 

\section{R-Matrix of DIM algebra}
\label{sec:R-Matrix}

\subsection{Explicit calculation of R-Matrix}
\label{sec:Explicit cal of R}

In general, 
a bialgebra $H$ is called quasi-cocommutative 
if there exists an invertible element $\mathcal{R} \in H \otimes H$ 
such that for all $x \in H$, 
\begin{equation}
\Delta^{\mathrm{op}} (x) = \mathcal{R} \Delta (x) \mathcal{R}^{-1}. 
\end{equation}
This $\mathcal{R}$ is called the universal R-matirx. 
The DIM algebra is quasi-cocommutative \cite{FJMM:2015:Quantum}. 
In this section, 
we explicitly calculate the representation of $\mathcal{R}$. 
Moreover, 
the expression of its representation matrix 
is generally conjectured.%
\footnote{
The results in this Section are contribution of the author 
in the collaborations \cite{AKMMMOZ:2016:Toric, AKMMMOZ:2016:Anomaly}. 
}

The point of calculation is to make use of the condition 
that the generalized Macdonald functions are the eigenfunctions of $\Xo_0$, 
to reduce the degree of freedom of the matrix in advance. 
In this section, 
we formally write the universal $\mathcal{R}$-matrix 
as $\mathcal{ R} = \sum_i a_i \otimes b_i$ 
and set $\mathcal{R}_{12}= \sum_i  a_i \otimes b_i \otimes 1$, 
$\mathcal{R}_{23}= \sum_i  1 \otimes a_i \otimes b_i$, 
$\mathcal{R}_{13}= \sum_i  a_i \otimes 1 \otimes b_i$. 
Occasionally, 
we explicitly write the variable of the generalized Macdonald functions 
like 
\begin{equation}
\ket{P_{\vl}}=
\ket{P_{\vl}\left(u_1,\ldots,u_N\Big|q,t\Big|a^{(1)},\ldots, a^{(N)}\right)}, 
\end{equation}
and the PBW basis of the bosons is written as 
\begin{equation}
\ket{a_{\vl}} \seteq 
a^{(1)}_{-\lo_1}a^{(1)}_{-\lo_2} \cdots 
a^{(1)}_{-\lo_1}a^{(1)}_{-\lo_2} \cdots 
a^{(N)}_{-\lN_1}a^{(N)}_{-\lN_2} \cdots\ket{\vu}. 
\end{equation}

Firstly, by definition of $\mathcal{R}$,  
we have 
\begin{align}
\rho_{u_1u_2} \left( \Delta^{\mathrm{op}}(x^+_0) \, \mathcal{R} \right) \ket{P_{AB}} 
&=\rho_{u_1u_2} \left( \mathcal{R} \, \Delta(x^+_0)  \right) \ket{P_{AB}} \\
&=\epsilon_{AB}\, \rho_{u_1u_2} \left( \mathcal{R} \right) \ket{P_{AB}} \nonumber 
\end{align}
Here $\rho_{u_1u_2}\seteq \rho_{u_1} \otimes \rho_{u_2}$. 
The formula for the coproduct of the DIM algebra and 
the definition of the representation $\rho_{u}$ are 
given in Appendix \ref{sec:Def of DIM}. 
Hence, 
$\rho_{u_1u_2} \left( \mathcal{R} \right) \ket{P_{AB}}$ are eigenfunctions 
of $\rho_{u_1u_2} \left( \Delta^{\mathrm{op}}(x^+_0) \right)$. 
By comparing the forms of 
$\rho_{u_1u_2} \left( \Delta^{\mathrm{op}}(x^+_0) \right)$ 
and $\Xo_0 = \rho_{u_1u_2} \left( \Delta(x^+_0) \right)$, 
the eigenfunctions of $\rho_{u_1u_2} \left( \Delta^{\mathrm{op}}(x^+_0) \right)$ are obtained by replacing $a^{(1)}$ with $a^{(2)}$
and $u_1$ with $u_2$. 
Moreover, by checking their eigenvalues, 
it can be seen that $\rho_{u_1u_2} \left( \mathcal{R} \right) \ket{P_{AB}}$ 
are proportional to 
$\ket{P_{BA}\left(u_2,u_1\Big|q,t\Big|a^{(2)},a^{(1)}\right)}$: 
\begin{equation}\label{eq:propo rel.}
\rho_{u_1u_2} \left( \mathcal{R} \right) \, 
\ket{P_{AB}\left(u_1,u_2\Big|q,t\Big|a^{(1)},a^{(2)}\right)}
= k_{AB} \ket{P_{BA}\left(u_2,u_1\Big|q,t\Big|a^{(2)},a^{(1)}\right)}, 
\end{equation}
where $k_{AB}=k_{AB} (u_1,u_2| q,t)$ are proportionality constants. 
By this property, 
the representation matrix of $\rho_{u_1u_2} \left( \mathcal{R} \right)$
is block-diagonalized at each level of the boson $a^{(i)}_n$. 
The proportionality constants $k_{AB}$ can be calculated by 
using the generalized Macdonald functions in the $N=3$ case.

Similarly to the $N=2$ case, 
from the relation 
\begin{equation}
(\Delta^{\mathrm{op}} \otimes id) \circ \Delta (x^+_0) 
= \mathcal{R}_{12} 
 ({\Delta} \otimes id) \circ \Delta (x^+_0)
 \mathcal{R}^{-1}_{12},
\end{equation}
we have 
\begin{equation}
\rho_{u_1u_2u_3}(\mathcal{R}_{12}) \ket{P_{ABC} }
=k^{(12)}_{ABC} \ket{P_{BAC}(u_2,u_1,u_3|q,t|\at,\ao,a^{(3)})}, 
\end{equation}
where $\rho_{u_1u_2u_3}\seteq \rho_{u_1}\otimes \rho_{u_2} \otimes \rho_{u_3}$ 
and $k^{(12)}_{ABC}$ are constants. 
Since the generalized Macdonald functions satisfy 
\begin{equation}
\ket{P_{AB \emptyset}\left(u_1,u_2,u_3\Big|q,t\Big|a^{(1)},a^{(2)}, a^{(3)}\right)}
=\ket{P_{AB}\left(u_1,u_2\Big|q,t\Big|a^{(1)},a^{(2)}\right)}, 
\end{equation}
the proportionality constants have the relation 
$k^{(12)}_{AB\emptyset}= k_{AB}$.

For example
let us describe the calculation of the representation matrix at level 1. 
The following are examples of $P_{ABC}$ at level 1:
\begin{equation}
\left(
\begin{array}{c}
 \tket{P_{\emptyset,\emptyset,[1]} }\\ 
\tket{P_{\emptyset,[1],\emptyset}}\\
\tket{P_{[1],\emptyset,\emptyset}}
\end{array}
\right)
=A(u_1,u_2,u_3)
\left(
\begin{array}{c}
 \tket{a_{\emptyset,\emptyset,[1]}} \\ 
\tket{a_{\emptyset,[1],\emptyset}}\\
\tket{a_{[1],\emptyset,\emptyset}}
\end{array}
\right),
\end{equation}
\begin{equation}
A(u_1,u_2,u_3):=\left(
\begin{array}{ccc}
 1 & -\frac{(q-t) u_3}{\sqrt{\frac{q}{t}} t (u_2-u_3)} & -\frac{(q-t) u_3 (q u_3-t u_2)}{q t (u_1-u_3) (u_3-u_2)} \\
 0 & 1 & -\frac{(q-t) \sqrt{\frac{q}{t}} u_2}{q (u_1-u_2)} \\
 0 & 0 & 1 \\
\end{array}
\right).
\end{equation}
By the above discussion, 
if we set the matrix 
\begin{equation}
B^{(12)}:=
\left(\begin{array}{ccc}
 k^{(12)}_{\emptyset,\emptyset,[1]} & 0 & 0 \\
 0 &  k^{(12)}_{\emptyset,[1],\emptyset} & 0 \\
 0 & 0 &  k^{(12)}_{[1],\emptyset,\emptyset} \\
\end{array}\right)
A^{(12)}(u_1,u_2,u_3)
A^{-1}(u_1,u_2,u_3),
\end{equation}
\begin{equation}
A^{(12)}(u_1,u_2,u_3):=
\left(\begin{array}{ccc}
 1 & 0 & 0 \\
 0 & 0 & 1 \\
 0 & 1 & 0 \\
\end{array}\right)
A(u_2,u_1,u_3)
\left(\begin{array}{ccc}
 1 & 0 & 0 \\
 0 & 0 & 1 \\
 0 & 1 & 0 \\
\end{array}\right),
\end{equation}
then the representation matrix of $\rho_{u_1u_2u_3}(\mathcal{R}_{12})$ 
in the basis of the generalized Macdonald functions 
is the transposed matrix of $B^{(12)}$:
\begin{equation}
\rho_{u_1u_2u_3}(\mathcal{R}_{12}) 
\left(\begin{array}{ccc}
 \ket{P_{\emptyset,\emptyset,[1]}} & \ket{P_{\emptyset,[1],\emptyset}} & \ket{P_{[1]\emptyset,\emptyset}} \\
\end{array}\right)
=
\left(\begin{array}{ccc}
 \ket{P_{\emptyset,\emptyset,[1]} }& \ket{P_{\emptyset,[1],\emptyset}} & \ket{P_{[1]\emptyset,\emptyset}} \\
\end{array}\right)
{}^t B^{(12)}.
\end{equation}
The constants $k^{(12)}_{\vl}$ are determined as follows. 
At first, 
since scalar multiples of R-matrices are also R-matrices, 
we can normalize as $k^{(12)}_1=1$. 
This means that $\rho_{u_1u_2} (\mathcal{R}) 
(\ket{\vu})=\ket{\vu}$. 
Next, 
we consider the base change from $\ket{P_{ABC}}$ 
to the bosons $\ket{a_{ABC}}$: 
\begin{equation}
\widetilde{B}^{(12)}:= {}^tA(u_1,u_2,u_3)\, {}^tB^{(12)}\, {}^tA^{-1}(u_1,u_2,u_3).
\end{equation}
Then $\widetilde{B}^{(12)}$ is in the form
\begin{equation}\label{eq:B12 containing k}
\widetilde{B}^{(12)}=
\left(\begin{array}{ccc}
 1 & 0 & 0 \\
 b^{(12)}_2 & * & * \\
 b^{(12)}_3 & * & * \\
\end{array}\right), 
\end{equation}
where $b^{(12)}_n$ is a function of $k^{(12)}_l$. 
Since for the action of $\mathcal{R}_{12}$ to $a^{(3)}_1$, 
variables $a^{(1)}_1$ and $a^{(2)}_1$ should not appear, 
we get equations $b^{(12)}_2=b^{(12)}_3=0$. 
By solving these equations, 
we can see that 
\begin{align}
&k^{(12)}_1=1,\quad
k^{(12)}_2=-\frac{\sqrt{\frac{q}{t}} (q u_2-t u_1)}{q (u_1-u_2)},\quad
k^{(12)}_3=\frac{t (u_1-u_2) \sqrt{\frac{q}{t}}}{q u_1-tu_2}. 
\end{align}
Substituting this value into the matrix  
(\ref{eq:B12 containing k}), we have 
\begin{equation}
\widetilde{B}^{(12)}=
\left(
\begin{array}{ccc}
 1 & 0 & 0 \\
 0 & \frac{\sqrt{\frac{q}{t}} t (u_1-u_2)}{q u_1-t u_2} & \frac{(q-t) u_1}{q u_1-t u_2} \\
 0 & \frac{(q-t) u_2}{q u_1-t u_2} & \frac{\sqrt{\frac{q}{t}} t (u_1-u_2)}{q u_1-t u_2} \\
\end{array}
\right). 
\end{equation}
In this way, 
we obtain the explicit expression $\widetilde{B}^{(12)}$ 
of representation matrix of universal $\mathcal{R}$ at level 1. 
Of course, 
it is possible to calculate the representation matrix of $\mathcal{R}_{23}$
in the same way, 
but by using symmetry with respect to $a^{(i)}$ at different $i$, 
we can easily understand the forms of 
$\mathcal{R}_{23}$ and $\mathcal{R}_{13}$: 
\begin{equation}
\widetilde{B}^{(23)}=
\left(
\begin{array}{ccc}
 \frac{\sqrt{\frac{q}{t}} t (u_2-u_3)}{q u_2-t u_3} & \frac{(q-t) u_2}{q u_2-t u_3} & 0 \\
 \frac{(q-t) u_3}{q u_2-t u_3} & \frac{\sqrt{\frac{q}{t}} t (u_2-u_3)}{q u_2-t u_3} & 0 \\
 0 & 0 & 1 \\
\end{array}
\right), 
\quad 
\widetilde{B}^{(13)}=
\left(
\begin{array}{ccc}
 \frac{\sqrt{\frac{q}{t}} t (u_1-u_3)}{q u_1-t u_3} & 0 & \frac{(q-t) u_1}{q u_1-t u_3} \\
 0 & 1 & 0 \\
 \frac{(q-t) u_3}{q u_1-t u_3} & 0 & \frac{\sqrt{\frac{q}{t}} t (u_1-u_3)}{q u_1-t u_3} \\
\end{array}
\right).
\end{equation}
Indeed, we can check that they satisfy the Yang-Baxter equation
\begin{equation}
\widetilde{B}^{(12)}\widetilde{B}^{(13)}\widetilde{B}^{(23)}=\widetilde{B}^{(23)}\widetilde{B}^{(13)}\widetilde{B}^{(12)}.
\end{equation}
Incidentally, in the basis of the generalized Macdonald functions, 
\begin{equation}
{}^tB^{(12)}=
\left(
\begin{array}{ccc}
 1 & 0 & 0 \\
 \frac{u_3 (q-t) \left(-t u_1 \sqrt{\frac{q}{t}}+q u_3 \sqrt{\frac{q}{t}}+q u_1-q u_3\right)}{q t (u_1-u_3)
   (u_2-u_3) \sqrt{\frac{q}{t}}} & -\frac{\sqrt{\frac{q}{t}} (q u_2-t u_1)}{q (u_1-u_2)} & \frac{u_1
   (q-t)}{q u_1-t u_2} \\
x & -\frac{u_2 (q-t) (q
   u_2-t u_1)}{q t (u_1-u_2)^2} & \frac{\sqrt{\frac{q}{t}} \left(q^2 u_1 u_2+q t u_1^2+q t u_2^2-4 q t
   u_1 u_2+t^2 u_1 u_2\right)}{q (u_1-u_2) (q u_1-t u_2)} \\
\end{array}
\right),
\end{equation}
{\footnotesize 
\begin{equation}
x= \frac{u_3 (q-t) \left(q^2 u_2 u_3-t^2 u_2^2 \sqrt{\frac{q}{t}}+t^2 u_2 u_3 \sqrt{\frac{q}{t}}+q t u_2^2+q
   t u_1 u_2 \sqrt{\frac{q}{t}}-2 q t u_1 u_2-q t u_1 u_3 \sqrt{\frac{q}{t}}+q t u_1 u_3-2 q t u_2
   u_3+t^2 u_1 u_2\right)}{q t^2 (u_1-u_2) (u_1-u_3) (u_2-u_3) \sqrt{\frac{q}{t}}} .
\end{equation}
} 
The representation matrix of $(\rho_{u_1} \otimes \rho_{u_2}) (\mathcal{R})$ is the $2 \times 2$ matrix block at the lower right corner of $B^{(12)}$ or $\widetilde{B}^{(12)}$. 
For example, 
in the basis of generalized Macdonald functions, 
its representation matrix is 
\begin{equation}
\left(
\begin{array}{cc}
 -\frac{\sqrt{\frac{q}{t}} (q u_2-t u_1)}{q (u_1-u_2)} & \frac{(q-t) u_1}{q u_1-t u_2} \\
 -\frac{(q-t) u_2 (q u_2-t u_1)}{q t (u_1-u_2)^2} & \frac{\sqrt{\frac{q}{t}} \left(u_1 u_2 q^2+t u_1^2 q+t
   u_2^2 q-4 t u_1 u_2 q+t^2 u_1 u_2\right)}{q (u_1-u_2) (q u_1-t u_2)} \\
\end{array}
\right). 
\end{equation}

Next, let us explain the case at level 2. 
The generalized Macdonald functions at level 2 
in the $N=3$ case are expressed as 
{\small 
\begin{align}
&\left(
\begin{array}{ccccccccc}
 \tket{P_{\emptyset,\emptyset,[2]}  }
& \tket{P_{\emptyset,\emptyset,[1,1]} }
& \tket{P_{\emptyset,[1],[1]} }
& \tket{P_{[1],\emptyset,[1]} }
& \tket{P_{\emptyset,[2],\emptyset} }
& \tket{P_{\emptyset,[1,1],\emptyset} }
& \tket{P_{[1],[1],\emptyset} }
&\tket{P_{[2],\emptyset,\emptyset}}
&\tket{P_{[1,1],\emptyset,\emptyset}}
\end{array}
\right) \nonumber \\
&={}^t\mathcal{A}
\left(
\begin{array}{ccccccccc}
 \tket{P'_{\emptyset,\emptyset,[2]} }
& \tket{P'_{\emptyset,\emptyset,[1,1]} }
& \tket{P'_{\emptyset,[1],[1]} }
& \tket{P'_{[1],\emptyset,[1]} }
& \tket{P'_{\emptyset,[2],\emptyset} }
& \tket{P'_{\emptyset,[1,1],\emptyset} }
& \tket{P'_{[1],[1],\emptyset} }
& \tket{P'_{[2],\emptyset,\emptyset}}
& \tket{P'_{[1,1],\emptyset,\emptyset}}
\end{array}
\right),
\end{align}}where 
$\tket{P'_{ABC}}$ denotes the product of ordinary Macdonald functions $P_A(a^{(1)})P_B(a^{(2)})P_C(a^{(3)})\ket{\vu}$, 
and the matrix $\mathcal{A}$ is given in Appendix \ref{sec:Ex of R-matrix}. 
In the same manner, 
we can get the representation matrix of $\mathcal{R}$. 
At first, $B^{(12)}$ at level 2  is in the form
\begin{equation}
\widetilde{B}^{(12)}=
\left(
\begin{array}{ccccccccc}
1& 0 & 0 & 0 & 0 & 0 & 0 & 0 & 0 \\
0& 1 & 0 & 0 & 0 & 0 & 0 & 0 & 0 \\
b_{31}& b_{32} & * & * & 0 & 0 & 0 & 0 & 0 \\
b_{41}& b_{42} & * & * & 0 & 0 & 0 & 0 & 0 \\
b_{51}& b_{52} & b_{53} & b_{54} & * & * & * & * & * \\
b_{61}& b_{62} & b_{63} & b_{64} & * & * & * & * & * \\
b_{71}& b_{72} & b_{73} & b_{74} & * & * & * & * & * \\
b_{81}& b_{82} & b_{83} & b_{84} & * & * & * & * & * \\
b_{91}& b_{92} & b_{93} & b_{94} & * & * & * & * & * \\
\end{array}
\right).
\end{equation}
Then we can find the proportionality constants 
such that all $b_{ij}$ are zero 
just by solving equations $b_{i1}=0$ ($i=3,4,\ldots ,9$). 
We have also checked that 
the representation matrix $\widetilde{B}^{ij}$ obtained in this way 
satisfies the Yang-Baxter equation up to level 3.  
The explicit expressions of $\mathcal{R}$ at level 2 
are written in Appendix \ref{sec:Ex of R-matrix}.

\subsection{General formula for R-matrix}
\label{sec:general form of R-matrix}

The proportionality constants are in the form
\begin{align}
&k_{\emptyset,(1)}=-\frac{\sqrt{\frac{q}{t}} (q u_2-t u_1)}{q (u_1-u_2)}, 
\qquad 
k_{(1),\emptyset}=\frac{t (u_1-u_2) \sqrt{\frac{q}{t}}}{q u_1-t u_2},\\
&k_{\emptyset,(2)}=-\frac{(q u_2-t u_1) \left(q^2 u_2-t u_1\right)}{q t (u_1-u_2) (q u_2-u_1)}, 
\qquad 
k_{\emptyset,(1,1)}=\frac{(q u_2-t u_1) \left(q u_2-t^2 u_1\right)}{q t (u_1-u_2) (t u_1-u_2)},\\
&k_{(1),(1)}=-\frac{(q u_2-u_1) (t u_1-u_2)}{(q u_1-u_2) (u_1-t u_2)}, 
\qquad 
k_{(2),\emptyset}=\frac{q t (u_1-u_2) (q u_1-u_2)}{(q u_1-t u_2) \left(q^2 u_1-t u_2\right)},\\
&k_{(1,1),\emptyset}=-\frac{q t (u_1-u_2) (t u_2-u_1)}{(q u_1-t u_2) \left(q u_1-t^2 u_2\right)}. 
\end{align}
These proportionality constants 
can be simplified by using the integral forms of the generalized Macdonald functions 
$\ket{K_{AB}} = \ket{K_{AB}(u_1,u_2|q,t|a^{(1)},a^{(2)})}$, 
which are defined in Section \ref{sec:Reargument of DI alg and AGT}.  
Define its opposite version by 
\begin{equation}
\ket{K^{\mathrm{op}}_{AB}(u_1,u_2|q,t|a^{(1)},a^{(2)})}
:=\ket{K_{BA}(u_2,u_1|q,t|a^{(2)},a^{(1)}) }
\end{equation} 
and the constants $\mathcal{C}_{AB}=\mathcal{C}_{AB}(u_1,u_2|q,t)$ by 
\begin{equation}
\ket{K_{AB}}=:\mathcal{C}_{AB} \ket{P_{AB}}. 
\end{equation}
By these renormalized functions, 
the relation (\ref{eq:propo rel.}) can be written as 
\begin{align}
\rho_{u_1u_2}(\mathcal{R}) \left( \ket{K_{AB}} \right)  
=k_{AB} \frac{\mathcal{C}_{AB}(u_1,u_2|q,t)}{\mathcal{C}_{BA}(u_2,u_1|q,t)} 
\ket{K^{\mathrm{op}}_{AB}}. 
\end{align}
Then it is conjectured that 
\begin{equation}\label{eq:propo const conj}
k_{AB}\frac{ \mathcal{C}_{AB}(u_1,u_2|q,t)}{\mathcal{C}_{BA}(u_2,u_1|q,t)}\overset{?}{=} 1. 
\end{equation}
This equation has been checked at $|A|+|B| \leq 2$. 
Therefore, 
the representation matrix $R_{\vl, \vm}$ of $\mathcal{R}$ 
in the basis of the integral forms 
can be expressed as the following conjecture.

\begin{conj}\label{conj:R-Matrix by int form}
\begin{equation}\label{eq:formula for R Matrix}
R_{\vl, \vm} \overset{?}{=} 
\frac{1}{\braket{K_{\vm} }{K_{\vm}}}
\braket{K_{\vm}}{K^{\mathrm{op}}_{\vl}}. 
\end{equation}
\end{conj}

Since the formula (\ref{eq:formula for R Matrix}) 
means the expansion coefficients of $\ket{K^{\mathrm{op}}}$ 
in front of $\ket{K_{\vl}}$, 
by using the transition matrix defined by 
\begin{equation}
\ket{K_{\vl}}=\sum_{\vm} \mathcal{A}_{\vl,\vm}(u_1,u_2) \ket{P'_{\vm}}
\end{equation}
and its opposite version 
$\mathcal{A}^{\mathrm{op}}_{(\lo,\lt), (\mo, \mt)}(u_1, u_2)
:=\mathcal{A}_{(\lt,\lo), (\mt, \mo)}(u_2, u_1)$, 
the R-matrix can be calculated by the matrix operation 
\begin{equation}
R_{\vl, \vm}=
\sum_{\vn} \mathcal{A}^{\mathrm{op}}_{\vn,\vl}(u_1, u_2) 
\mathcal{A}^{-1}_{\vm, \vn}(u_1, u_2). 
\end{equation}
This formula gives a much simpler way to get explicit expressions 
as compared with deducing them from the universal R-matrix \cite{FJMM:2015:Quantum}. 
Incidentally, the proportionality constants are conjectured to be 
\begin{equation}\label{eq:conj of prop const}
k_{AB} 
\overset{?}{=} \left( \frac{q}{t} \right)^{\frac{1}{2}(|A| + |B|)}
\frac{N_{AB}\left( \frac{u_1}{u_2} \right)} {N_{AB}\left( \frac{qu_1}{tu_2} \right)}
= \left( \frac{t}{q} \right)^{\frac{1}{2}(|A| + |B|)}
\frac{N_{BA}\left( \frac{qu_2}{tu_1} \right)} {N_{BA}\left( \frac{u_2}{u_1} \right)}, 
\end{equation}
where $N_{AB}(Q)$ is the Nekrasov factor defined 
in (\ref{eq:def of Nek factor}), Section \ref{sec:Review of Simplest 5D AGT}. 
The second equality follows from
the formula (eq.(2.34) in \cite{AK:2008Refined}, 
eq.(102) in \cite{MZ:2015:Decomposing})
\begin{equation}
N_{AB}\left( \sqrt{\frac{q}{t}} Q ; q,t \right)
= N_{BA}\left( \sqrt{\frac{q}{t}} Q^{-1} ; q,t \right) 
Q^{|A| + |B|} \frac{f_A(q,t)}{f_B(q,t)},
\end{equation}
where $f_A(q,t)$ is the framing factor \cite{Taki:2007:Refined}.
The equation (\ref{eq:conj of prop const}) has been checked up to level 3.

\section{Properties of generalized Macdonald functions}
\label{sec:Properties of Gn Mac}

\subsection{Partial orderings}\label{sec:partial orderings}

The existence theorem of generalized Macdonald functions can be stated 
by the ordering $\overstar{>}$ in Definition \ref{def:ordering1}. 
In this subsection, 
we introduce a more elaborated ordering. 
Using this ordering, 
we can find more elements 
which is $0$ in the transition matrix $c_{\vl, \vm}$, 
where $\ket{P_{\vl}} = \sum_{\vm} c_{\vl \vm} \prod_iP_{\mu^{(i)}}(a_{-n}^{(i)}) \ket{\vu} $, 
and get more strict condition to existence theorem.

\begin{df}\label{df:ordering elaborated version}
For $N$-tuples of partitions $\vl$ and $\vm$, 
\begin{align}
\vl \overstar{\succ} \vm \quad \overset{\mathrm{def}}{\Leftrightarrow} \quad
& \vl \overstar{>} \vm \quad \mathrm{and} 
& \{ \nu \mid \nu \supset \lambda^{(\alpha)}, \nu \supset \mu^{(\alpha)},  |\nu|=|\lambda^{(\alpha)}|+\sum_{\beta=\alpha+1}^N (|\lambda^{(\beta)}|-|\mu^{(\beta)}|) \} \neq \nonumber \emptyset
\end{align}
for all $\alpha$. 
Here $\lambda \supset \mu$ denote that $\lambda_i \geq \mu_i$ for all $i$. 
\end{df}

\begin{ex}
If $N=3$ and the number of boxes is $3$, 
then 
\[\xymatrix@!C=26pt{
  & (\emptyset, \emptyset, (3))  \ar[rd] & & (\emptyset, \emptyset, (2,1)) \ar[ld]\ar[rd] & & (\emptyset, \emptyset, (1,1,1))\ar[ld] & \\
  & & (\emptyset, (1), (2))\ar[lld]\ar[d]\ar[rrrrd] & & (\emptyset, (1), (1,1))\ar[lllld]\ar[d]\ar[rrd] & & \\
(\emptyset, (2), (1))\ar[d]\ar[rrd]\ar[rrrrd] & & ((1), \emptyset, (2))\ar[d] & & ((1), \emptyset, (1,1))\ar[lld] & & (\emptyset, (1,1), (1))\ar[d]\ar[lld]\ar[lllld] \\
(\emptyset, (3), \emptyset)\ar[d] & & ((1), (1), (1))\ar[lld]\ar[d]\ar[rrd]\ar[rrrrd] & & (\emptyset, (2,1), \emptyset)\ar[lllld]\ar[rrd]  & & (\emptyset, (1,1,1), \emptyset)\ar[d] \\
((1), (2), \emptyset)\ar[rrd]\ar[rrrrd] & & ((2), \emptyset, (1))\ar[d]  & & ((1,1), \emptyset, (1))\ar[d] & & ((1), (1,1), \emptyset)\ar[lllld]\ar[lld] \\
  & & ((2), (1), \emptyset)\ar[ld]\ar[rd] & & ((1,1), (1), \emptyset)\ar[ld]\ar[rd] & & \\
  & ((3), \emptyset, \emptyset) & & ((2,1), \emptyset, \emptyset) & & ((1,1,1), \emptyset, \emptyset). &
}\]
Here $\vl \rightarrow \vm$ stands for $\vl \overstar{\succ} \vm$. 
\end{ex}

By using the following conjecture, 
we can state the existence theorem.

\begin{conj}\label{conj:action of eta_n}
Let $\displaystyle \eta_n^{(i)} \seteq \oint \frac{dz}{2 \pi \sqrt{-1}z} \eta^{(i)}(z) z^{n}$.  
In the action of $\eta_n^{(i)}$ ($n \geq 1$) on Macdonald functions 
$P_{\lambda}(a_{-n}^{(i)};q,t) \ket{u}$, 
there only appear partitions $\mu$ contained in $\lambda$, i.e.,  
\begin{equation}
\eta_{n}^{(i)} P_{\lambda}(a^{(i)}_{-n};q,t) \ket{u} = \sum_{\mu \subset \lambda} c_{\lambda, \mu} P_{\mu}(a^{(i)}_{-n};q,t) \ket{\vm}. 
\end{equation}
\end{conj}

\begin{thm}\label{thm: another existence thm}
Under the Conjecture \ref{conj:action of eta_n}, 
for an $N$-tuple of partitions $\vl$, 
there exists an unique vector $\ket{P_{\vl}} \in \mathcal{F}_{\vu}$ such that 
\begin{align}
 &\ket{P_{\vec{\lambda}}} 
  = \prod_{i=1}^N P_{\lambda^{(i)}}(a^{(i)}_{-n};q,t) \ket{\vu} 
  + \sum_{\vec{\mu} \overstar{\prec} \vec{\lambda}} c_{\vl, \vm} \prod_{i=1}^N P_{\mu^{(i)}}(a^{(i)}_{-n};q,t) \ket{\vu}, \\
 &X^{(1)}_0 \ket{P_{\vec{\lambda}}} = \epsilon_{\vec{\lambda}} \ket{P_{\vec{\lambda}}}. 
\end{align}
\end{thm}

\Proof
At first, 
$\eta_n^{(i)}$ satisfies 
\begin{align}
& \eta_0^{(i)} P_{\lambda}(a^{(i)}_{-n}) \ket{\vu}= e_{\lambda} P_{\lambda}(a^{(i)}_{-n}) \ket{\vu}, \quad 
\eta_0^{(j)} P_{\lambda}(a^{(i)}_{-n}) \ket{\vu}=  P_{\lambda}(a^{(i)}_{-n}) \ket{\vu}, \\
& \eta_n^{(j)} P_{\lambda}(a^{(i)}_{-n}) \ket{\vu}=0  \qquad  i \neq j, \quad n \geq 1. \nonumber
\end{align}
If we act $\displaystyle \Lambda_0^{i} \seteq \oint \frac{dz}{2 \pi \sqrt{-1} z} \Lambda^i(z)$ 
on the product of the Macdonald functions, then 
\begin{align}
\Lambda^{i}_0 \prod_{j=1}^{N} P_{\lambda^{(j)}}(a_{-n}^{(j)}) \ket{\vu} 
=&e_{\lambda^{(i)}} \prod_{j=1}^{N} P_{\lambda^{(j)}}(a_{-n}^{(j)}) \ket{\vu} \\
 &+ \sum_{\mu \subset \lambda^{(i)}} c'_{\lambda^{(i)}, \mu}(a_{-n}^{(1)},\ldots , a_{-n}^{(i-1)}) P_{\mu}(a_{-n}^{(i)}) \prod_{j \neq i} P_{\lambda^{(j)}}(a_{-n}^{(j)}) \ket{\vu},  \nonumber
\end{align}
where $c'_{\lambda^{(i)}, \mu}(a_{-n}^{(1)},\ldots , a_{-n}^{(i-1)})$ is a polynomial 
of degree
 $|\lambda^{(i)}| - |\mu|$ of $a_{-n}^{(1)},\ldots , a_{-n}^{(i-1)}$.%
\footnote{
 	That is to say, 
 	for the operator $\mathcal{O}$ such that 
 	$[\mathcal{O}, a^{(j)}_{-n} ]= n a^{(j)}_{-n}$, 
 	the polynomial $c'_{\lambda^{(i)}, \mu}$ satisfies 
 	$[\mathcal{O}, c'_{\lambda^{(i)}, \mu}] = (|\lambda^{(i)}| - |\mu|) c'_{\lambda^{(i)}, \mu}$. 
 }
Hence 
\begin{equation}
X_{0}^{(1)} \prod_{j=1}^{N} P_{\lambda^{(j)}}(a_{-n}^{(j)}) \ket{\vu} 
= \epsilon_{\vl} \prod_{j=1}^{N} P_{\lambda^{(j)}}(a_{-n}^{(j)}) \ket{\vu} 
 + \sum_{\vm \overstar{\prec} \vl} c_{\vl, \vm} \prod_{j=1}^{N} P_{\mu^{(j)}}(a_{-n}^{(j)}) \ket{\vu}. 
\end{equation}
Therefore one can easily diagonalize it and we have this theorem. 
\qed

In the basis of monomial symmetric functions 
$\ket{m_{\vl}} \seteq \prod_{i=1}^N m_{\lambda^{(i)}}(a_{-n}^{(i)}) \ket{\vu}$, 
we have 
\begin{align}
\Xo_0 \ket{m_{\vl}} 
 &= \Xo_0 \sum_{\vl \geq \vm} d_{\vl \vm} \prod_{i=1}^N P_{\mu^{(i)}}(a_{-n}^{(i)}) \ket{\vu} \\
 &= \sum_{\vl \geq \vm} d_{\vl \vm} \sum_{\vm \overstar{\succeq} \vn} d'_{\vl \vm} \prod_{i=1}^N P_{\nu^{(i)}}(a_{-n}^{(i)}) \ket{\vu}  \nonumber \\
 &= \sum_{\vl \geq \vm} \sum_{\vm \overstar{\succeq} \vn} \sum_{\vn \geq \vec{\rho}} d_{\vl \vm} \,  d'_{\vm \vn} \, d''_{\vn \vec{\rho}} \,  \ket{m_{\vec{\rho}}},   \nonumber 
\end{align}
where 
\begin{equation}
\vec{\lambda} \geq \vec{\mu} \quad \overset{\mathrm{def}}{\Leftrightarrow} \quad 
(|\lambda^{(1)}|, \ldots , |\lambda^{(N)}|) = (|\mu^{(1)}|, \ldots , |\mu^{(N)}|) 
\quad \mathrm{and} \quad
\lambda^{(\alpha)} \geq \mu^{(\alpha)}
\end{equation}
($1 \leq \forall \alpha \leq N$).
Thus the partial ordering $\overwstar{\succeq}$ defined as follows also triangulates $\Xo_0$. 
\begin{equation}
\vl \overwstar{\succeq} \vec{\rho} \quad \overset{\mathrm{def}}{\Leftrightarrow} \quad 
\mbox{there exsist $\vm$ and $\vn$ such that 
$\vec{\lambda} \geq \vec{\mu} \overstar{\succeq} \vec{\nu} \geq \vec{\rho}$ }. 
\end{equation}
It can be shown that the partial ordering $\overwstar{\succeq}$ 
is equivalent to the ordering $\geq^{\mathrm{L}}$ introduced in \cite{awata2011notes}. 
Therefore Theorem \ref{thm: another existence thm} supports 
the existence theorem in \cite[Proposition3.8]{awata2011notes}.


\subsection{Realization of rank $N$ representation by generalized Macdonald function}
\label{sec:realization of rank N rep}

A representation of the DIM algebra called rank $N$ representation 
is provided in \cite{Bourgine:2016Coherent} 
in terms of a basis $\tket{\vu, \vl}$ called AFLT basis. 
This rank $N$ representation corresponds 
to the $N$-fold tensor product of the level (0,1) representation 
described in Appendix \ref{sec:Def of DIM}. 
The level (0,1) representation can be considered as the 
spectral dual to the level (1,0) representation 
which is realized by the Heisenberg algebra. 
In this subsection, based on this spectral duality 
we present conjectures for explicit expressions 
of the action of $x^{+}_{\pm 1}$ on the generalized Macdonald functions,
which are defined to be eigenfunctions of the Hamiltonian $\Xo_0$.
We can also conjecture the eigenvalues of higher rank Hamiltonians on 
the generalized Macdonald functions from those of the spectral 
dual generators provided in \cite{Bourgine:2016Coherent}.
Our conjectures mean that 
the generalized Macdonald functions concretely realize 
the spectral dual basis to $\tket{\vu, \vl}$ in \cite{Bourgine:2016Coherent}.

\subsubsection*{Action of $x_{\pm 1}^{+}$ on generalized Macdonald function}

Although we already define the integral forms $\ket{K_{\vl}}$ 
of the generalized Macdonald functions, 
let us use another renormalization $\tket{\widetilde{M}_{\vl}}$ 
of them, 
which is defined by 
\begin{equation}
\tket{\widetilde{M}_{\vl}} =  \ket{P_{\vl}} \times
\prod_{1\leq i<j \leq N} N_{\lambda^{(j)}, \lambda^{(i)}}(u_j/u_i) 
\prod_{k=1}^N \prod_{(i,j)\in \lambda^{(k)}} 
(1-q^{\lambda^{(k)}_i-j} t^{\lambda^{(k)'}_j-i+1}), 
\end{equation}
where $N_{\lambda,\mu}(u)$ is the Nekrasov factor. 
This renormalization is the almost same as $\ket{K_{\vl}}$. 
Their difference is conjectured to be the scalar multiplication 
of only monomials in parameter $q$, $t$ and $u_i$.

It is expected that 
the basis $\tket{\widetilde{M}}_{\vl}$ corresponds to 
the AFLT basis\footnote{Originally, the AFLT basis is defined by the property that 
their inner products and matrix elements of vertex operators 
reproduce the Nekrasov factor. In \cite{awata2011notes}
the integral forms $\widetilde{M}_{\vl}$ were already 
conjectured to be the AFLT basis in this original sense.}
 in \cite{Bourgine:2016Coherent} 
and realizes the rank $N$ representation 
through the spectral duality $\cal S$. 
That is to say, for any generator $a$ in the DIM algebra, 
the action of $\rho^{(N)}_{\vu} \circ {\cal S}(a)$
on the integral forms $\tket{\widetilde{M}_{\vl}}$ is 
in the same form as one of $\rho^{\mathrm{rank} N} (a)$ on the basis $\tket{\vu,\vl}$
\cite{Bourgine:2016Coherent}, 
where $\rho^{\mathrm{rank} N}\seteq \rho^{(0,1)}_{u_1} \otimes \cdots 
\otimes \rho^{(0,1)}_{u_N} \circ \Delta^{(N)}$. 
Indeed, we can check that 
the action of $x^+_{\pm 1}$ on the generalized Macdonald functions is 
as the following conjecture.
Let us denote adding a box to or removing it
from the Young diagram $\vl$ through $A(\vl)$ and $R(\vl)$ respectively.
We also use the notation $\chi_{(\ell, i,j)}=u_{\ell} t^{-i+1}q^{j-1}$
for the triple $x=(\ell,i,j)$, where $(i,j) \in \lambda^{(\ell)}$ is 
the coordinate of the box of the Young diagram $\lambda^{(\ell)}$.
\begin{conj}\label{conj:Action of X}
\begin{equation}
\Xo_1 \tket{\tM_{\vl}} \overset{?}{=} 
\sum_{\substack{|\vm| = |\vl|-1 \\ \vl \supset \vm}} 
\tilde{c}^{(+)}_{\vl,\vm} \tket{\tM_{\vm}}, \quad  
\Xo_{-1} \tket{\tM_{\vl}} \overset{?}{=} 
\sum_{\substack{|\vm| = |\vl|+1\\ \vl \subset \vm}} 
\tilde{c}^{(-)}_{\vl,\vm} \tket{\tM_{\vm}}, 
\end{equation}
where 
\begin{align}
&\tilde{c}^{(+)}_{\vl,\vm} = 
\xi^{(+)}_x 
\frac{\prod_{y \in A(\vl)} (1-\chi_x \chi_y^{-1}  (q/t) )}
{\prod_{\substack{y \in R(\vl)\\ y \neq x}} (1-\chi_x \chi_y^{-1})}, 
\quad x \in \vl \setminus \vm, \\
& \tilde{c}^{(-)}_{\vl,\vm} = 
\xi^{(-)}_x 
\frac{\prod_{y \in R(\vl)} (1-\chi_y \chi_x^{-1}  (q/t) )}
{\prod_{\substack{y \in A(\vl)\\ y \neq x}} (1-\chi_y \chi_x^{-1})}, 
\quad x \in \vm \setminus \vl, 
\end{align}
and for the triple $(\ell, i,j)$, we put 
\begin{equation}
\xi^{(+)}_{(\ell, i,j)} = (-1)^{N+\ell} p^{-\frac{\ell+1}{2}} t^{(N-\ell)i} q^{(\ell-N+1)j} \frac{\prod_{k=1}^{N-\ell}u_{\ell+k}}{u_{\ell}^{N-\ell-1}}, \quad 
\xi^{(-)}_{(\ell, i,j)} = (-1)^{\ell} p^{\frac{\ell-1}{2}} t^{(\ell-2)i} q^{(1-\ell)j} \frac{\prod_{k=1}^{\ell -1}u_k}{u_{\ell}^{\ell-2}}.
\end{equation}
\end{conj}

These actions of $\Xo_{\pm 1}$  in this conjecture come from the corresponding actions of 
the generators $f_1$ and $e_1$ in \cite{Bourgine:2016Coherent} respectively, i.e., 
$x^{-}_1$ and $x^+_{1}$ in our notation, which are the spectral duals of $x^{+}_1$ and $x^{+}_{-1}$.  
Incidentally, introducing the coefficients 
$c^{(\pm)}_{\vl,\vm}=c^{(\pm)}_{\vl,\vm}(q,t|u_1,\ldots,u_N)$ by
\begin{equation}
c^{(\pm)}_{\vl,\vm}= 
\prod_{1\leq i<j\leq N} 
\frac{N_{\mu^{(j)},\mu^{(i)}}(u_j/u_i)}
     {N_{\lambda^{(j)},\lambda^{(i)}}(u_j/u_i)}
\prod_{k=1}^N
\frac{ \prod_{(i,j)\in \mu^{(k)}} 
(1-q^{\mu^{(k)}_i-j} t^{\mu^{(k)\mathrm{T}}_j-i+1})}
{ \prod_{(i,j)\in \lambda^{(k)}} 
(1-q^{\lambda^{(k)}_i-j} t^{\lambda^{(k)\mathrm{T}}_j-i+1})}
\times 
\tilde{c}^{(\pm)}_{\vl,\vm}, 
\end{equation}
i.e., $\Xo_{\pm 1} \ket{P_{\vl}} 
= \sum_{\vm} c^{(\pm)}_{\vl,\vm} \ket{P_{\vm}}$, 
we can further conjecture that 
\begin{equation}\label{eq:rel between cplus and cminus}
c^{(+)}_{\vl,\vm}(q,t|u_1,\ldots , u_N) 
\overset{?}{=} 
-c^{(-)}_{(\mu^{(N)\prime} ,\ldots,\mu^{(1)\prime}),(\lambda^{(N)\prime},\ldots,\lambda^{(1)\prime})}
(t^{-1},q^{-1}|p^{(N-1)/2}u_N,\ldots,p^{(N-1)/2}u_1). 
\end{equation}
Conjecture \ref{conj:Action of X} with respect to $X^{(1)}_1$ and 
the formula (\ref{eq:rel between cplus and cminus}) 
are checked on a computer for $|\vl| \leq 5$ for $N=1$, 
for $|\vl| \leq 3$ for $N=2, 3$ and  
for $|\vl| \leq 2$ for $N=4$. 
Conjecture \ref{conj:Action of X} with respect to $\Xo_{-1}$ is also checked 
for the same size of $\vm$.

\subsubsection*{Higher Hamiltonian}

For each integer $k\geq 1$, 
the spectral dual of $\psi^+_k$ is $H_k$ defined by $H_1 = \Xo_0$ and 
\begin{equation}
H_k=[\Xo_{-1},\underbrace{[\Xo_0,\cdots ,[\Xo_0}_{k-2},\Xo_{1}]\cdots]],
\qquad k \geq 2.
\end{equation}
According to \cite{Miki:2007}, $H_k$ are spectral dual to $\psi^+_{k}$ and 
consequently mutually commuting; $[H_k,H_l]=0$. 
Thus the generalized Macdonald functions $\ket{P_{\vl}}$ are automatically 
eigenfunctions of all $H_k$, 
i.e.,  
$H_k \ket{P_{\vl}} = \epsilon^{(k)}_{\vl} \ket{P_{\vl}}$,  
and $H_k$ can be regarded as higher Hamiltonians 
for the generalized Macdonald functions. 
Since $H_k$ are the spectral duals to $\psi^+_{k}$; $H_k = {\cal S}(\psi^+_{k})$,
their eigenvalues are expected to be

\begin{conj}
\begin{equation}\label{eq:higher eigenvlue}
\epsilon^{(k)}_{\vl} \overset{?}{=} 
\frac{(1-q)^{k-1}(1-t^{-1})^{k-1}}{1-p^{-1}}
\oint \frac{dz}{2 \pi \sqrt{-1}z }
\prod_{i=1}^N B^{+}_{\lambda^{(i)}}(u_i z)z^{-k},  
\end{equation} 
where $B^{+}_{\lambda}(z)$ is defined in (\ref{eq:B+}). 
\end{conj}

The eigenvalues $\epsilon^{(k)}_{\vl}$ correspond to those of the rank $N$ representation of the generators $\psi^+_{k}$
in \cite{Bourgine:2016Coherent}. 
In the $k=1$ case, the conjecture (\ref{eq:higher eigenvlue}) can be proved. 
We have checked it  
for $|\vl| \leq 5$ for $N=1$, 
for $|\vl| \leq 3$ for $N=2$, 
for $|\vl| \leq 2$ for $N=3$ and
for $|\vl| \leq 1$ for $N=4$ in the $k \leq 5$ case.



\subsection{Limit to $\beta$ deformation}
\label{sec:Limit to beta}
In this subsection, 
we show that the generalized Macdonald functions are reduced to 
the generalized Jack functions introduced by 
Morozov and Smirnov in \cite{morozov2013finalizing} 
in the $q \rightarrow 1$ limit (see also the sub-thesis \cite{Ohkubo:2014:existence}). 
Although the scenario of proof of AGT correspondence is given 
in \cite{morozov2013finalizing}, 
the orthogonality of the generalized Jack functions are non-trivial 
since there are degenerate eigenvalues. 
The Cauchy formula used in the scenario of proof 
can not be proved without the orthogonality.  
However, 
the eigenvalues of generalized Macdonald functions are non-degenerate. 
Hence, 
we can prove the orthogonality of the generalized Jack functions 
by using the limit in this section.  
When taking this limit, 
we set $u_i=q^{u_i'}$ 
$( i=1, \ldots , N )$, 
$t=q^{\beta}$, 
$q=e^{\hbar}$ 
and take the limit $\hbar\rightarrow 0$ with $\beta$ fixed. 
To expose the $\hbar$ dependence of the generators $a^{(i)}_n$ in the Heisenberg algebra, 
they are realized in terms of $N$ kinds of power sum symmetric functions 
$p^{(i)}_{\lambda} \seteq \prod_{k\geq 1} p^{(i)}_{\lambda_k}$ 
with $p^{(i)}_n=\sum_{l} \left( x_l^{(i)} \right)^n$ 
in the ring of symmetric functions $\Lambda^{\otimes N}$ 
in the variables $x^{(i)}_k$ ($i=1,\ldots , N$, $k\in \mathbb{N}$):
\begin{eqnarray}
a^{(i)}_n \mapsto n \frac{1-q^n}{1-t^n} \frac{\partial }{\partial p^{(i)}_n }, \qquad 
a^{(i)}_{-n} \mapsto p^{(i)}_n \qquad (n>0). 
\end{eqnarray} 
Since the operators $U_i$ become parameter $u_i$ in the representation 
space $\mathcal{F}_{\vu}$, 
they are transformed as 
\begin{eqnarray}
U_i \mapsto u_i 
\end{eqnarray}
from the beginning in this section. 
With the above transformation, 
the generator $\Xo_0$ can be regarded as the operator over the ring of symmetric functions $\Lambda^{\otimes N}$. 
Define the isomorphism 
$\iota : \mathcal{F}_{\vu} \rightarrow \Lambda^{\otimes N}$ by 
\begin{equation}
\ket{a_{\vl}} \overset{\iota}{\mapsto} p_{\vl} \seteq \prod_{i=1}^{N}p^{(i)}_{\lambda^{(i)}},  
\end{equation}
where $\ket{a_{\vl}}$ is defined in Section \ref{sec:Explicit cal of R}. 
The inner product $\left\langle -,- \right\rangle_{q,t}$ 
over $\Lambda^{\otimes N}$ is defined by 
\begin{equation}
\left\langle p_{\vec{\lambda}}, p_{\vec{\mu}} \right\rangle_{q,t}
=\delta_{\vec{\lambda}, \vec{\mu}} \prod_{i=1}^{N} z_{\lambda^{(i)}}
 \prod_{k=1}^{\ell (\lambda^{(i)})}\frac{1-q^{\lambda_k^{(i)}}}{1-t^{\lambda_k^{(i)}}} , \qquad
z_{\lambda^{(i)}} \mathbin{:=} \prod_{k \geq 1}k^{m_k} \, m_k !. 
\end{equation}
This inner product naturally realize the one over the Fock module $\mathcal{F}_{\vu}$, i.e., 
\begin{equation}
\left\langle p_{\vl}, p_{\vm} \right\rangle_{q,t} = 
\braket{a_{\vl}}{ a_{\vm}}.
\end{equation}
Let $X^{(1) \perp}_{0}$ be the adjoint operator of $\Xo_0$ 
with respect to the inner product. 
Since the generalized Macdonald functions 
$P_{\vl} \seteq \iota (\ket{P_{\vl}})$ have non-degenerate eigenvalues 
the orthogonality clearly follows: 
\begin{eqnarray}\label{eq:orthogonality of gn Mac}
\vec{\lambda} \neq \vec{\mu} \quad  \Longrightarrow  \quad 
\left\langle P^*_{\vec{\lambda}}, P_{\vec{\mu}} \right\rangle_{q,t} =0.
\end{eqnarray} 
where $P^*_{\vl}$ is defined to be the eigenfunctions 
of the adjoint operator $X^{(1) \perp}_{0}$ of the eigenvalue $\epsilon_{\vl}$.

Now let us take the $q\rightarrow 1$ limit. 
At first, consider the $\hbar$ expansion of $\Xo(z)$. 
By $(1-t^{-n})(1-t^n q^{-n})/n=\mathcal{O}(\hbar^2)$ and 
$(1-t^{-n})/n$, $(1-q^{n})=\mathcal{O}(\hbar)$, 
we have 
\begin{align}
\oint \frac{dz}{2\pi \sqrt{-1} z}\Lambda_i(z) 
 =&1+ \sum_{n=1}^{\infty} \left( -\frac{(1-t^{-n})(1-q^n)}{n} p_{n}^{(i)} \frac{\partial }{\partial p_{n}^{(i)}} \right) \\
  &+ \sum_{k=1}^{i-1} \sum_{n=1}^{\infty} \left( -\frac{(1-t^{-n})(1-t^n q^{-n})(1-q^n)}{n} (t/q)^{-\frac{(i-k-1)n}{2}} p_{n}^{(k)} \frac{\partial }{\partial p_{n}^{(i)}} \right) \nonumber \\
  &+ \frac{1}{2}\sum_{n,\, m} \left( -\frac{(1-t^{-n})(1-t^{-m})(1-q^{n+m})}{n\, m } p_{n}^{(i)} p_{m}^{(i)} \frac{\partial }{\partial p_{n+m}^{(i)}} \right) \nonumber \\
  &+ \frac{1}{2}\sum_{n,\, m} \left( \frac{(1-t^{-n-m})(1-q^{n})(1-q^{m})}{ (n+m)} p_{n+m}^{(i)} \frac{\partial }{\partial p_{n}^{(i)}} \frac{\partial }{\partial p_{m}^{(i)}} \right) + \mathcal{O}(\hbar^4).  \nonumber
\end{align}
Hence the $\hbar$ expansion is
\begin{align}
\oint \frac{dz}{2\pi \sqrt{-1} z}\Lambda_i(z) 
=& 1+ \hbar^2 \left\{ \beta \sum_{n=1}^{\infty} n \, p^{(i)}_n \frac{\partial }{\partial p^{(i)}_n} \right\} 
 + \hbar^3 \left\{ \beta (1-\beta) \sum_{k=1}^{i-1} \sum_{i=1}^{\infty} n^2 p^{(k)}_n \frac{\partial }{\partial p^{(i)}_n} \right.  \\
 & + \frac{\beta ^2}{2}\sum_{n, m} (n+m) p^{(i)}_n p^{(i)}_m \frac{\partial }{\partial p^{(i)}_{n+m}} 
 + \frac{\beta }{2}\sum_{n, m} n\, m\, p^{(i)}_{n+m} \frac{\partial^2 }{\partial p^{(i)}_n \partial p^{(i)}_m }  \nonumber \\
 &+ \left. \frac{\beta(1-\beta)}{2} \sum_{n=1}^{\infty} n^2 p^{(i)}_n \frac{\partial }{\partial p^{(i)}_n}  \right\} + \mathcal{O}(\hbar^4).  \nonumber
\end{align}
Thus we get
\begin{align}
u_i \oint \frac{dz}{2\pi \sqrt{-1} z}\Lambda_i(z) 
=& 1+ u_i' \, \hbar + \hbar^2 \left\{ \beta \sum_{n=1}^{\infty} n \, p^{(i)}_n \frac{\partial }{\partial p^{(i)}_n} +\frac{1}{2}u_i'^2 \right\} \\
 & + \hbar^3 \left\{ \beta\,  \mathcal{H}_{\beta}^{(i)} + \beta \sum_{k=1}^{i-1} \mathcal{H}_{\beta}^{(i,k)} + \frac{u_i'^3}{6}  \right\} + \mathcal{O}(\hbar^4),  \nonumber
\end{align}
where
\begin{equation}
\mathcal{H}_{\beta}^{(i)}
\mathbin{:=} \frac{1}{2}\sum_{n,m} \left( \beta (n+m) p^{(i)}_n p^{(i)}_m \frac{\partial }{\partial p^{(i)}_{n+m}}+n\, m\, p^{(i)}_{n+m} \frac{\partial^2 }{\partial p^{(i)}_n \partial p^{(i)}_m } \right) 
 +\sum_{n=1}^{\infty} \left( u_i' + \frac{1-\beta}{2} n \right) n p^{(i)}_n \frac{\partial }{\partial p^{(i)}_n}, 
\end{equation}
\begin{equation}
\mathcal{H}_{\beta}^{(i,k)} \mathbin{:=} (1-\beta)\sum_{n=1}^{\infty} n^2 p^{(k)}_n \frac{\partial }{\partial p^{(i)}_n}.
\end{equation}
For $k=0,1,2,\ldots$ ,
we define operators $H_k$ by 
\begin{equation}
X_0^{(1)} \mathbin{=:} \sum_{k=0}^{\infty} \hbar^k H_k .
\end{equation}
With respect to  $H_0$, $H_1$ and $H_2$, 
all homogeneous symmetric functions 
belong to the same eigenspace. 
Hence, 
the eigenfunctions of $\sum_{k=3}^{\infty} \hbar^k H_k$ 
are the same to those of $X_0$. 
In addition, we have
\begin{equation}
\lim_{\hbar \rightarrow 0} \left( \frac{X_0-(H_0+\hbar H_1+ \hbar^2 H_2)}{(t-1)(q-1)^2}\right) =\mathcal{H}_{\beta} +\frac{1}{6 \beta}\sum_{i=1}^{N}u_i'^3 , 
\end{equation}
\begin{equation}
\mathcal{H}_{\beta} \mathbin{:=} \sum_{i=1}^{N} \mathcal{H}_{\beta}^{(i)} + \sum_{i>j}\mathcal{H}_{\beta}^{(i,j)}. 
\end{equation}
Consequently the limit $q \rightarrow 1$ of the generalized Macdonald functions 
are eigenfunctions of the differential operator $\mathcal{H}_{\beta}$. 
As a matter of fact, 
$\mathcal{H}_{\beta}$ 
plus the momentum $(\beta-1)\sum_{i=1}^{N} \sum_{n =1}^{\infty} n p^{(i)}_n \frac{\partial }{\partial p^{(i)}_n}$
corresponds to the differential operator of \cite{morozov2013finalizing, SU(3)GnJack}, 
the eigenfunctions of which are called generalized Jack symmetric functions.%
\footnote{
To be adjusted to the notation of \cite{morozov2013finalizing, SU(3)GnJack}, 
we need to transform the subscripts:
$u_i' \rightarrow u_{N-i+1}'$, $p^{(i)} \rightarrow p^{(N-i+1)}$.
}

As in \cite[Proposition 3.7]{awata2011notes},  
we can triangulate $\mathcal{H}_{\beta}$ similarly. 
Moreover 
if $\vec{\lambda} \geq^{\mathrm{L}} \vec{\mu}$  
\footnote{
We write $\vec{\lambda} \geq^{\mathrm{L}} \vec{\mu}$ 
(resp. $\vec{\lambda} \geq^{\mathrm{R}} \vec{\mu}$) 
if and only if  $|\vec{\lambda}|=|\vec{\mu}|$ and
\begin{equation}
|\lambda^{(N)}|+ \cdots + |\lambda^{(j+1)}|+\sum_{k=1}^i \lambda^{(j)}_k 
\geq |\mu^{(N)}|+ \cdots + |\mu^{(j+1)}|+\sum_{k=1}^i \mu^{(j)}_k
\end{equation}
\begin{equation}
\left( \mathrm{resp. } \quad  |\lambda^{(1)}|+ \cdots + |\lambda^{(j-1)}|+\sum_{k=1}^i \lambda^{(j)}_k 
\geq |\mu^{(1)}|+ \cdots + |\mu^{(j-1)}|+\sum_{k=1}^i \mu^{(j)}_k \right)
\end{equation}
for all $i\geq 1$ and $1 \leq j \leq N$. 
}
and $\beta$ is generic, 
then $e'_{\vec{\lambda}} \neq e'_{\vec{\mu}}$. 
($e'_{\vec{\lambda}}$, $e'_{\vec{\mu}}$ are eigenvalues of $\mathcal{H}_{\beta}$.)
Therefore we get the existence theorem of the generalized Jack symmetric functions.

\begin{prop} 
There exists a unique symmetric function $J_{\vec{\lambda}}$ 
satisfying the following two conditions: 
\begin{align}
 &J_{\vec{\lambda}}=m_{\vec{\lambda}}+\sum_{\vec{\mu} <^{\mathrm{L}} \vec{\lambda}} d'_{\vec{\lambda} \vec{\mu}}\, m_{\vec{\mu}} , \qquad d'_{\vec{\lambda} \vec{\mu}} \in \mathbb{Q}(\beta, u'_1,\ldots ,u'_N); \\
 &\mathcal{H}_{\beta} J_{\vec{\lambda}}= e'_{ \vec{\lambda}} J_{\vec{\lambda}} ,\qquad e'_{ \vec{\lambda}} \in \mathbb{Q}(\beta , u'_1,\ldots ,u'_N), 
\end{align}
where $m_{\vec{\lambda}}$ denotes 
the product of monomial symmetric functions 
$\prod_{i=1}^{N} m_{\lambda^{(i)}}^{(i)}$. 
($ m_{\lambda^{(i)}}^{(i)}$ is the usual monomial symmetric function 
of variables $ \{x^{(i)}_n \mid n \}$.)
\end{prop}

From the above argument and the uniqueness in this proposition
we get the following important result.

\begin{prop}\label{prop:limit of GnMac}
The limit of the generalized Macdonald symmetric functions $P_{\vec{\lambda}}$ 
to $\beta$-deformation coincide with the generalized Jack symmetric functions $J_{\vec{\lambda}}$. 
That is 
\begin{equation}
P_{\vec{\lambda}} \underset{\substack{\hbar \rightarrow 0,\\ u_i=q^{u_i'},\, t=q^{\beta},\, q=e^{\hbar}}}{\longrightarrow} J_{\vec{\lambda}}.
\end{equation}
\end{prop}

\begin{rem}
For the dual functions $P_{\vec{\lambda}}^*$ and $J_{\vec{\lambda}}^*$, 
a similar proposition holds. 
\end{rem}

By the orthogonality (\ref{eq:orthogonality of gn Mac}), 
Proposition \ref{prop:limit of GnMac} and 
the fact that the scalar product $\left\langle -,- \right\rangle_{q,t}$
reduces to the scalar product $\left\langle -,- \right\rangle_{\beta}$
which is defined by 
\begin{equation}
\left\langle p_{\vec{\lambda}}, p_{\vec{\mu}} \right\rangle_{\beta} 
=\delta_{\vec{\lambda}, \vec{\mu}} \prod_{i=1}^{N} z_{\lambda^{(i)}} \beta^{-\ell (\lambda^{(i)})}, 
\end{equation} 
we obtain the orthogonality of the generalized Jack symmetric functions.

\begin{prop}
If $\vec{\lambda} \neq \vec{\mu}$, 
then 
\begin{equation}
\left\langle J_{\vec{\lambda}}^* , J_{\vec{\mu}}  \right\rangle_{\beta} = 0. 
\end{equation}
\end{prop}

By this proposition, 
we can prove the Cauchy formula for generalized Jack symmetric functions 
in the usual way. 
For example, in the $N=2$ case, 
we have 
\begin{equation}
\sum_{\vec{\lambda}}
\frac{J_{\vec{\lambda}}(x^{(1)},x^{(4)})\, J^*_{\vec{\lambda}}(x^{(2)},x^{(3)})}{v_{\vec{\lambda}}}
=\exp \left( \beta\sum_{n \geq 1} \frac{1}{n} p_n^{(1)}p_n^{(2)}  \right)
\exp \left( \beta\sum_{n \geq 1} \frac{1}{n} p_n^{(3)}p_n^{(4)}  \right),  
\end{equation}
where $v_{\vec{\lambda}}:
= \langle J_{\vec{\lambda}}^* , J_{\vec{\lambda}}  \rangle_{\beta}$. 
This is the necessary formula 
in the scenario of proof of the AGT conjecture \cite{morozov2013finalizing}. 

We give examples of Proposition \ref{prop:limit of GnMac} in the case $N=2$.
The generalized Macdonald symmetric functions of level 1 and 2
have the forms:
\begin{equation}
\left(
\begin{array}{l}
P_{(0),(1)}  \\
P_{(1),(0)} 
\end{array}
\right)
=
M_{q,t}^1 
\left(
\begin{array}{l}
m_{(0),(1)}  \\
m_{(1),(0)} 
\end{array}
\right), \qquad
M_{q,t}^1 
\mathbin{:=}
\left(
\begin{array}{cc}
 1 & (t/q)^{\frac{1}{2}} \frac{(t-q)u_2}{t(u_1-u_2)} \\
 0 & 1
\end{array}
\right), 
\end{equation}
\begin{equation}
\left(
\begin{array}{l}
P_{(0),(2)}  \\
P_{(0),(1,1)} \\
P_{(1),(1)}  \\
P_{(2),(0)} \\
P_{(1,1),(0)}  
\end{array}
\right)
=
M_{q,t}^2
\left(
\begin{array}{l}
m_{(0),(2)}  \\
m_{(0),(1,1)} \\
m_{(1),(1)}  \\
m_{(2),(0)} \\
m_{(1,1),(0)} 
\end{array}
\right), 
\qquad M_{q,t}^2 \mathbin{:=}
\end{equation}
{\footnotesize 
\begin{equation*}
 \left(
\begin{array}{ccccc}
1 & \frac{(1+q)(t-1)}{q t-1} &  \frac{(t/q)^{-\frac{1}{2}} (1+q)(q-t)(t-1)u_2}{(1-q t)(u_1-q u_2)} & \frac{(q-t)((1-q^2)t u_1 - q(t^2 -q(1+q)t+q)u_2 ) u_2}{q t (q t -1)(u_1 - u_2 )(u_1- q u_2)} & \frac{(1+q)(q-t)(t-1)((q-1)t u_1 +q(q-t)u_2)u_2}{qt (qt-1)(u_1-u_2)(u_1-q u_2)}  \\
0 & 1 & (t/q)^{\frac{1}{2}} \frac{(t-q)u_2}{t(t u_1 -u_2)} & \frac{(q-t)u_2}{q(t u_1 -u_2)} & \frac{(q-t)(q u_2 - t((t-1)u_1+u_2))u_2}{q t (u_1 -u_2)(tu_1 -u_2)} \\
0 & 0 & 1 & (t/q)^{\frac{1}{2}} \frac{(t-q)u_2}{t(qu_1-u_2)} & (t/q)^{\frac{1}{2}} \frac{(q-t)((1+q+(q-1)t)u_1-2t u_2)}{t(q u_1- u_2)(-u_1 +t u_2)} \\
0 & 0 & 0 & 1 & \frac{(1+q)(t-1)}{q t-1} \\
0 & 0 & 0 & 0 & 1
\end{array}
\right)
\end{equation*}
}

\noindent Also the generalized Jack symmetric functions have the forms: 
\begin{equation}
\left(
\begin{array}{l}
J_{(0),(1)}  \\
J_{(1),(0)} 
\end{array}
\right)
=
M_{\beta}^1 
\left(
\begin{array}{l}
m_{(0),(1)}  \\
m_{(1),(0)} 
\end{array}
\right), \qquad 
M_{\beta}^1 
\mathbin{:=}
\left(
\begin{array}{cc}
 1 & \frac{1-\beta}{-u'_1+u'_2} \\
 0 & 1
\end{array}
\right), 
\end{equation}
\begin{equation}
\left(
\begin{array}{l}
J_{(0),(2)}  \\
J_{(0),(1,1)} \\
J_{(1),(1)}  \\
J_{(2),(0)} \\
J_{(1,1),(0)}  
\end{array}
\right)
=
M_{\beta}^2
\left(
\begin{array}{l}
m_{(0),(2)}  \\
m_{(0),(1,1)} \\
m_{(1),(1)}  \\
m_{(2),(0)} \\
m_{(1,1),(0)} 
\end{array}
\right),
\end{equation}
\begin{equation}
M_{\beta}^2 \mathbin{:=}
 \left(
\begin{array}{ccccc}
 1 &\frac{2\beta}{1+\beta} & \frac{2\beta(1-\beta)}{(1+\beta)(1-u'_1+u'_2)} & \frac{(1-\beta)(2+\beta-\beta^2-2u'_1+2u'_2)}{(1+\beta)(u'_1-u'_2)(-1+u'_1-u'_2)} & \frac{2\beta(2-3\beta+\beta^2)}{(1+\beta)(u'_1-u'_2)(-1+u'_1-u'_2)}  \\
 0 &1 &\frac{1-\beta}{-\beta-u'_1+u'_2} &\frac{1-\beta}{\beta+u'_1-u'_2} &\frac{-1+3\beta-2\beta^2}{(u'_1-u'_2)(-\beta-u'_1+u'_2)}  \\
 0 &0 &1 & \frac{1-\beta}{-1-u'_1+u'_2} & \frac{2(1-\beta)(-1+\beta-u'_1+u'_2)}{(-1-u'_1+u'_2)(\beta-u'_1+u'_2)} \\
 0 &0 &0 &1 &\frac{2\beta}{1+\beta}  \\
 0 &0 &0 &0 &1  
\end{array}
\right). 
\end{equation}
If we take the limit $q \rightarrow 1$ of $M_{q,t}^i$, 
then $M_{\beta}^i$ appears. 

\appendix
\section*{Appendix}

\section{Macdonald functions and Hall-Littlewood functions}\label{sec: Macdonald and HL}

In this subsection, 
we briefly review some properties of Hall-Littlewood functions and Macdonald functions following \cite[Chap.\ III, VI]{Macdonald}.

Let $\Lambda_{N} \seteq \mathbb{Q}(q,t)[x_1, \ldots, x_N]^{S_N}$ 
be the ring of symmetric polynomials of $N$ variables and
$\Lambda \seteq \varprojlim \Lambda_N$ be the ring of symmetric functions. 
The inner product $\langle-,- \rangle_{q,t}$ over $\Lambda$ is defined such that 
for power sum symmetric functions $p_{\lambda} = \prod_{k\geq 1} p_{\lambda_k}$ 
($p_n =\sum_{i\geq 1} x_i^n$), 
\begin{equation}
\langle p_{\lambda}, p_{\mu} \rangle_{q,t} 
= z_{\lambda} \prod_{k=1}^{\ell(\lambda)} \frac{1-q^{\lambda_k}}{1-t^{\lambda_k}} \delta_{\lambda, \mu}, \quad 
z_{\lambda} \seteq \prod_{i \geq 1} i^{m_i} m_i !,
\end{equation}
where $m_i=m_i (\lambda)$ is the number of entries in $\lambda$ equal to $i$. 
For a partition $\lambda$, 
Macdonald functions $P_{\lambda} \in \Lambda$ are uniquely determined 
by the following two conditions \cite{Macdonald}:
\begin{align}
 &\lambda  \neq \mu \quad \Rightarrow \quad \langle P_{\lambda}, P_{\mu} \rangle_{q,t}  = 0;  \\
 &P_{\lambda} = m_{\lambda} + \sum_{\mu < \lambda} c_{\lambda \mu} m_{\mu}.  
\end{align}
Here $m_{\lambda}$ is the monomial symmetric function and 
$<$ is the ordinary dominance partial ordering, 
which is defined as follows: 
\begin{equation}
\lambda \geq \mu \quad \overset{\mathrm{def}}{\Longleftrightarrow}  \quad 
\sum_{i=1}^k \lambda_i \geq \sum_{i=1}^k \mu_i \quad (\forall k) \quad \mathrm{and} \quad |\lambda| = |\mu|.
\end{equation}
In this paper, 
we regard power sum symmetric functions $p_n$ ($n \in \mathbb{N}$) 
as the variables of Macdonald functions, 
i.e., $P_{\lambda}= P_{\lambda}(p_n; q,t)$. 
Here $P_{\lambda}(p_n; q,t)$ is an abbreviation for $P_{\lambda}(p_1, p_2, \ldots; q,t)$. 
In this paper, we often use the symbol $P_{\lambda}(a_{-n}; q,t)$, 
which is the polynomial of bosons $a_{-n}$ obtained by replacing $p_n$ in Macdonald functions with $a_{-n}$.

Next, 
let the Hall-Littlewood function $P_{\lambda}(p_n;t)$ be given 
by $P_{\lambda}(p_n;t) \seteq P_{\lambda}(p_n;0,t)$. 
If $x_{N+1} = x_{N+2} = \cdots =0$, then 
for a partition $\lambda$ of length $\leq N$, 
the Hall-Littlewood polynomial $P_{\lambda}(p_n;t)$ with $p_n=\sum_{i=1}^N x_i^n$
is expressed by 
\begin{equation}
P_{\lambda}(p_n;t) = \frac{1}{v_{\lambda}(t)} \sum_{w \in S_n} w \left( x_1^{\lambda_1} \cdots x_N^{\lambda_N} \prod_{i<j}\frac{x_i-tx_j}{x_i-x_j} \right), 
\end{equation}
where $v_{\lambda}(t) = \prod_{i\geq 0} \prod_{k=1}^{m_i(\lambda)}\frac{1-t^k}{1-t}$. 
Note that $m_0=N-\ell(\lambda)$. 
The action of the symmetric group $S_N$ of degree $N$ is defined by 
$w(x_1^{\alpha_1} \cdots x_N^{\alpha_N})= x_{w(1)}^{\alpha_1} \cdots x_{w(N)}^{\alpha_N}$ for $w \in S_N$.

It is convenient to introduce functions $Q_{\lambda}(p_n;t)$, 
which are defined by scalar multiples of $P_{\lambda}$ as follows:  
\begin{equation}
Q_{\lambda}(p_n;t) \seteq b_{\lambda}(t) P_{\lambda}(p_n;t), 
\end{equation}
where $b_{\lambda}(t) \seteq \prod_{i\geq 1} \prod_{k=1}^{m_i} (1-t^k)$. 
They are diagonalized as 
\begin{equation}\label{eq:inner prod of HL poly}
\langle Q_{\lambda}, Q_{\mu} \rangle_{0,t} = b_{\lambda}(t) \delta_{\lambda, \mu}. 
\end{equation}
These functions $Q_{\lambda}(p_n;t)$ can be constructed  
by using Jing's operators $H_n$ and $H^{\dagger}_n$ 
\cite{J:1991:Vertex}, 
which is defined by 
\begin{align}
H(z) &\seteq \exp \left\{ \sum_{n\geq 1} \frac{1-t^n}{n} b_{-n} z^n \right\} 
       \exp \left\{ -\sum_{n\geq 1} \frac{1-t^n}{n} b_{n} z^{-n} \right\} 
\rseteq \sum_{n \in \mathbb{Z}} H_n z^{-n}, \label{eq:Jing's op}\\
H^{\dagger}(z) &\seteq \exp \left\{ -\sum_{n\geq 1} \frac{1-t^n}{n} b_{-n} z^n \right\} 
       \exp \left\{ \sum_{n\geq 1} \frac{1-t^n}{n} b_{n} z^{-n} \right\}
\rseteq \sum_{n \in \mathbb{Z}} H^{\dagger}_n z^{-n}, \label{eq:dual Jing's op}
\end{align}
where $b_n$ is the bosons realized by 
\begin{equation}
b_{-n}=p_n, \qquad b_{n}= \frac{n}{1-t^n} \frac{\partial}{\partial p_n} \quad (n>0).
\end{equation}

\begin{fact}[\cite{J:1991:Vertex}] \label{fact:Jing's operator}
Let $\ket{0}$ be the vector such that $b_n \ket{0}=0$ ($n>0$). 
Then for a partition $\lambda$, we have
\begin{eqnarray}
&H_{-\lambda_1}H_{-\lambda_2} \cdots \ket{0} &= Q_{\lambda}(b_{-n};t)\ket{0}, \\
&\bra{0} \cdots H^{\dagger}_{\lambda_2}H^{\dagger}_{\lambda_1}  &= \bra{0} Q_{\lambda}(b_{n};t). 
\end{eqnarray}
\end{fact}

Furthermore, the following specialization formula is known.

\begin{fact}[Chap.\ III, \S 4, Example 3 in \cite{Macdonald}]\label{fact:spetialization of HL poly}
Let $r$ be indeterminate. Under the specialization 
\begin{equation}
p_n = \frac{1-r^n}{1-t^n}, 
\end{equation}
The Hall-Littlewood function $Q_{\lambda} \in \Lambda$ is specialized as 
\begin{equation}
Q_{\lambda}(\frac{1-r^n}{1-t^n}; t) = t^{n(\lambda)} \prod_{i=1}^{\ell(\lambda)} (1-t^{1-i} r). 
\end{equation}
\end{fact}

\section{Definition of DIM algebra and level $N$ representation}
\label{sec:Def of DIM}

In this section, 
we recall the definition of the DIM algebra and the level $N$ representation. 
For the notations, we follow \cite{FHHSY}. 
The DIM algebra has two parameters $q$ and $t$. 
Let $g(z)$ be the formal series 
\begin{equation}
g(z)\seteq \frac{G^+(z)}{G^-(z)}, \quad 
G^{\pm}(z)\seteq (1-q^{\pm 1}z)(1-t^{\mp 1}z)(1-q^{\mp 1}t^{\pm 1}z). 
\end{equation}
Then this series satisfies $g(z)=g(z^{-1})^{-1}$.

\begin{df}
Define the algebra $\mathcal{U}$ to be the
unital associative algebra over $\mathbb{Q}(q,t)$ generated by 
 the currents 
$x^\pm(z)=\sum_{n\in \mathbb{Z}}x^\pm_n z^{-n}$,
$\psi^\pm(z)=\sum_{\pm n\in \mathbb{Z}_{\ge0}}\psi^\pm_n z^{-n}$
and the central element $\gamma^{\pm 1/2}$ 
satisfying the defining relations
\begin{align}
&\psi^\pm(z) \psi^\pm(w) = \psi^\pm(w) \psi^\pm(z),
 \qquad\qquad\qquad
 \psi^+(z)\psi^-(w) =
 \dfrac{g(\gamma^{+1} w/z)}{g(\gamma^{-1}w/z)}\psi^-(w)\psi^+(z),
\\
&\psi^+(z)x^\pm(w) = g(\gamma^{\mp 1/2}w/z)^{\mp1} x^\pm(w)\psi^+(z), \\
& \psi^-(z)x^\pm(w) = g(\gamma^{\mp 1/2}z/w)^{\pm1} x^\pm(w)\psi^-(z),
\\
&[x^+(z),x^-(w)]
 =\dfrac{(1-q)(1-1/t)}{1-q/t}
 \big( \delta(\gamma^{-1}z/w) \psi^+(\gamma^{1/2}w)-
 \delta(\gamma z/w) \psi^-(\gamma^{-1/2}w) \big),
\\
&G^{\mp}(z/w)x^\pm(z)x^\pm(w)=G^{\pm}(z/w)x^\pm(w)x^\pm(z). 
\end{align}
\end{df}

Note that $\psi^{\pm}_0$ are central elements in $\mathcal{U}$. 
Let us contain the invertible elements $(\psi^+_0)^{1/2}$ 
and $(\psi^-_0)^{1/2}$ in the definition of $\mathcal{U}$. 
Further, set $\gamma^{\perp} \seteq (\psi^+_0)^{1/2} (\psi^-_0)^{-1/2}$. 
This algebra $\mathcal{U}$ is an example of the family 
topological Hopf algebras introduced by Ding and Iohara \cite{Ding-Iohara}. 
This family is a sort of generalization 
of the Drinfeld realization of the quantum affine algebras. 
However, 
Miki introduce a deformation of the $W_{1+\infty}$ algebra 
in \cite{Miki:2007}, 
which is the quotient of the algebra $\mathcal{U}$ 
by the Serre-type relation. 
Hence we call the algebra $\mathcal{U}$ 
the Ding-Iohara-Miki algebra (DIM algebra). 
Since the algebra $\mathcal{U}$ has a lot of background, 
there are a lot of other names 
such as quantum toroidal $\mathfrak{gl}_1$ algebra 
\cite{FJMM:2015:Quantum, FJMM:2016:Finite}, 
quantum $W_{1+\infty}$ algebra \cite{Bourgine:2016Coherent}, 
elliptic Hall algebra \cite{BS:2012:I} and so on.  
This algebra has a Hopf algebra structure. 
The formulas for its coproduct are 
\begin{align}
&\Delta (\psi^\pm(z))=
 \psi^\pm (\gamma_{(2)}^{\pm 1/2}z)\otimes \psi^\pm (\gamma_{(1)}^{\mp 1/2}z),
\\
&\Delta (x^+(z))=
  x^+(z)\otimes 1+
  \psi^-(\gamma_{(1)}^{1/2}z)\otimes x^+(\gamma_{(1)}z),\\
&\Delta (x^-(z))=
  x^-(\gamma_{(2)}z)\otimes \psi^+(\gamma_{(2)}^{1/2}z)+1 \otimes x^-(z),
\end{align}
and $\Delta(\gamma^{\pm 1/2})=\gamma^{\pm 1/2} \otimes \gamma^{\pm 1/2}$, 
where $\gamma_{(1)}^{\pm 1/2} \seteq \gamma^{\pm 1/2}\otimes 1$
and $\gamma_{(2)}^{\pm 1/2} \seteq 1\otimes \gamma^{\pm 1/2}$.
Since we do not use the antipode and the counit in this thesis, 
we omit them.
The DIM algebra $\mathcal{U}$ can be realized by 
the Heisenberg algebra defined in Section \ref{sec:Review of Simplest 5D AGT}.

\begin{fact}[\cite{FHHSY, FHSSY}]\label{fact:lv. 1 rep of DIM}
The morphism $\rho_u(\cdot)$ defined as follows is a representation of the 
DIM algebra: 
\begin{align}
&\rho_u(x^+(z))=u\, \eta(z),\quad 
 \rho_u(x^-(z))=u^{-1} \xi(z),\\
&\rho_u(\psi^\pm(z))=\varphi^\pm(z), 
\quad \rho_u(\gamma^{\pm 1/2})=(t/q)^{\pm 1/4}, 
\end{align}
where
\begin{align}
&\eta(z)\seteq
\exp\Big( \sum_{n=1}^{\infty} \dfrac{1-t^{-n}}{n} z^{n} a_{-n} \Big)
\exp\Big(-\sum_{n=1}^{\infty} \dfrac{1-t^{n} }{n} z^{-n} a_n \Big),\\
&\xi(z)\seteq
\exp\Big(-\sum_{n=1}^{\infty} \dfrac{1-t^{-n}}{n}(t/q)^{n/2} z^{n}a_{-n}\Big)
\exp\Big( \sum_{n=1}^{\infty} \dfrac{1-t^{n}}{n} (t/q)^{n/2} z^{-n}a_n \Big),\\
&\varphi_{+}(z)\seteq
\exp\Big(
 -\sum_{n=1}^{\infty} \dfrac{1-t^{n}}{n} (1-t^n q^{-n})(t/q)^{-n/4} z^{-n}a_n
    \Big),
\\
&\varphi_{-}(z)\seteq
\exp\Big(
 \sum_{n=1}^{\infty} \dfrac{1-t^{-n}}{n} (1-t^n q^{-n})(t/q)^{-n/4} z^{n}a_{-n}
    \Big).
\end{align}
\end{fact}

Not that the zero mode $\eta_0$ of $\eta(z)=\sum_n \eta_n z^{-n}$
can be essentially identified with the Macdonald difference operator \cite{Macdonald, SKAO:1995:quantum}. 
By using the coproduct of $\mathcal{U}$, 
we can consider its tensor representations. 
For an $N$-tuple of parameters $\vu=(u_1,u_2,\ldots,u_N)$,
define the morphism $\rho_{\vu}^{(N)}$ by
\begin{align}
\rho_{\vu}^{(N)}\seteq
(\rho_{u_1}\otimes\rho_{u_2}\otimes\cdots \otimes\rho_{u_N})
\circ\Delta^{(N)},
\end{align}
where $\Delta^{(N)}$ is inductively defined by
$\Delta^{(1)}\seteq \mathrm{id} $, $\Delta^{(2)}\seteq \Delta$ and
$\Delta^{(N)}\seteq( \mathrm{id} \otimes \cdots 
\otimes{\rm id}\otimes \Delta)\circ \Delta^{(N-1)}$.
The representation $\rho_{\vu}^{(N)}$ is called 
the level $N$ representations. 
In Section \ref{sec:Reargument of DI alg and AGT}, 
for simplicity, 
we write the $i$-th bosons as 
\begin{equation}
a^{(i)}_n \seteq \underbrace{1 \otimes \cdots \otimes 1 \otimes a_n}_i \otimes 1 \otimes \cdots \otimes 1. 
\end{equation}
The generator $X^{(1)}(z)$ is defined by 
\begin{equation}
\Xo(z)\seteq \rho^{(N)}_{\vu} (x^{+}(z)). 
\end{equation}

The $\rho_{u}$ is also called the level $(1,0)$ representation 
or the horizontal representation in 
\cite{awata2012quantum, AKMMMMOZ:2016Explicit, AKMMMOZ:2016:Toric, AKMMMOZ:2016:Anomaly} 
in order to distinguish another representation of the DIM algebra, 
which is called the level $(0,1)$ representation or 
the vertical representation. 
To define the level $(0,1)$ representation, 
we introduce some notations. 
For a partition $\lambda=(\lambda_1, \lambda_2, \ldots )$ and a number $i \in \mathbb{Z}_{\geq 0}$, 
we set 
\begin{align}
&A^+_{\lambda,i}\seteq (1-t)
\prod_{j=1}^{i-1}
{(1-q^{\lambda_i-\lambda_j}t^{-i+j+1})
(1-q^{\lambda_i-\lambda_j+1}t^{-i+j-1})\over 
(1-q^{\lambda_i-\lambda_j}t^{-i+j})
(1-q^{\lambda_i-\lambda_j+1}t^{-i+j})},\label{eq:A+}\\
&A^-_{\lambda,i}\seteq (1-t^{-1})
{1-q^{\lambda_{i+1}-\lambda_i} \over 1-q^{\lambda_{i+1}-\lambda_i+1} t^{-1}}
\prod_{j=i+1}^{\infty}
{(1-q^{\lambda_j-\lambda_i+1}t^{-j+i-1})
(1-q^{\lambda_{j+1}-\lambda_i}t^{-j+i})\over 
(1-q^{\lambda_{j+1}-\lambda_i+1}t^{-j+i-1})
(1-q^{\lambda_j-\lambda_i}t^{-j+i})} ,\label{eq:A-}\\
&
B^+_\lambda(z)\seteq 
{1-q^{\lambda_{1}-1} t z \over 1-q^{\lambda_{1}} z}
\prod_{i=1}^{\infty}
{(1-q^{\lambda_i}t^{-i}z)
(1-q^{\lambda_{i+1}-1}t^{-i+1}z)\over 
(1-q^{\lambda_{i+1}}t^{-i}z)
(1-q^{\lambda_i-1}t^{-i+1}z)},\label{eq:B+}\\
&
B^-_\lambda(z)\seteq 
{1-q^{-\lambda_{1}+1} t^{-1} z \over 1-q^{-\lambda_{1}} z}
\prod_{i=1}^{\infty}
{(1-q^{-\lambda_i}t^{i}z)
(1-q^{-\lambda_{i+1}+1}t^{i-1}z)\over 
(1-q^{-\lambda_{i+1}}t^{i}z)
(1-q^{-\lambda_i+1}t^{i-1}z)},  \label{eq:B-}  \\
&\lambda+{\bf 1}_i \seteq 
(\lambda_1,\lambda_2,\ldots,\lambda_{i-1},\lambda_i+1,\lambda_{i+1},\ldots), \\
&\lambda-{\bf 1}_i \seteq 
(\lambda_1,\lambda_2,\ldots,\lambda_{i-1},\lambda_i-1,\lambda_{i+1},\ldots). 
\end{align}
Here, $B^+_\lambda(z)$ and $B^-_\lambda(z)$ are 
considered as elements in $\mathbb{Q}(q,t)[[z]]$.

\begin{fact}[\cite{FFJMM:2011:semiinfinite, FT:2011:Equivariant}]
Let $u$ be an indeterminate parameter and 
$\mathcal{F}$ be the Fock module generated 
by the highest weight vector $\ket{0}$.  
The morphism 
$\rho^{(0,1)}_u :\mathcal{U} \rightarrow \mathrm{End}(\mathcal{F}) $
defined as follows gives a representation of the DIM algebra: 
\begin{align}
&
\rho^{(0,1)}_u( x^+(z) ) \ket{\lambda}
=\sum_{i=1}^{\ell(\lambda)+1} A^+_{\lambda,i}\,
\delta(q^{\lambda_i}t^{-i+1}u/z) 
\ket{\lambda+{\bf 1}_i},\\
&
\rho^{(0,1)}_u( x^-(z) ) \ket{\lambda}
=q^{1/2}t^{-1/2}
\sum_{i=1}^{\ell(\lambda)} A^-_{\lambda,i}\,
\delta(q^{\lambda_i-1}t^{-i+1}u/z)
\ket{\lambda-{\bf 1}_i},\\
&
\rho^{(0,1)}_u( \psi^+(z) )\ket{\lambda}
=q^{1/2}t^{-1/2}
B^+_\lambda(u/z)\ket{\lambda},\\
&
\rho^{(0,1)}_u( \psi^-(z) )\ket{\lambda}
=q^{-1/2}t^{1/2}
\,B^-_\lambda(z/u)\ket{\lambda}
\end{align}
and $\rho^{(0,1)}_u( \gamma^{1/2}) =1$, 
where $\ket{\lambda}$ denotes the ordinary Macdonald function 
$P_\lambda(a_{-n};q,t)\ket{0}$ 
associated with the partition $\lambda$. 
\end{fact}

In this thesis, 
the vectors $\ket{\lambda}$ 
are realized by the Macdonald functions $P_\lambda(a_{-n};q,t)$ 
along $\cite{awata2012quantum}$. 
However, 
they can also be regarded as 
the abstract vectors labeled by Young diagrams. 
Note that the factors (\ref{eq:A+})-(\ref{eq:B-}) can be written 
by the contribution from the edges of the Young diagram $\lambda$, 
i.e., the positions which we can add a box in or remove it from. 
The tensor representation of $\rho^{(0,1)}$ 
is also called the rank $N$ representation, 
which is given in \cite{Bourgine:2016Coherent} 
in terms of the edge contribution. 
This representation is connected to the level $(1,0)$ representation 
under the change of basis and the automorphism of the DIM algebra, 
that is defined as follows.  
This connection is called the spectral duality
\cite[Section 5]{AKMMMMOZ:2016Explicit}.

\begin{fact}[\cite{Miki:2007}]
There exists an automorphism 
$\mathcal{S}:\mathcal{U} \rightarrow \mathcal{U}$ 
such that 
\begin{equation}
x^+_0 \mapsto \psi^-_{-1}, \quad 
\psi^-_{-1} \mapsto x^-_0, \quad 
x^-_0 \mapsto \psi^+_{1}, \quad
\psi^+_{1} \mapsto x^+_0, 
\end{equation}
$\gamma \mapsto \gamma^{\perp} $ and $\gamma^{\perp} \mapsto \gamma^{-1}$. 
\end{fact}

Although the algebra in \cite{Miki:2007} slightly differs 
from the algebra $\mathcal{U}$ in the Serre-type relation, 
this fact holds.  
This automorphism is of order four. 
By using this automorphism, 
we can check the correspondence 
between two representations of the DIM algebra. 
In Section \ref{sec:realization of rank N rep}, 
we briefly explain the spectral duality and 
check it 
with respect to the generators $x^{+}_{\pm 1}$ and $\psi^+_k$. 

\section{Proofs and checks in Section \ref{sec:Crystallization}}
\label{sec:proofs and chekc of crystal}

\subsection{Other proofs of Lemma \ref{lem:exlicit form of inv. shapovalov}}
\label{seq:Another proof of Lemma }

In this subsection, 
let us explain other proofs of Lemma \ref{lem:exlicit form of inv. shapovalov} 
by the method of contour integrals.

The generating function of elementary symmetric functions and Jing's operator 
makes the equation
\begin{align}
& (-1)^s \left\langle e_{s}(-p_n), Q_{\lambda}(p_n;t) \right\rangle_{0,t}  \\
&= \oint \frac{dz}{\twopii z}\frac{dw}{\twopii w} 
    \prod_{i=1}^{l}\left( \frac{1}{1-zw_i}\right) 
    \prod_{1\leq i <j \leq \ell(\lambda)} \left( \frac{w_i-w_j}{w_i-tw_j} \right) 
    z^{-s}w^{-\lambda} \nonumber \\
 &\rseteq F_{\lambda_1, \lambda_2,\ldots , \lambda_{l}}, \nonumber
\end{align}
where we put $s=|\lambda|$, $l=\ell(\lambda)$, 
and $|1/z|>|w_1|>\cdots > |w_l|$. 
It suffices to show that $F_{\lambda_1, \lambda_2,\ldots , \lambda_{l}} = t^{n(\lambda)}$, 
which is proved by a recursive relation of $F$ as follows. 
The contour integral $\oint \frac{dw_1}{\twopii w_1}$ surrounding origin is represented 
as that surrounding $\infty$. 
Since $\lambda_1>0$, the residue of $w_1$ at $w_1=\infty$ is $0$. 
Hence, 
the only residue at $w_1=\frac{1}{z}$ is left, 
and it is
\begin{equation}
F_{\lambda_1, \lambda_2,\ldots , \lambda_{l}} 
 = \oint \frac{dz}{\twopii z} z^{-\lambda_2 \cdots -\lambda_l} 
   \prod_{i=2}^l \frac{dw_i}{\twopii w_i}w^{-\lambda_i} \prod_{i=2}^l \left( \frac{1}{1-tw_iz} \right)
   \prod_{2 \leq i <j \leq l} \left( \frac{w_i-w_j}{w_i-tw_j} \right). 
\end{equation}
By change of variable $t z \mapsto z$, 
\begin{equation}
F_{\lambda_1, \lambda_2,\ldots , \lambda_{l}} 
= \prod_{i=2}^{l} t^{\lambda_i} \cdot F_{\lambda_2, \ldots , \lambda_l}
\end{equation}
Therefore 
\begin{equation}
F_{\lambda_1, \lambda_2,\ldots , \lambda_{l}} 
= \left( \prod_{i=2}^{l} t^{\lambda_i}\right) \left(\prod_{i=3}^{l} t^{\lambda_i}\right) \cdots  t^{\lambda_l} 
\oint\frac{dz}{\twopii z} 
= t^{n(\lambda)}. 
\end{equation}
Thus the Lemma \ref{lem:exlicit form of inv. shapovalov} is proved.

Although it is slightly hard, 
one can also prove this lemma by reversing the order of integration, 
i.e., first perform over a variable $w_l$ surrounding origin.  
Indeed $w_l$ has the pole only at $w_l=0$, 
and its residue satisfies 
\begin{equation}
F_{\lambda_1, \ldots, \lambda_l}
=\sum_{\substack{\alpha_0, \alpha_1, \ldots , \alpha_{l-1} \geq 0  \\ \alpha_0+\cdots \alpha_{l-1}=\lambda_l }} 
 \left( \prod_{i=1}^{l-1} \mathcal{A}_{\alpha_i} \right) F_{\lambda_1+\alpha_1, \ldots , \lambda_{l-1}+\alpha_{l-1}}, 
\end{equation}
where for $n > 0$,  $\mathcal{A}_n \seteq (t-1)t^{n-1}$ and 
$\mathcal{A}_0 \seteq 1$. 
By the assumption that $F_{\beta} = t^{n(\beta)}$  for $\beta =(\beta_1, \beta_2, \ldots )$ 
with $\ell(\beta)=l-1$, 
we inductively  get
\begin{equation}
F_{\lambda_1, \ldots, \lambda_l}
=t^{n((\lambda_1, \ldots, \lambda_{l-1}))} 
\sum_{0 \leq k \leq \lambda_l} 
\sum_{\substack{\alpha_1, \ldots , \alpha_{l-1} \geq 0  
      \\ \alpha_1 +\cdots + \alpha_{l-1}= k }} 
 \left( \prod_{i=1}^{l-1} \mathcal{A}_{\alpha_i} \right) t^{n(\alpha)}. 
\end{equation}
By virtue of the equation 
\begin{equation}
\sum_{\substack{\alpha_1, \ldots , \alpha_{l} \geq 0  
      \\ \alpha_1 +\cdots + \alpha_{l}= k }} 
 \left( \prod_{i=1}^{l} \mathcal{A}_{\alpha_i} \right)
 t^{n(\alpha)}
= 
\left\{
\begin{array}{ll}
t^{lk}-t^{l(k-1)} ,  &\quad k\geq 1, \\
1 , &\quad k=0, 
\end{array}
\right.
\end{equation}
which is also proved by induction with respect to $l$, 
it can be seen that $F_{\lambda_1, \ldots , \lambda_l} = t^{n(\lambda)}$.  

\subsection{Explicit form of $\left\langle Q_{(s)}(-p_n), Q_{\lambda}(p_n;t) \right\rangle_{0,t}$}
\label{sec:explicit form of <Q,Q>} 

The formula for $S^{\lambda, (1^s)}$ is given 
in the Lemma \ref{lem:exlicit form of inv. shapovalov} and the last subsection. 
We also have an explicit form of $S^{\lambda, (s)}$ and $S_{\lambda, (s)}$. 
By the Proposition \ref{prop:Shapovalov in terms of HL poly}, 
it suffices to give the explicit form of 
$\left\langle Q_{(s)}(-p_n), Q_{\lambda}(p_n;t) \right\rangle_{0,t}$.

\begin{prop}
\begin{equation}
\left\langle Q_{(s)}(-p_n;t), Q_{\lambda}(p_n;t) \right\rangle_{0,t}
= t^{|\lambda|+n(\lambda)} \prod_{k=1}^{\ell(\lambda)} (1-t^{-k}). 
\end{equation}
\end{prop}

\Proof
The proof is similar to the previous subsection. 
Set 
\begin{equation}
G^k_{\lambda_1, \lambda_2, \ldots , \lambda_l}
\seteq \oint \frac{dz}{\twopii z} \frac{dw}{\twopii w} \prod_{i=1}^l \left( \frac{z-w_i}{t^{-k}z-tw_i} \right) 
\prod_{1\leq i<j \leq \ell(\lambda)} \left( \frac{w_i-w_j}{w_i-tw_j} \right)z^{|\lambda|} w^{-\lambda}. 
\end{equation}
Then 
$\left\langle Q_{(s)}(-p_n), Q_{\lambda}(p_n;t) \right\rangle_{0,t} 
= G^0_{\lambda_1, \lambda_2, \ldots , \lambda_l}$ 
by Jing's operator. 
Integration of $w_l$ around $\infty$ makes recursive relation
\begin{equation}
G^k_{\lambda_1, \lambda_2, \ldots , \lambda_l}
= t^{\lambda_1(k+1)-\ell(\lambda)}(t^{k+1}-1) G^{k+1}_{\lambda_2, \ldots , \lambda_l}, 
\end{equation}
and leads this Proposition. 
\qed

For general partitions $\lambda$ and $\mu$, 
we can get the integral representation of 
$\left\langle Q_{\lambda}(-p_n), Q_{\mu}(p_n;t) \right\rangle_{0,t}$. 
However, 
it is very hard to give their explicit formula. 

\subsection{Check of (\ref{eq:N=1 conjecture})}\label{sec:check of N=1 conjecture}

The integral formula (\ref{eq:N=1 conjecture}) can be checked
by the similar way to subsections \ref{seq:Another proof of Lemma } and \ref{sec:explicit form of <Q,Q>}. 

Let us set 
\begin{align}
\mathfrak{F}_{\lambda_1, \ldots, \lambda_l} (u) 
& \seteq 
\oint \prod_{i=1}^l \frac{dw_i}{\twopii w_i} w_i^{\lambda_i} 
\prod_{i=1}^l \left( \frac{w_i}{w_i-ux} \right) \prod_{1 \leq j< i \leq l} \frac{w_i-w_j}{w_i-tw_j},   \\
\mathfrak{G}_{\mu_1, \ldots, \mu_m} (u) 
& \seteq 
\oint \prod_{i=1}^m\frac{dz_i}{\twopii z_i} z_i^{-\mu_i} 
\prod_{1 \leq i< j\leq m} \left( \frac{z_i-z_j}{z_i-tz_j} \right)
\prod_{1 \leq i \leq m} \left( \frac{x-(t/v)z_i}{x-(t/u)z_i} \right). \nonumber
\end{align}
Then 
$\mathfrak{F}_{\lambda_1, \ldots, \lambda_l} (u) 
= (-u)^{-|\lambda|} t^{n(\lambda)} \tbra{\tK_{\lambda}}\tPhi(z) \tket{\tK_{\emptyset}}$, 
$\mathfrak{G}_{\mu_1, \ldots, \mu_l} (u) 
= (-u)^{-|\mu|} t^{n(\mu)+|\mu|} \tbra{\tK_{\emptyset}}\tPhi(z) \tket{\tK_{\mu}}$. 
The integration of $w_1$ around $0$ give the relation
\begin{equation}
\mathfrak{F}_{\lambda_1, \ldots, \lambda_l} (u) 
= (ux)^{\lambda_1} \mathfrak{F}_{\lambda_2, \ldots, \lambda_l} (tu). 
\end{equation}
On the other hand, 
the integration of $z_1$ around $\infty$ makes 
\begin{equation}
\mathfrak{G}_{\mu_1, \ldots, \mu_m} (u) 
= \left( 1-(u/v) \right) (ux/v)^{\mu_1} \mathfrak{G}_{\mu_2 \ldots, \mu_m} (u/t). 
\end{equation}
Thus 
\begin{align}
\mathfrak{F}_{\lambda_1, \ldots, \lambda_l} (u)  
&= (ux)^{\lambda_1} (utx)^{\lambda_2} \cdots (ut^{l-1}x)^{\lambda_l},  \\
\mathfrak{G}_{\mu_1, \ldots, \mu_m} (u)  
&= \left(1-\frac{u}{v}\right) \left(1-\frac{u}{vt}\right) \cdots \left(1-\frac{u}{vt^{m-1}}\right)
\left( \frac{u}{xt} \right)^{-\mu_1} \left( \frac{u}{xt^2} \right)^{-\mu_2} \cdots \left( \frac{u}{xt^m} \right)^{-\mu_m}.    \nonumber
\end{align}
These agree with the right hand side of (\ref{eq:N=1 conjecture}). 

\subsection{Comparison of formulas  (\ref{eq:expansion by AFLT}) and (\ref{eq:expansion formula by PBW})}
\label{Comparison of two formula}

In this subsection, we compare two formulas (\ref{eq:expansion by AFLT}) 
and (\ref{eq:expansion formula by PBW}) 
which are obtained by the other basis.

Comparing the coefficients of $\frac{z_1}{z_2}$, we have the equation 
\begin{equation}\label{eq:comparison of two formula}
\sum_{|\vl |=n} 
\prod_{i,j=1}^{2} 
\frac{\tN_{\emptyset , \lambda^{(j)}} (w_i/v_j) }{\tN_{\lambda^{(i)}, \lambda^{(j)}} (v_i/v_j)}
\overset{?}{=}
\sum_{|\lambda|=n}
\frac{ \prod_{k=1}^{\ell(\lambda)} \left(1- t^{k-1} \frac{w_1w_2}{v_1v_2} \right) }{t^{2n(\lambda)} b_{\lambda}(t^{-1}) }. 
\end{equation}
Note that the left hand side is the summation with respect to pairs of partitions $\vl=(\lo, \lt)$ 
and the right hand side is the summation with respect to single partitions $\lambda$. 
The right hand side depends only on the ratio $\frac{w_1 w_2}{v_1 v_2}$ 
though the left hand side doesn't look that way.

For a single partition $\lambda$, 
let us define $\left\langle \lambda \right\rangle$ to be 
the set of all pairs of partitions $(\lo, \lt)$ such that 
a permutation of the sequence $(\lo_1, \lo_2, \ldots, \lt_1, \lt_2, \ldots)$ 
coincides with $\lambda$. 
For example, if $\lambda=(2,1,1)$, 
\begin{equation}
\left\langle \lambda \right\rangle 
= \{ ((2,1,1),\emptyset), ((2,1),(1)), ((2),(1,1)), ((1,1),(2)),((1),(2,1)), (\emptyset,(2,1,1)) \}. 
\end{equation}
Then we obtain a strange factorization formula  
with respect to the partial summation of left hand side in (\ref{eq:comparison of two formula})
\begin{equation}\label{eq:strange facorization formula}
\sum_{\vl \in \langle \lambda \rangle} 
\prod_{i,j=1}^{2} 
\frac{\tN_{\emptyset , \lambda^{(j)}} (w_i/v_j) }{\tN_{\lambda^{(i)}, \lambda^{(j)}} (v_i/v_j)}
\overset{?}{=}
\frac{ \prod_{k=1}^{\ell(\lambda)} \left(1- t^{k-1} \frac{w_1w_2}{v_1v_2} \right) }{t^{2n(\lambda)} b_{\lambda}(t^{-1}) }
\left(\frac{w_1w_2}{v_1v_2} \right)^{|\lambda|-\ell(\lambda)} t^{2 n(\lambda)-I_{\lambda}}, 
\end{equation}
where 
$I_{\lambda} \seteq \sum_{s\in \check{\lambda}}( L_{\emptyset}(s) -L_{\lambda}(s) )
= \sum_{(i,j)\in \check{\lambda}} \lambda'_j$ 
and $\check{\lambda}$ is introduced 
in Definition \ref{def:crystal Nek factor}. 
If we prove that left hand side of (\ref{eq:strange facorization formula}) 
depends only on $\frac{w_1w_2}{v_1v_2}$, 
(\ref{eq:strange facorization formula}) is easily seen 
by checking the case of $w_2=v_2$. 
This equation almost reproduces each term 
of the right hand side in (\ref{eq:comparison of two formula}). 
Hence (\ref{eq:comparison of two formula}) may be proved by this equation. 
If (\ref{eq:comparison of two formula}) holds, 
the AGT conjecture at $q \rightarrow 0$ with the help of the AFLT basis 
(\ref{eq:expansion by AFLT}) is completely proved. 

\begin{landscape}
{\footnotesize 

\section{Examples of R-matrix}
\label{sec:Ex of R-matrix}

Examples of the generalized Macdonald functions at level 3 in the $N=3$ case: 
\begin{equation}
\mathcal{A}=
\left(
\begin{array}{ccccccccc}
 1 & 0 & -\frac{q (q+1) (q-t) (t-1) u_3}{\sqrt{\frac{q}{t}} t (q t-1) (u_2-q
   u_3)} & \frac{(q+1) (t-1) (t-q) u_3 \left(t u_2-q^2 u_3\right)}{t (q t-1)
   (q u_3-u_1) (q u_3-u_2)} & -\frac{(q-t) u_3 \left(t u_3 q^3-t u_2 q^2+t
   u_3 q^2-u_3 q^2-t^2 u_3 q+t u_2\right)}{q t (q t-1) (u_2-u_3) (q
   u_3-u_2)} & -\frac{(q-1) (q+1) (q-t) (t-1) (t+1) u_3}{(q t-1)^2 (q
   u_3-u_2)} \\
 0 & 1 & -\frac{(q-t) u_3}{\sqrt{\frac{q}{t}} t (t u_2-u_3)} & \frac{(q-t) u_3
   \left(q u_3-t^2 u_2\right)}{q t (t u_1-u_3) (t u_2-u_3)} & \frac{(q-t)
   u_3}{q (t u_2-u_3)} & \frac{(q-t) u_3 \left(-q u_2 t^3+u_3 t^2+q u_2
   t+q^2 u_3 t-q u_3 t-q u_3\right)}{q t (q t-1) (u_2-u_3) (t u_2-u_3)} \\
 0 & 0 & 1 & -\frac{(q-t) \sqrt{\frac{q}{t}} u_2}{q (u_1-u_2)} & -\frac{(q-t)
   u_3}{\sqrt{\frac{q}{t}} t (q u_2-u_3)} & -\frac{(q-1) (q-t) (t+1)
   u_3}{\sqrt{\frac{q}{t}} (q t-1) (u_2-t u_3)} \\
 0 & 0 & 0 & 1 & 0 & 0 \\
 0 & 0 & 0 & 0 & 1 & 0 \\
 0 & 0 & 0 & 0 & 0 & 1 \\
 0 & 0 & 0 & 0 & 0 & 0 \\
 0 & 0 & 0 & 0 & 0 & 0 \\
 0 & 0 & 0 & 0 & 0 & 0 \\
\end{array}
\right.
\end{equation}
\begin{equation}
\left.
\begin{array}{ccccccccc}
  -\frac{(q+1) (q-t)^2 (t-1) u_3^2 \left(q^2 u_3-t
   u_2\right)}{\sqrt{\frac{q}{t}} t^2 (q t-1) (u_2-u_3) (u_1-q u_3) (u_2-q
   u_3)} & -\frac{(q-t) u_3 (q u_3-t u_2) \left(q^2 u_3-t u_2\right) \left(t
   u_3 q^3-t u_1 q^2+t u_3 q^2-u_3 q^2-t^2 u_3 q+t u_1\right)}{q^2 t^2 (q
   t-1) (u_1-u_3) (u_3-u_2) (q u_3-u_1) (q u_3-u_2)} & \frac{(q-1) (q+1)
   (q-t) (t-1) (t+1) u_3 (q u_3-t u_2) \left(q^2 u_3-t u_2\right)}{q t (q
   t-1)^2 (u_2-u_3) (q u_3-u_1) (q u_3-u_2)} \\
   -\frac{(q-t)^2 \sqrt{\frac{q}{t}} u_3^2 \left(q u_3-t^2 u_2\right)}{q^2 t (t
   u_1-u_3) (u_2-u_3) (t u_2-u_3)} & -\frac{(q-t) u_3 (q u_3-t u_2)
   \left(q u_3-t^2 u_2\right)}{q^2 t (t u_1-u_3) (t u_2-u_3) (u_3-u_2)} &
   -\frac{(q-t) u_3 (q u_3-t u_2) \left(q u_3-t^2 u_2\right) \left(-q u_1
   t^3+u_3 t^2+q u_1 t+q^2 u_3 t-q u_3 t-q u_3\right)}{q^2 t^2 (q t-1)
   (u_1-u_3) (t u_1-u_3) (t u_2-u_3) (u_3-u_2)} \\
 a_{37} & a_{38} & a_{39} \\
 -\frac{(q-t) u_3}{\sqrt{\frac{q}{t}} t (u_2-u_3)} &
   -\frac{(q-t) u_3 (q u_3-t u_2)}{q t (q u_1-u_3) (u_3-u_2)} &
   -\frac{(q-1) (q-t) (t+1) u_3 (q u_3-t u_2)}{q (q t-1) (u_2-u_3) (t
   u_3-u_1)} \\
 -\frac{q (q+1) (q-t) (t-1) u_2}{\sqrt{\frac{q}{t}} t (q
   t-1) (u_1-q u_2)} & -\frac{(q-t) u_2 \left(t u_2 q^3-t u_1 q^2+t u_2
   q^2-u_2 q^2-t^2 u_2 q+t u_1\right)}{q t (q t-1) (u_1-u_2) (q u_2-u_1)} &
   -\frac{(q-1) (q+1) (q-t) (t-1) (t+1) u_2}{(q t-1)^2 (q u_2-u_1)} \\
 -\frac{(q-t) u_2}{\sqrt{\frac{q}{t}} t (t u_1-u_2)} &
   \frac{(q-t) u_2}{q (t u_1-u_2)} & \frac{(q-t) u_2 \left(-q u_1 t^3+u_2
   t^2+q u_1 t+q^2 u_2 t-q u_2 t-q u_2\right)}{q t (q t-1) (u_1-u_2) (t
   u_1-u_2)} \\
 1 & -\frac{(q-t) \sqrt{\frac{q}{t}} u_2}{q (q
   u_1-u_2)} & -\frac{(q-1) (q-t) \sqrt{\frac{q}{t}} t (t+1) u_2}{q (q t-1)
   (u_1-t u_2)} \\
 0 & 1 & 0 \\
 0 & 0 & 1 \\
\end{array}
\right),
\end{equation}
\begin{align}
&a_{37}=\frac{u_3 (q-t) \left(q^2 u_2 ((t+1) u_2 u_3-u_1 (u_2+u_3))+q \left(t^2
   (-u_1) u_2 u_3+t \left(u_1 \left(2 u_2^2+2 u_3 u_2+u_3^2\right)-u_2
   \left(u_2^2+2 u_3 u_2+2 u_3^2\right)\right)+u_2^2 u_3\right)+t u_2 u_3
   (t (u_2+u_3)-(t+1) u_1)\right)}{q t (u_1-u_2) (u_1-u_3) (q u_2-u_3) (t
   u_3-u_2)},\\
&a_{38}=-\frac{(q-t)^2 u_2 u_3 \left(u_2 u_3 t^2-q
   u_2^2 t-q u_3^2 t+q u_1 u_2 t-u_1 u_2 t+q u_1 u_3 t-u_1 u_3 t-q u_2
   u_3 t+u_2 u_3 t+q u_2 u_3\right)}{q \sqrt{\frac{q}{t}} t^2 (u_1-u_2)
   (u_1-u_3) (q u_2-u_3) (u_2-t u_3)},\\
&a_{39}=-\frac{(q-1) (q-t)^2 (t+1) u_2 u_3
   \left(u_2 u_3 q^2-u_2^2 q-u_3^2 q-t u_1 u_2 q+u_1 u_2 q-t u_1 u_3
   q+u_1 u_3 q+t u_2 u_3 q-u_2 u_3 q+t u_2 u_3\right)}{q
   \sqrt{\frac{q}{t}} t (q t-1) (u_1-u_2) (u_1-u_3) (q u_2-u_3) (u_2-t
   u_3)}.
\end{align}

The representation matrix of $\mathcal{R}$ in the basis of the generalized Macdonald functions at level 2:
\begin{align}
&\left(
\begin{array}{ccccc}
 -\frac{(q-Q t) \left(q^2-Q t\right)}{q (q-Q) (Q-1) t} & 0 & \frac{Q (q-t)
   \sqrt{\frac{q}{t}} (Q t-1)}{q (q Q-1) (Q-t)} & \\
 0 & \frac{(q-Q t) \left(q-Q t^2\right)}{q (Q-1) t (Q t-1)} & -\frac{(q-1) (q-Q) Q
   (q-t) \sqrt{\frac{q}{t}} t (t+1)}{q (q Q-1) (Q-t) (q t-1)}  \\
 \frac{(q+1) (q-t) (t-1) (q-Q t) \left(q^2-Q t\right)}{(q-Q)^2 (Q-1)
   \sqrt{\frac{q}{t}} t^2 (q t-1)} & \frac{(q-t) (q-Q t) \left(q-Q t^2\right)}{q
   (Q-1) \sqrt{\frac{q}{t}} t^2 (Q t-1)^2} &v_{33}& \\
 -\frac{(q-t) (q-Q t) \left(q^2-Q t\right) \left(Q t q^3-q^2-t^2 q-Q t q+t
   q+t\right)}{q^2 (q-Q) (Q-1)^2 (q Q-1) t^2 (q t-1)} & -\frac{(q-t) (q-Q t)
   \left(q-Q t^2\right)}{q (Q-1) (q Q-1) t^2 (Q t-1)} & v_{43} \\
 -\frac{(q-1) (q+1) (q-t) (t-1) (t+1) (q-Q t) \left(q^2-Q t\right)}{q (q-Q) (Q-1)
   t (t-Q) (q t-1)^2} & -\frac{(q-t) (q-Q t) \left(q-Q t^2\right) \left(-q t^3-q
   t^2+q Q t^2+t^2+q^2 t-q Q\right)}{q^2 (Q-1)^2 t^2 (t-Q) (q t-1) (Q t-1)} &
  v_{53} \\
\end{array}
\right. \\
& \qquad
\left.
\begin{array}{ccccc}
\frac{Q (q-t) \left(Q t q^3-Q
   q^2+Q t q^2-t q^2-Q t^2 q+t\right)}{(q Q-t) \left(q^2 Q-t\right) (q t-1)} &
   -\frac{(Q-1) Q (q-t) t}{(q Q-t) \left(q Q-t^2\right)} \\
 -\frac{(q-1) q
   (q+1) (Q-1) Q (q-t) (t-1) t (t+1)}{(q Q-t) \left(q^2 Q-t\right) (q t-1)^2} &
   \frac{Q (q-t) \left(-q t^3+Q t^2+q t+q^2 Q t-q Q t-q Q\right)}{(q Q-t) (q t-1)
   \left(q Q-t^2\right)} \\
 v_{34} & v_{35} \\
   v_{44}& -\frac{Q
   (q-t)^2 \left(Q q^2+Q t q^2-t^2 q-Q t q+t q-t^2\right)}{q (q Q-1) (q Q-t) t
   \left(q Q-t^2\right)} \\
 -\frac{(q-1) (q+1) Q (q-t)^2 (t-1) (t+1) \left(Q q^2+Q q-Q t q+t
   q-t^2-t\right)}{(Q-t) (q Q-t) \left(q^2 Q-t\right) (q t-1)^2} & 
v_{55} \\
\end{array}
\right),
\end{align}
\begin{align}
&\textstyle v_{33}= \frac{Q q^4+Q t q^4+Q^2 t^3 q^3+Q^3
   t^2 q^3-4 Q t^2 q^3+Q q^3-Q^3 t q^3-4 Q^2 t q^3-3 Q t q^3+t q^3-4 Q^3 t^3 q^2+Q
   t^3 q^2+3 Q^3 t^2 q^2}{q (q-Q) (q Q-1) t (t-Q) (Q t-1)} \\
&\textstyle \qquad +\frac{12 Q^2 t^2 q^2+3 Q t^2 q^2+Q^3 t q^2-4 Q t q^2+Q^3 t^4
   q+Q^4 t^3 q-3 Q^3 t^3 q-4 Q^2 t^3 q-Q t^3 q-4 Q^3 t^2 q+Q t^2 q+Q^2 t q+Q^3
   t^4+Q^3 t^3}{q (q-Q) (q Q-1) t (t-Q) (Q t-1)}, \\
&\textstyle v_{43}=\frac{(q-t)
   \sqrt{\frac{q}{t}} \left(-Q q^3+Q^3 t q^3+Q^2 t q^3-Q t q^3-Q^3 t^2 q^2-2 Q^2
   t^2 q^2+2 Q t^2 q^2-Q^3 t q^2+3 Q t q^2-t q^2+2 Q^3 t^2 q-2 Q^2 t^2 q-Q t^2
   q-Q^2 t q+2 Q t q+Q^2 t^3-Q t^2\right)}{q^2 (Q-1) (q Q-1)^2 (Q-t) t},\\
&\textstyle v_{53}=-\frac{(q-1) (t+1) (q-t) \sqrt{\frac{q}{t}} \left(q^3 Q+q^2 \left(Q^3 (t-1)+Q^2
   \left(t^2-1\right)+Q \left(-2 t^2-3 t+1\right)+t\right)+q Q t \left(Q^2 (1-2
   t)+2 Q (t+1)+t^2+t-2\right)-Q^2 t^3\right)}{q^2 (Q-1) (q Q-1) (q t-1) (Q-t)^2},\\
&\textstyle v_{44}=\frac{q^7 Q^3 t-q^6 Q^2 \left(2 Q t^2-Q t+t+1\right)+q^5 Q t \left(Q^3 t^2-Q^2
   \left(t^2+t+2\right)+Q \left(t^2-3 t+5\right)+2 t-1\right)-q^4 Q t \left(Q^3
   t+Q^2 \left(3 t^2-7 t+2\right)+Q \left(-13 t^2+11 t-5\right)\right)}
{q (Q-1) t (q Q-1) (q t-1) (q Q-t) \left(q^2Q-t\right)} \\ 
&\textstyle \qquad +\frac{-q^4 Q t(6 t^2-4t+1)+q^3 t^2 \left(Q^3 \left(t^2-4 t+6\right)+Q^2 \left(-5 t^2+11
   t-13\right)+Q \left(2 t^2-7 t+3\right)+t\right)
+q Q t^3 \left(Qt^2+(Q-1) t+2\right)-Q t^4 +q^2 t^2 \left(Q^3 (t-2) t+Q^2
   \left(-5 t^2+3 t-1\right)+Q \left(2 t^2+t+1\right)-1\right)}
{q (Q-1) t (q Q-1) (q t-1) (q Q-t) \left(q^2Q-t\right)}, \\
&\textstyle v_{34}=-\frac{q (q+1) Q (q-t) (t-1)
   \left(-Q q^3+Q^2 t q^3-2 Q^2 t^2 q^2+Q t^2 q^2+Q^2 t q^2+2 Q t q^2-2 t q^2+Q^3
   t^2 q-3 Q^2 t^2 q+t^2 q-2 Q^2 t q+2 Q t q+t q+Q^2 t^3+Q^2 t^2-Q
   t^2-t^2\right)}{(q-Q) (q Q-t) \left(q^2 Q-t\right) \sqrt{\frac{q}{t}} t (q t-1)
   (Q t-1)},\\
&\textstyle v_{35}=\frac{Q (q-t) \left(Q q^3-Q^2 q^2+2 Q^2 t^2 q^2-2 Q t^2 q^2-t^2
   q^2-Q^2 t q^2-2 Q t q^2+2 t q^2-Q^2 t^3 q+Q t^3 q+t^3 q-Q^3 t^2 q+3 Q^2 t^2
   q-t^2 q+2 Q^2 t q-Q t q-Q^2 t^3\right)}{(q-Q) (q Q-t) \sqrt{\frac{q}{t}} t (Q
   t-1) \left(q Q-t^2\right)},\\
&\textstyle v_{55}=\frac{q^5 Q^2 t+q^4 Q \left(Q^2 \left(2 t^3+2 t^2-t-1\right)+Q \left(-5 t^3-5
   t^2+t\right)+t^2 (t+1)\right)+q^3 t \left(Q^4 t^2+Q^3 \left(-6 t^3-7
   t^2+t+2\right)+Q^2 t \left(t^3+13 t^2+11 t+3\right)-Q t \left(t^3+3 t^2+4
   t+2\right)+t^4\right)}{q(Q-1) t (q t-1) (Q-t) (q Q-t) \left(q Q-t^2\right)}\\
& \textstyle \qquad +\frac{-q^2 t^2 \left(Q^4-Q^3 \left(2 t^3+4 t^2+3 t+1\right)+Q^2
   \left(3 t^3+11 t^2+13 t+1\right)+Q t \left(2 t^3+t^2-7 t-6\right)+t^2\right)-q
   Q t^4 \left(Q^2 (t+1)+Q \left(t^2-5 t-5\right)-t^3-t^2+2 t+2\right)-Q^2 t^6}
{q(Q-1) t (q t-1) (Q-t) (q Q-t) \left(q Q-t^2\right)}.
\end{align}

Representation Matrix of $\mathcal{R}$ in the basis of bosons $\ket{a_{\vl}}$ 
at level 2:
\begin{equation}
\left(
\begin{array}{ccccc}
r_{11} & -\frac{(q-1) q
   (Q-1) Q (q-t) t (t+1)}{2 (q Q-t) \left(q^2 Q-t\right) \left(q Q-t^2\right)} &
   \frac{(q-1) (Q-1) Q (q-t) \sqrt{\frac{q}{t}} t^2 (t+1)}{2 (q Q-t) \left(q^2
   Q-t\right) \left(q Q-t^2\right)} & 
r_{14}&
   -\frac{(q-1) (Q-1) Q (q-t) t^2 (t+1)}{2 (q Q-t) \left(q^2 Q-t\right) \left(q
   Q-t^2\right)} \\
 -\frac{q (q+1) (Q-1) Q (q-t) (t-1) t}{2 (q Q-t) \left(q^2 Q-t\right) \left(q
   Q-t^2\right)} & r_{22}& \frac{(Q-1) Q (q-t) \sqrt{\frac{q}{t}} t \left(2 Q q^2-t^2 q+t
   q-t^2-t\right)}{2 (q Q-t) \left(q^2 Q-t\right) \left(q Q-t^2\right)} &
   \frac{(q+1) (Q-1) Q (q-t) (t-1) t^2}{2 (q Q-t) \left(q^2 Q-t\right) \left(q
   Q-t^2\right)} & r_{25} \\
 \frac{q^2 (q+1) (Q-1) Q (q-t) (t-1)}{(q Q-t) \left(q^2 Q-t\right)
   \sqrt{\frac{q}{t}} \left(q Q-t^2\right)} & \frac{q (Q-1) (q-t) \left(Q q^2+Q t
   q^2+Q q-Q t q-2 t^2\right)}{(q Q-t) \left(q^2 Q-t\right) \sqrt{\frac{q}{t}}
   \left(q Q-t^2\right)} & r_{33}& -\frac{q
   (q+1) (Q-1) Q (q-t) (t-1) t}{(q Q-t) \left(q^2 Q-t\right) \sqrt{\frac{q}{t}}
   \left(q Q-t^2\right)} & \frac{q (Q-1) Q (q-t) \left(2 Q q^2-t^2 q+t
   q-t^2-t\right)}{(q Q-t) \left(q^2 Q-t\right) \sqrt{\frac{q}{t}} \left(q
   Q-t^2\right)} \\
r_{41}& \frac{(q-1) q^2 (Q-1) Q (q-t) (t+1)}{2 (q
   Q-t) \left(q^2 Q-t\right) \left(q Q-t^2\right)} & -\frac{(q-1) q (Q-1) Q (q-t)
   \sqrt{\frac{q}{t}} t (t+1)}{2 (q Q-t) \left(q^2 Q-t\right) \left(q
   Q-t^2\right)} & 
r_{44}& \frac{(q-1) q (Q-1) Q (q-t) t (t+1)}{2 (q Q-t) \left(q^2
   Q-t\right) \left(q Q-t^2\right)} \\
 -\frac{q^2 (q+1) (Q-1) Q (q-t) (t-1)}{2 (q Q-t) \left(q^2 Q-t\right) \left(q
   Q-t^2\right)} &
 r_{52}& \frac{(Q-1) (q-t)
   \sqrt{\frac{q}{t}} t \left(Q q^2+Q t q^2+Q q-Q t q-2 t^2\right)}{2 (q Q-t)
   \left(q^2 Q-t\right) \left(q Q-t^2\right)} & \frac{q (q+1) (Q-1) Q (q-t) (t-1)
   t}{2 (q Q-t) \left(q^2 Q-t\right) \left(q Q-t^2\right)} & r_{55} \\
\end{array}
\right),
\end{equation}
\begin{align}
&r_{11}= -\frac{q (Q-1) t \left(-2 Q^2 q^2-Q q^2+Q t q^2+Q t^2 q+Q q+2 Q t q-Q t^2-2 t^2+Q
   t\right)}{2 (q Q-t) \left(q^2 Q-t\right) \left(q Q-t^2\right)}, \\
&r_{41}= \frac{(q-t) \left(Q^2 q^3+Q q^3+Q^2 t q^3-Q t q^3+Q^2 q^2-2 Q t^2 q^2-Q q^2+Q^2 t
   q^2-Q t q^2-2 Q t^2 q+2 t^2 q-2 Q t q+2 t^3\right)}{2 (q Q-t) \left(q^2
   Q-t\right) \left(q Q-t^2\right)}, \\
&r_{22}=-\frac{q (Q-1) t \left(-2 Q^2 q^2+Q q^2+Q t q^2+Q t^2 q+Q q-2 Q
   t q+Q t^2-2 t^2+Q t\right)}{2 (q Q-t) \left(q^2 Q-t\right) \left(q
   Q-t^2\right)}, \\
&r_{52}=-\frac{(q-t) \left(-Q^2 q^3-Q q^3+Q^2 t q^3-Q t q^3+Q^2 q^2+2 Q
   t^2 q^2-Q q^2-Q^2 t q^2+Q t q^2-2 Q t^2 q+2 t^2 q+2 Q t q-2 t^3\right)}{2 (q
   Q-t) \left(q^2 Q-t\right) \left(q Q-t^2\right)}, \\
&r_{33}=\frac{Q^2 q^4-Q^2 t^2 q^3+Q^3 t q^3-3 Q^2 t q^3+Q t
   q^3+Q t^3 q^2+2 Q^2 t^2 q^2-2 Q t^2 q^2-Q^2 t q^2-Q^2 t^3 q+3 Q t^3 q-t^3 q+Q
   t^2 q-Q t^4}{(q Q-t) \left(q^2 Q-t\right) \left(q Q-t^2\right)}, \\
&r_{14}=\frac{Q (q-t) \left(2 Q^2 q^3-2 Q t^2 q^2+2
   Q^2 t q^2-2 Q t q^2-Q t^3 q+t^3 q-Q t^2 q+t^2 q-2 Q t q+Q t^3+t^3-Q
   t^2+t^2\right)}{2 (q Q-t) \left(q^2 Q-t\right) \left(q Q-t^2\right)}, \\
&r_{44}=-\frac{q (Q-1) t \left(-2 Q^2 q^2-Q q^2+Q t q^2+Q t^2 q+Q q+2 Q
   t q-Q t^2-2 t^2+Q t\right)}{2 (q Q-t) \left(q^2 Q-t\right) \left(q
   Q-t^2\right)}, \\
&r_{25}=\frac{Q (q-t) \left(2 Q^2 q^3-2 Q t^2 q^2-2 Q^2 t q^2+2 Q t
   q^2+Q t^3 q-t^3 q-Q t^2 q+t^2 q-2 Q t q+Q t^3+t^3+Q t^2-t^2\right)}{2 (q Q-t)
   \left(q^2 Q-t\right) \left(q Q-t^2\right)}, \\
&r_{55}=-\frac{q (Q-1) t
   \left(-2 Q^2 q^2+Q q^2+Q t q^2+Q t^2 q+Q q-2 Q t q+Q t^2-2 t^2+Q t\right)}{2 (qQ-t) \left(q^2 Q-t\right) \left(q Q-t^2\right)},
\end{align}
where $Q=\frac{u_1}{u_2}$.

}
\end{landscape}

\providecommand{\href}[2]{#2}\begingroup\raggedright\endgroup


\end{document}